\documentclass[prd,amsmath,amssymb,floatfix,superscriptaddress,notitlepage,nofootinbib,preprintnumbers]{revtex4-1}
\usepackage{amsfonts,amssymb,amsmath,graphicx,color,bm,enumitem}
\usepackage[utf8]{inputenc}
\usepackage{amsthm}
\usepackage[english]{babel}
\definecolor{ultramarine}{rgb}{0.07, 0.04, 0.56}
\definecolor{cadmiumgreen}{rgb}{0.0, 0.42, 0.24}
\definecolor{indigo(dye)}{rgb}{0.0, 0.25, 0.42}
\usepackage[linktocpage=true]{hyperref}
\hypersetup{
colorlinks=true,
citecolor=blue,
linkcolor=red,
urlcolor=indigo(dye),
}

\newcommand{\be}{\begin{equation}}  
\newcommand{\ee}{\end{equation}}
\newcommand{\bem}{\begin{pmatrix}}
\newcommand{\eem}{\end{pmatrix}}

\newcommand{\F}{\mathcal{F}}
\newcommand{\G}{\mathcal{G}}

\renewcommand{\P}{\mathcal{P}}
\newcommand{\Q}{\mathcal{Q}}
\newcommand{\R}{\mathcal{R}}
\newcommand{\W}{\mathcal{W}}
\newcommand{\D}{\mathcal{D}}

\begin{document}

\title{Disformal transformation of stationary and axisymmetric solutions in modified gravity}

\author{Masato Minamitsuji}
\affiliation{Centro de Astrof\'{\i}sica e Gravita\c c\~ao  - CENTRA, Departamento de F\'{\i}sica, Instituto Superior T\'ecnico - IST,
Universidade de Lisboa - UL, Av. Rovisco Pais 1, 1049-001 Lisboa, Portugal}

\begin{abstract}
The extended scalar-tensor and vector-tensor theories admit black hole solutions with the nontrivial profiles of the scalar and vector fields, respectively. The disformal transformation maps a solution in a class of the scalar-tensor or vector-tensor theories to that in another class, and hence it can be a useful tool to construct a new nontrivial solution from the known one. First, we investigate how the stationary and axisymmetric solutions in the vector-tensor theories without and with the $U(1)$ gauge symmetry are disformally transformed. We start from a stationary and axisymmetric solution satisfying the circularity conditions, and show that in both the cases the metric of the disformed solution in general does not satisfy the circularity conditions. Using the fact that a solution in a class of the vector-tensor theories with the vanishing field strength is mapped to that in a class of the shift-symmetric scalar-tensor theories, we derive the disformed stationary and axisymmetric solutions in a class of these theories, and show that the metric of the disformed solutions does not satisfy the circularity conditions if the scalar field depends on the time or azimuthal coordinate. We also confirm that in the scalar-tensor theories without the shift symmetry, the disformed stationary and axisymmetric solutions satisfy the circularity conditions. Second, we investigate the disformal transformations of the stationary and axisymmetric black hole solutions in the generalized Proca theory with the nonminimal coupling to the Einstein tensor, the shift-symmetric scalar-tensor theory with the nonminimal derivative coupling to the Einstein tensor, the Einstein-Maxwell theory, and the Einstein-conformally coupled scalar field theory. We show that the disformal transformations modify the causal properties of the spacetime, in terms of the number of horizons, the position of the black hole event horizon, and the appearance of the singularities at the finite radii.
\end{abstract}
\maketitle

\section{Introduction}
\label{sec1}

Black hole solutions have played very important role in modern astrophysics, cosmology,
and high energy physics.
The first black hole solutions found in general relativity
under the assumption of the static and spherically symmetric spacetime
are the Schwarzschild solution \cite{Sch} in the vacuum
and 
the Reissner-Nordstr\"{o}m (RN) solution \cite{RN1,RN3}
in the presence of the electric/ magnetic fields.
The rotating black hole solutions in general relativity 
obtained under the assumption of the stationary and axisymmetric spacetime
are
the Kerr solution \cite{Kerr} in the vacuum 
and 
the Kerr-Newman solution \cite{Newman}
in the presence of the electric/magnetic fields.
In general relativity,
the uniqueness of these black hole solutions under the given symmetries
has been proven \cite{uni1,uni2,uni3},
and hence 
the spacetime geometry is uniquely specified
only by the three parameters,
i.e.,
the mass, the angular momentum, and the electric/magnetic charge.

Black hole solutions exist
not only in general relativity 
but also in the other gravitational theories \cite{Herdeiro:2015waa}.
Many gravitational theories share the same black hole solutions with general relativity.
In several classes of the scalar-tensor and vector-tensor theories under the certain conditions,
the black hole no-hair theorems have been proven
in Refs.~\cite{nh1,nh2,nh3,nh4,nh5,nh6,nh7,nh8,nh9,nh10},
stating that
the Schwarzschild or Kerr solution with the trivial profile of the scalar or vector field
is the unique vacuum black hole solution
in these classes.
On the other hand,
black hole solutions
with the nontrivial profiles of the scalar and vector fields
have been found 
in the scalar-tensor and vector-tensor theories which evade the conditions for the no-hair theorems.
The examples of the nontrivial black hole solutions
are
the Bocharova-Bronnikov-Melnikov-Bekenstein (BBMB) solution in the Einstein-conformally coupled scalar field theory
\cite{bbmb1,bbmb2},
the hairy black hole solutions in the Einstein-scalar-Gauss-Bonnet theory 
\cite{edgb1,edgb2,edgb3,edgb4,edgb5,edgb6},
the stealth black hole solutions
in the shift-symmetric scalar-tensor theories
\cite{Babichev:2013cya,Appleby:2015ysa,Babichev:2016rlq,Rinaldi:2012vy,Anabalon:2013oea,Minamitsuji:2013ura,bhbh1,bhbh2,dhostbh1,dhostbh2,dhostbh3,dhostbh4,dhostbh5,sz1,sz2}
and
the generalized Proca theories
\cite{Chagoya:2016aar,Minamitsuji:2016ydr,Babichev:2017rti,Heisenberg:2017xda,Heisenberg:2017hwb}.
Beyond the assumption of the static and spherically symmetric solutions, 
the nontrivial
stationary and axisymmetric black hole solutions  
have been explored 
within the Hartle-Thorne slow rotation approximation
(see Refs. \cite{Hartle:1967he,Hartle:1968si})
in Refs. \cite{edgb7,edgb4,Maselli:2015yva,Chagoya:2016aar,Minamitsuji:2016ydr}
and 
by the direct approaches 
in Refs. \cite{edgb8,edgb9,dhostbh6,dhostbh7,dhostbh8,dhostbh9,sz3,kh1,kh2}.
In this paper, we will investigate
how the stationary and axisymmetric solutions are mapped under the disformal transformation 
and 
what are the general properties of the stationary and axisymmetric solutions
obtained after the disformal transformation
\cite{Bekenstein:1992pj,Zumalacarregui:2013pma,Liberati:2013xla,Kimura:2016rzw,Domenech:2018vqj,Gumrukcuoglu:2019ebp,DeFelice:2019hxb}.

The frame transformation is
a useful tool to generate the new nontrivial black hole solutions 
in the gravitational theories beyond general relativity.
In the general gravitational theories 
including the scalar fields $\phi_{(a)}$ and the vector fields $A_{(b)\mu}$,
where
the indices $\left(\mu,\nu,\cdots\right)$ run the four-dimensional spacetime
and the indices $(a)$ and $(b)$ run all the species of the scalar and vector fields,
respectively,
the general transformation
from one frame to another is given by 
\begin{eqnarray}
\label{disformal0}
{\tilde g}_{\mu\nu}
= \Phi_{\mu\nu}{}^{\alpha\beta}
\left[g_{\rho\sigma},\phi_{(a)},A_{(b)\rho}\right] 
\times
g_{\alpha\beta},
\end{eqnarray}
where
$\Phi_{\mu\nu}{}^{\alpha\beta}\left[g_{\rho\sigma},\phi_{(a)},A_{(b)\rho}\right]$
represents any covariant tensor constructed 
by the metric $g_{\mu\nu}$, the fields $\phi_{(a)}$ and $A_{(b)\mu}$, and their derivatives
(e.g., the field strengths $F_{(b)\mu\nu}:=\partial_\mu A_{(b)\nu}-\partial_\nu A_{(b)\mu}$),
which is symmetric with respect to the exchanges of 
the spacetime indices
$(\mu \leftrightarrow\nu)$ and $(\alpha \leftrightarrow \beta)$.
In the case that 
the transformation $\Phi_{\mu\nu}{}^{\alpha\beta}\left[g_{\rho\sigma},\phi_{(a)},A_{(b)\rho}\right]$
is invertible, 
namely,
there exists the inverse transformation such that 
\begin{eqnarray}
\label{disformal00}
g_{\mu\nu}
= \left(\Phi^{-1}\right)_{\mu\nu}{}^{\alpha\beta}
\left[{\tilde g}_{\rho\sigma},\phi_{(a)},A_{(b)\rho}\right] 
\times
{\tilde g}_{\alpha\beta},
\end{eqnarray}
a known solution $\left({g}_{\mu\nu},\phi_{(a)},A_{(b)\mu}\right)$ 
in a class of the theories ${S}\left[{g}_{\mu\nu},\phi_{(a)},A_{(b)\mu}\right]$
is uniquely mapped to a solution $\left({\tilde g}_{\mu\nu},\phi_{(a)},A_{(b)\mu}\right)$ 
in a class of the new theories ${\tilde S}\left[{\tilde g}_{\mu\nu},\phi_{(a)},A_{(b)\mu}\right]$,
obtained by 
\begin{eqnarray}
\label{rel_action}
{\tilde S}
\left[
{\tilde g}_{\mu\nu},\phi_{(a)},A_{(b)\mu}
\right]
=
S
\left[
g_{\mu\nu} 
\left(
{\tilde g}_{\rho\sigma},
\phi_{(a)},
A_{(b)\rho}
\right),
\phi_{(a)},
A_{(b)\mu}
\right].
\end{eqnarray}
In the vacuum spacetime, 
the two theories $S$ and $\tilde{S}$ are equivalent in the
sense that they share the equivalent space of the solutions.
However, 
in the presence of the matter sector minimally coupled to the metrics
$g_{\mu\nu}$ in the action $S$ and ${\tilde g}_{\mu\nu}$ in the action ${\tilde S}$,
respectively,
the two theories
$S$ and ${\tilde S}$ are no longer physically equivalent.
In this paper, 
although we will neglect the contribution of matter for obtaining the solutions,
we will assume the existence of the matter sector and 
regard all the metrics related by the general transformation \eqref{disformal0}
as the physically independent ones.
Since
we assume that 
matter is minimally coupled to the metric in each frame,
the two frames can be distinguished observationally;
e.g., 
the trajectories of particles freely falling into a black hole are different.

The disformal transformations
\cite{Bekenstein:1992pj,Zumalacarregui:2013pma,Liberati:2013xla,Kimura:2016rzw,Domenech:2018vqj,Kimura:2016rzw,Gumrukcuoglu:2019ebp,DeFelice:2019hxb}
(see Eqs. \eqref{disformal}, \eqref{disformal3}, and \eqref{disformal2})
map a class of the scalar-tensor or vector-tensor theories
without the Ostrogradsky ghosts \cite{Woodard:2015zca}
to another class,
without any change of the number of the degrees of freedom.
The conventional scalar-tensor theories
are mapped 
to 
the degenerate higher-order scalar-tensor
(DHOST) theories \cite{Langlois:2015cwa,Achour:2016rkg,BenAchour:2016fzp,Langlois:2018dxi},
and
similarly 
the conventional vector-tensor  theories
without/ with the $U(1)$ gauge symmetry
are also mapped
to the extended vector-tensor theories
without the Ostrogradsky ghosts
\cite{Kimura:2016rzw,Gumrukcuoglu:2019ebp,DeFelice:2019hxb}.
Refs. \cite{dhostbh6,dhostbh7}
analyzed 
how the Kerr solution 
with the nontrivial profile of the scalar field and the constant scalar kinetic term
is transformed 
under the scalar disformal transformation,
and showed that
the {\it disformed}~\footnote{
In this paper, we will call the solutions 
after the disformal transformations \eqref{disformal}, \eqref{disformal3}, and \eqref{disformal2} 
the disformed solutions.}
Kerr  solution
does not satisfy the circularity conditions
(see Appendix \ref{app1}).
The stationary and axisymmetric black hole solutions 
which violate the circularity conditions 
have also been obtained in the scalar-tensor theories with
the nonminimal couplings to the quadratic order curvature invariants~\cite{Nakashi:2020phm}.
In the present paper,
we will investigate
how a general stationary and axisymmetric solution
in a class of the vector-tensor and scalar-tensor theories
is disformally mapped to that in another class of them.

Here,
we will not explore the explicit form of the disformed theories ${\tilde S}$
shown in Eq. \eqref{rel_action}
and the concrete form of the coupling functions,
but rather focus on the general properties of the disformed stationary and axisymmetric solutions,
by imposing that the disformal transformation is invertible.
We would like to emphasize that 
as long as the disformal transformation is invertible,
there should be one-to-one correspondence
between the original theory 
and the theory obtained after the disformal transformation,
and they should share the equivalent solution space,
in the case that the contribution of matter is absent
\cite{Zumalacarregui:2013pma,Achour:2016rkg,Kimura:2016rzw,Gumrukcuoglu:2019ebp}.
Even in the case that the theory obtained via the disformal transformation
is apparently higher derivative
while the original theory gives rise to the second-order Euler-Lagrange equations,
the higher-derivative Euler-Lagrange equations in the disformed theory
should be degenerate
and reduce to the second-order system 
after the suitable redefinition of the variables.
Thus, the disformal transformation should not modify the number of the degrees of freedom,
and the two theories should be the two different mathematical descriptions of the equivalent theory,
in the case that the contribution of matter is absent.
In such a case, 
the order of the disformal transformation
and the variation of the action 
should be able to be exchanged,
and 
the disformed solutions
should satisfy the Euler-Lagrange equations 
obtained 
by varying the action of the theory 
derived via the disformal transformation
(see Appendix \ref{app3} for the explicit examples).
In other words,
we should be able to obtain the disformed solutions
without the derivation of the detailed form of the action of the theory 
derived via the disformal transformation.
If the future astrophysical observations detect any deviation from the predictions of general relativity
in the strong gravity regimes,
the type of the deviations indicated by the disformal transformation 
may allow us to test the existence of the scalar or vector fields
in the vicinity of the black hole.
Therefore,
in this paper,
instead of the derivation of the explicit action 
of the scalar-tensor and vector-tensor theories
obtained via the disformal transformations,
we will focus on the disformal transformations
at the level of the solutions.

First, we will derive the disformal transformation of the metric of 
the most general stationary and axisymmetric solutions satisfying the circularity conditions,
which ensure the existence of the two-surfaces orthogonal to 
the two commuting Killing fields
$\xi^\mu$ and $\sigma^\mu$ (see Appendix \ref{app2}),
and 
show that the disformed solutions do not necessarily satisfy the circularity conditions.
Using the fact that the solutions in a class of the vector-tensor theories with the vanishing field strength,
$F_{\mu\nu}=\partial_\mu A_\nu-\partial_\nu A_\mu=0$,
after the replacement of $A_\mu\to\partial_\mu \phi$ and the integration of it,
are mapped to those in the corresponding class of 
the shift-symmetric scalar-tensor theories with the scalar field $\phi$,
we will show that
the disformed stationary and axisymmetric solutions 
in the shift-symmetric scalar-tensor theories
do not satisfy the circularity conditions
in the case that the scalar field depends on the time or azimuthal coordinate.
We will then consider the disformal transformation of the vector-tensor theory 
with the $U(1)$ gauge symmetry, 
and find that the disformed solution satisfies the circularity conditions,
in the case that the azimuthal component of the magnetic field vanishes.
Second, we will investigate 
the disformal transformation of the slowly rotating black hole solutions
in the generalized Proca, the shift-symmetric scalar-tensor,
and the Einstein-conformally coupled scalar field theories,
and 
the disformal transformation of 
the Kerr-Newman solution with an arbitrary spin dimensionless parameter
in the Einstein-Maxwell theory,
which will be discussed separately in details.

The paper is organized as follows:
in Sec. \ref{sec2},
we review the general properties of the stationary and axisymmetric solutions
in the scalar-tensor and vector-tensor theories.
In Sec. \ref{sec3},
we investigate 
the properties of the disformed stationary and axisymmetric solutions.
In Secs. \ref{sec4}-\ref{sec6},
we derive the disformal transformation
of 
the stealth Schwarzschild,
the Schwarzschild-(anti-) de Sitter,
and 
the locally asymptotically anti-de Sitter black hole solutions,
respectively,
with the leading order slow rotation corrections
in the generalized Proca theory
with the nonminimal coupling of the vector field to the Einstein tensor
and the shift-symmetric scalar-tensor theory with the nonminimal derivative coupling
to the Einstein tensor,
and argue that the disformal transformation can modify
the causal properties of the spacetime
in terms of the position of the black hole event horizon and the number of the horizons.
In Sec. \ref{sec7},
we consider the disformal transformation 
of the Kerr-Newman solution with an arbitrary spin parameter
and discuss the general properties of the disformed Kerr-Newman solution
in terms of the position of the horizons and the existence of the singularities.
In Sec. \ref{sec8},
we derive disformal transformation
of the slowly rotating BBMB solution
in the Einstein-conformally coupled scalar field theories.
Sec. \ref{sec9} is devoted to giving a brief summary and conclusion.

\section{The stationary and axisymmetric solutions}
\label{sec2}

\subsection{The metric}

The most general stationary and axisymmetric spacetime
which possesses the two commuting Killing vector fields
$\xi^\mu=(\partial/\partial t)^\mu$
and 
$\sigma^\mu=(\partial/\partial \varphi)^\mu$,
satisfying 
$\nabla_{(\mu} \xi_{\nu)}=\nabla_{(\mu} \sigma_{\nu)}=0$
and
$\left[\xi,\sigma\right]^\nu
=\xi^\mu \partial_\mu \sigma^\nu-\sigma^\mu \partial_\mu \xi^\nu=0$,
is given by 
\begin{eqnarray}
\label{stationary0}
ds^2=g_{\mu\nu}(r,\theta)dx^\mu dx^\nu,
\end{eqnarray}
where 
$x^\mu= (t,r,\theta,\varphi)$ represents the time, radial, polar angular, and azimuthal angular coordinates, respectively.
The metric \eqref{stationary0} further reduces to the simpler form,
if the two Killing vector fields obey the integrability conditions
\cite{circ1,circ2,waldbook}
\begin{eqnarray}
\label{circularity2}
\sigma_{[\mu} \xi_{\nu}\nabla_{\alpha} \xi_{\beta]}=0,
\qquad 
\sigma_{[\mu} \xi_{\nu}\nabla_{\alpha} \sigma_{\beta]}=0.
\end{eqnarray}
Moreover, 
the conditions~\eqref{circularity2}
are satisfied in the case that the so-called circularity conditions 
\begin{eqnarray}
\label{circularity0}
\xi^{\mu} R_{\mu} {}^{[\nu} \xi^{\alpha} \sigma^{\beta]}=0,
\qquad 
\sigma^{\mu} R_{\mu} {}^{[\nu} \xi^{\alpha}  \sigma^{\beta]}=0,
\end{eqnarray}
are satisfied (see Appendix \ref{app2}),
where $R_{\mu\nu}$ is the Ricci tensor associated with the metric $g_{\mu\nu}$.
In the case that the circularity conditions are satisfied,
there exist the two-surfaces orthogonal to the two Killing vector fields $\xi^\mu$ and $\sigma^\mu$,
and then 
the general stationary and axisymmetric metric  \eqref{stationary0} 
reduces to the block-diagonal  form 
\begin{eqnarray}
\label{stationary_ast}
ds^2
= 
\left(
-A dt^2
+B d\varphi^2
+2 C dt d\varphi
\right)
+ 
\left(
D dr^2
+ E d\theta^2
+2F  drd\theta
\right),
\end{eqnarray}
where $A, B, C, D, E$, and $F$ are the functions of $r$ and $\theta$.
An appropriate coordinate transformation on the two-surfaces
allows us to set $F=0$,
then Eq. \eqref{stationary_ast} reduces to 
\begin{eqnarray}
\label{stationary}
ds^2
= 
\left(
-Adt^2
+B d\varphi^2
+2 C dt d\varphi
\right)
+\left(
 D dr^2
+ Ed\theta^2
\right).
\end{eqnarray}
In general relativity,
the vacuum spacetime where $R_{\mu\nu}=\lambda g_{\mu\nu}$ with $\lambda$ being constant
and
the spacetime filled with perfect fluids
with four-velocities spanned by $\xi^\mu$ and $\sigma^\mu$
satisfy the circularity conditions \eqref{circularity0}.
On the other hand, 
in the gravitational theories with the additional scalar or vector degrees of freedom,
the absence of matter
or
the perfect fluid form of the matter energy-momentum tensor
does not ensure the circularity conditions \eqref{circularity0}
and hence the integrability \eqref{circularity2}.

We assume that 
our starting gravitational theory possesses a 
stationary and axisymmetric solution given in the form of Eq. \eqref{stationary}.
In general relativity, 
it is well known that the Kerr solution is the unique, asymptotically flat,
stationary, and axisymmetric black hole solution in vacuum
\cite{uni1,uni2,uni3},
which can be written into the form of  Eq. \eqref{stationary}.
The parametrized `non-Kerr' metrics~\cite{Bambi:2011zs,Bambi:2012zg,Konoplya:2012df, Glampedakis:2005cf,Johannsen:2010xs,Johannsen:2010ru,Vigeland:2011ji,Medeiros:2019cde}
employed to test the Kerr hypothesis observationally
also belong to the form of Eq. \eqref{stationary}.
In the next section,
we will show that the disformal transformations,
such as Eqs. \eqref{disformal}, \eqref{disformal3}, and \eqref{disformal2} below,
map a stationary and axisymmetric metric solution
satisfying the circularity conditions \eqref{stationary}
to a noncircular one given by the general metric \eqref{stationary0}.

\subsection{In the vector-tensor theories}

For a given stationary and axisymmetric spacetime \eqref{stationary}
satisfying the circularity conditions \eqref{circularity0},
the most general Ansatz of the vector field is given by 
\begin{eqnarray}
\label{vector_Ansatz}
A_\mu dx^\mu
= A_t(r,\theta) dt + A_r (r,\theta) dr
+ A_\theta (r,\theta) d\theta 
+ A_\varphi (r,\theta) d\varphi,
\end{eqnarray}
for which the norm of the vector field can be written as 
\begin{eqnarray}
\label{y}
{\cal Y}
:=
g^{\mu\nu} A_\mu A_\nu
=
\frac{A_\varphi^2 A-A_t^2 B+2A_t A_\varphi  C}
            {AB+C^2}
+\frac{A_r^2}{D}
+\frac{A_\theta^2}{E}.
\end{eqnarray}
In the vector-tensor theories with the $U(1)$ gauge symmetry, 
there is an ambiguity about the gauge transformation $A_\mu \to A_\mu +\partial _\mu \chi$,
where $\chi$ is an arbitrary function of the spacetime coordinates $(r,\theta)$.

\subsection{In the shift-symmetric scalar-tensor theories}

A solution in a class of the vector-tensor theories with $F_{\mu\nu}=0$
can be mapped to that in a class of the shift-symmetric scalar-tensor theories
with the scalar field $\phi$,
via the substitution of $A_\mu\to \partial_\mu\phi$ and the integration of it \cite{Minamitsuji:2016ydr}.
The conditions $F_{\mu\nu}=0$ 
for the Ansatz \eqref{vector_Ansatz}
lead to 
\begin{eqnarray}
&&
\partial_r A_t=0,
\quad
\partial_\theta A_t=0,
\quad
\partial_r A_\theta-\partial_\theta A_r=0,
\quad
\partial_r A_\varphi
=0,
\quad
\partial_\theta A_\varphi=0,
\end{eqnarray}
which are integrated as 
\begin{eqnarray} 
\label{qm}
A_t= q,
\qquad
A_\varphi=m,
\qquad 
A_r= \partial_r \psi (r,\theta),
\qquad
A_\theta= \partial_\theta \psi (r,\theta),
\end{eqnarray}
where $q$ and $m$ are constants
and $\psi$ is the function of $(r,\theta)$.
Through the replacement of $A_\mu\to \partial_\mu \phi$
and its further integration,
we obtain the most general form of the scalar field 
in the stationary and axisymmetric spacetime \eqref{stationary}
in terms of the shift-symmetric scalar-tensor theories,
\begin{eqnarray}
\label{scalar_Ansatz}
\phi= \psi (r,\theta)+ q t+ m \varphi.
\end{eqnarray}
The norm of the vector field ${\cal Y}$
is then promoted to the kinetic term of the scalar field
\begin{eqnarray}
\label{kin}
{\cal X}
:=
g^{\mu\nu} \partial_\mu \phi \partial_\nu \phi
=
\frac{m^2 A-q^2B+2mq C}
            {AB+C^2}
+\frac{(\partial_r\psi)^2}{D}
+\frac{(\partial_\theta \psi)^2}{E}.
\end{eqnarray}

\subsection{In the scalar-tensor theories without the shift symmetry}

In the case of the scalar-tensor theories without the shift symmetry,
the explicit dependence on $t$ and $\varphi$ in Eq. \eqref{scalar_Ansatz}
contradicts with the stationarity and axisymmetry of the spacetime,
since the equations of motion explicitly would depend on $t$ and $\varphi$.
Hence the most general Ansatz of the scalar field 
in the stationary and axisymmetric spacetimes
is given by
\begin{eqnarray}
\label{scalar_Ansatz2}
\phi= \psi (r,\theta),
\end{eqnarray}
and hence,
${\cal X}
=(\partial_r\psi)^2/D+(\partial_\theta \psi)^2/E$.

\subsection{The slowly rotating solutions}

In the latter part of this paper,
we will also focus on the disformal transformation of 
the slowly rotating solutions
within the first order of the Hartle-Thorne approximation \cite{Hartle:1967he,Hartle:1968si}.
In this approach,
the metric functions $A$, $B$, $C$, $D$, and $E$ in Eq. \eqref{stationary}
are expanded 
with respect to the small parameter $\epsilon (\ll 1)$
which is of  the order of the dimensionless angular speed of the black hole rotation
$M\Omega\sim J/M^2 (\ll 1)$ with $M$ and $J$ being the mass
and the angular momentum of the black hole, respectively.
At ${\cal O} (\epsilon^0)$,
we prepare for the reference static and spherically symmetric solutions.
At ${\cal O} (\epsilon^1)$,
the diagonal components of the metric $A$, $B$, $D$, and $E$ in Eq. \eqref{stationary}
remain the same as those in the reference solutions and the pure function of $r$,
and the leading slow rotation corrections appear only in the function $C$.
Thus, 
the general metric Ansatz \eqref{stationary}
reduces to the form of 
\begin{eqnarray}
\label{stationary2}
ds^2
&=&
g_{\mu\nu}dx^\mu dx^\nu
= 
-A_{(0)}(r) dt^2
+ D_{(0)}(r) dr^2
+r^2 
\left(
d\theta^2
+
\sin^2\theta 
d\varphi^2
\right)
+2 C_{(1)} (r,\theta)
dt d\varphi
+
{\cal O} \left(\epsilon^2\right),
\end{eqnarray}
with
\begin{eqnarray}
\label{1stc}
C_{(1)}(r,\theta):= -r^2 \omega_{(1)} (r) \sin^2\theta,
\end{eqnarray}
where  $A_{(0)} (r)$ and $D_{(0)} (r)$ are of ${\cal O} (\epsilon^0)$,
and 
$\omega_{(1)}(r)={\cal O} (\epsilon)$ represents the frame-dragging function.
The slow rotation corrections to the diagonal components
$A$, $B$, $D$, and $E$
appear at ${\cal O}(\epsilon^2)$,
where we denote the $n$th order quantities of the slow rotation approximation  
by ${\cal O} (\epsilon^n)$.
Going beyond ${\cal O} (\epsilon)$
would be much more involved,
as one has to solve for all the components of the metric and matter fields.
Thus,  for simplicity, 
we will focus only on the leading order slow rotation corrections of ${\cal O} (\epsilon)$.

On the other hand,
the slow rotation corrections to $A_t$ and $A_r$ in Eq. \eqref{vector_Ansatz}
appear at ${\cal O} (\epsilon^2)$, 
and $A_\theta$ in Eq. \eqref{vector_Ansatz} is the quantity of ${\cal O} (\epsilon^2)$.
Thus, 
the general Ansatz of the vector field \eqref{vector_Ansatz}
reduces to 
\begin{eqnarray}
\label{vector2}
A_\mu dx^\mu
=
A_{t(0)} (r)dt 
+ A_{r(0)} (r)
 dr
+ 
A_{\varphi(1)} (r,\theta)
d\varphi
+{\cal O} \left(\epsilon^2\right),
\end{eqnarray}
with 
\begin{eqnarray}
\label{1stat}
A_{\varphi(1)}
(r,\theta):= r^2 a_{3(1)}(r) \sin^2\theta, 
\end{eqnarray}
where  $A_{t(0)} (r)$ and $A_{r(0)} (r)$ are of ${\cal O} (\epsilon^0)$,
and $a_{3(1)}(r)$, 
which is of ${\cal O} (\epsilon)$,
represents the radial profile of the magnetic component of the vector field
\cite{Chagoya:2016aar,Minamitsuji:2016ydr}.

In order to obtain a slowly rotating solution in a class of the vector-tensor theories, 
we substitute the Ansatz of Eqs.~\eqref{stationary2} and \eqref{vector2}
into the action
and then expand it with respect to $\epsilon$.
At ${\cal O}(\epsilon^0)$,
the variations of the action with respect to
$A_{(0)}(r)$, $D_{(0)}(r)$, $A_{t(0)}(r)$, and $A_{r(0)}(r)$
yield the Euler-Lagrange
equations 
which give rise to the reference static and spherically symmetric solutions.
Then, 
at ${\cal O} (\epsilon^2)$,
the variations with respect to $C_{(1)}$ and $A_{\varphi(1)}$
provide the coupled equations,
which after the substitution of Eqs.~\eqref{1stc} and \eqref{1stat}
reduce to the equations to determine $\omega_{(1)}(r)$ and $a_{3(1)}(r)$.

A slowly rotating solution in a class of the vector-tensor theories
with $F_{\mu\nu}=0$ can be mapped to that in a class of the shift-symmetric scalar-tensor theories.
Up to ${\cal O} (\epsilon)$,
the conditions $F_{\mu\nu}=0$ for the Ansatz \eqref{vector2}
lead to
\begin{eqnarray}
&&
\partial_r A_{t(0)}=0,
\quad
a_{3(1)} =0.
\end{eqnarray}
After the replacement of $A_\mu\to \partial_\mu \phi$ and 
the integration of it,
the Ansatz for the scalar field in the slow rotation limit 
of the stationary and axisymmetric solutions 
in the shift-symmetric scalar-tensor theories
is obtained as \cite{Maselli:2015yva}.
\begin{eqnarray} 
\label{qt}
\phi= qt+\psi_{(0)}(r)+{\cal O}(\epsilon^2).
\end{eqnarray}
In the scalar-tensor theories without the shift symmetry,
we have to set $q=0$,
otherwise the equations of motion would explicitly depend on the time.
Thus, 
up to ${\cal O} (\epsilon)$,
the scalar field remains the same as that in 
the static and spherically symmetric black hole solutions
\cite{Babichev:2013cya,Appleby:2015ysa,Maselli:2015yva}.

In order to obtain a slowly rotating solution in the scalar-tensor theories, 
we substitute the Ansatze \eqref{stationary2} and \eqref{qt} into the action
and follow the same way as in the case of the vector-tensor theories.
At ${\cal O} (\epsilon^2)$,
the variation with respect to $C_{(1)}$ provides the equation for it,
which after the substitution of Eq.~\eqref{1stc} 
reduces to the equation to determine $\omega_{(1)}(r)$.

\section{The disformal transformation of stationary and axisymmetric solutions}
\label{sec3}

\subsection{The vector disformal transformation without the $U(1)$ gauge symmetry}

We consider the disformal transformation in a vector-tensor theory
\begin{eqnarray}
\label{disformal}
{\tilde g}_{\mu\nu}
= \P g_{\mu\nu}
 +\Q A_\mu A_\nu,
\end{eqnarray}
which explicitly breaks the $U(1)$ gauge invariance,
where we assume that the dependence on $x^\mu$
in the functions $\P$ and $\Q$
comes through the norm of the vector field ${\cal Y}$
(see Eq.~\eqref{y}),
$\P:=\P({\cal Y} )$
and 
$\Q:=\Q({\cal Y})$,
so that the disformally transformed theory is covariant, 
where
the disformal transformation \eqref{disformal}
relates a class of
degenerate higher-order vector-tensor theories
to another \cite{Kimura:2016rzw}.
\footnote{
In principle, 
we may consider more general disformal transformations
which contain both $A_\mu A_\nu$, $g^{\alpha\beta}F_{\mu\alpha}F_{\nu\beta}$,
and the dependence on ${\cal Y}$ and ${\cal F}$ into $\P$ and $\Q$ in Eq. \eqref{disformal}.
Here, 
for simplicity,
for the vector-tensor theories without the $U(1)$ gauge symmetry,
we will focus on the transformation in the form of Eq. \eqref{disformal}
which minimally breaks the $U(1)$ gauge symmetry.}
The inverse disformally mapped metric of Eq. \eqref{disformal}
is given by 
\begin{eqnarray}
{\tilde g}^{\mu\nu}
= 
\frac{1}{\P}
\left(
g^{\mu\nu}
-\frac{\Q}
        {\P+{\cal Y}\Q} 
  A^\mu A^\nu
\right).
\end{eqnarray}
We choose the functions $\P$ and $\Q$ in Eq. \eqref{disformal},
so that the disformal transformation \eqref{disformal} is invertible,
namely,
$\P+ {\cal Y} \Q\neq 0$ and $\P\neq 0$ \cite{Kimura:2016rzw}.
As long as the disformal transformation \eqref{disformal}
is invertible,
any known solution $\left( {g}_{\mu\nu},A_\mu\right)$ 
in a class of the vector-tensor theories ${S}\left[{g}_{\mu\nu},A_\mu\right]$
is uniquely mapped to a solution $\left(g_{\mu\nu},A_\mu\right)$ 
in a new class of the vector-tensor theories given by 
${\tilde S}
\left[
{\tilde g}_{\mu\nu}, A_\mu
\right]
=
S
\left[
g_{\mu\nu}\left({\tilde g}_{\rho\sigma},A_\rho\right), A_\mu
\right]$.
The norm of the vector field for the disformed metric ${\tilde g}_{\mu\nu}$
is given by 
\begin{eqnarray}
{\tilde {\cal Y}}
:={\tilde g}^{\mu\nu}
A_\mu 
A_\nu
=
\frac{{\cal Y}}
       {\P+{\cal Y}\Q}.
\end{eqnarray}
In the case that ${\cal Y}={\cal Y}_0={\rm const}$, 
$\P({\cal Y}_0)$, $\Q({\cal Y}_0)$,
and ${\tilde {\cal Y}}_0:=\frac{ {\cal Y}_0}{\P ({\cal Y}_0)+{\cal Y}_0 \Q({\cal Y}_0)}$
are also constants.

Following Eq. \eqref{disformal},
we find that 
the stationary and axisymmetric spacetime metric \eqref{stationary}
is disformally mapped to 
\begin{eqnarray}
\label{vector_disformed}
d{\tilde s}^2
:=
{\tilde g}_{\mu\nu}
dx^\mu dx^\nu
&=&
- \left(A\P -\Q A_t^2\right)dt^2
+\left(\P B + \Q A_\varphi ^2\right)  d\varphi^2
+2
\left(
\P C+ \Q A_t A_\varphi
\right)
dt d\varphi
\nonumber\\
&+&
\left(
\P D
+ \Q A_r^2
\right)
dr^2
+ 
\left(
\P E
+ \Q A_\theta^2 
\right)
d\theta^2
+ 
2 \Q 
   A_r A_\theta dr d\theta
\nonumber\\
&+& 
2 \Q 
\left(
 A_t A_r dt dr
+ A_t  A_\theta dt d\theta
+ A_\varphi  A_r dr d\varphi
+ A_\varphi A_\theta d\theta d\varphi
\right),
\end{eqnarray}
where each component in the last line violates
the circularity conditions \eqref{circularity0}.
Imposing them leads to 
\begin{eqnarray}
\label{disform_circularity}
A_t A_r =0,
\qquad 
 A_t  A_\theta=0,
\qquad
A_\varphi  A_r =0,
\qquad 
A_\varphi A_\theta =0,
\end{eqnarray}
which further can be solved as 
\begin{eqnarray}
\label{circularity_case1}
A_t=A_\varphi=0,
\qquad 
{\rm or}
\qquad 
A_r=A_\theta=0.
\end{eqnarray}
In the former and latter cases,
the metric of Eq. \eqref{vector_disformed} reduces to 
\begin{eqnarray}
\label{former}
d{\tilde s}^2
&=&
- A\P dt^2
+\P B  d\varphi^2
+2
\P C
dt d\varphi
+
\left(
\P D
+ \Q A_r^2
\right)
dr^2
+
\left(
\P E
+ \Q A_\theta^2 
\right)
d\theta^2
+ 
2 \Q 
   A_r A_\theta dr d\theta,
\end{eqnarray}
and 
\begin{eqnarray}
\label{latter}
d{\tilde s}^2
&=&
- \left(A\P -\Q A_t^2\right)dt^2
+\left(\P B + \Q A_\varphi ^2\right)  d\varphi^2
+2
\left(
\P C+ \Q A_t A_\varphi
\right)
dt d\varphi
+
\P D
dr^2
+
\P E
d\theta^2.
\end{eqnarray}
The metrics \eqref{former} and \eqref{latter} belong to
the class of  Eqs. \eqref{stationary_ast} and \eqref{stationary},
respectively.

After the disformal transformation \eqref{disformal},
the metric of a slowly rotating solution \eqref{stationary2}
is given by 
\begin{eqnarray}
\label{ori}
d{\tilde s}^2
=
{\tilde g}_{\mu\nu} dx^\mu dx^\nu
&=&
-
\left(
\P_{(0)} A_{(0)}-\Q_{(0)} A_{t(0)}^2
\right)
dt^2
+ (\P_{(0)} D_{(0)}+ \Q_{(0)} A_{r(0)}^2)dr^2
+2\Q_{(0)} A_{t(0)} A_{r(0)} dt dr
\nonumber\\
&+&r^2 \P_{(0)}
\left(
d\theta^2
+
\sin^2\theta 
d\varphi^2
\right)
-
2r^2
\sin^2\theta
\left[ 
\left(
\P_{(0)}
\omega_{(1)}
- \Q_{(0)} a_{3(1)} A_{t(0)}
\right)
dt
-
\Q_{(0)} a_{3(1)}  A_{r(0)}
 dr
\right]
d\varphi
\nonumber\\
&+&
{\cal O} (\epsilon^2),
\end{eqnarray}
where 
since ${\cal Y}={\cal Y}_{(0)}(r)+{\cal O} (\epsilon^2)$
with ${\cal Y}_{(0)}(r):= A_{r(0)}^2/D_{(0)}-A_{t(0)}^2 /A_{(0)}$,
\begin{eqnarray}
\P ({\cal Y})
&=& \P ({\cal Y}_{(0)}(r)) +{\cal O} (\epsilon^2)
=:\P_{(0)} (r) +{\cal O} (\epsilon^2),
\nonumber\\
\Q ({\cal Y})
&=&
\Q ({\cal Y}_{(0)}(r)) +{\cal O} (\epsilon^2)
= :\Q_{(0)} (r) +{\cal O} (\epsilon^2).
\end{eqnarray}
Introducing the new time coordinate by 
\begin{eqnarray}
\label{tildet}
dt=d{\tilde t}
+\frac{\Q_{(0)} A_{t(0)} A_{r(0)}}
          {\P_{(0)} A_{(0)}-\Q_{(0)} A_{t(0)}^2}
dr,
\end{eqnarray}
the disformed metric \eqref{ori} is rewritten as 
\begin{eqnarray}
\label{trans}
d{\tilde s}^2
&=&
-\left(
\P_{(0)} A_{(0)}-\Q_{(0)}A_{t(0)}^2
\right)
d{\tilde t}^2
+
\P_{(0)}
\left[
D_{(0)}+\frac{\Q_{(0)} A_{(0)} A_{r(0)}^2}
            {\P_{(0)} A_{(0)}-\Q_{(0)} A_{t(0)}^2}
\right]dr^2
+r^2\P_{(0)}
\left(
d\theta^2
+
\sin^2\theta 
d\varphi^2
\right)
\nonumber\\
&-&
2r^2 
\sin^2\theta
\left[
\left(
\P_{(0)}
\omega_{(1)}- \Q_{(0)} a_{3(1)} A_{t(0)}
\right)
d{\tilde t}
-\P_{(0)} \Q_{(0)}  A_{r(0)} 
\frac{A_{(0)} a_{3(1)}-\omega_{(1)} A_{t(0)}}
        {\P_{(0)} A_{(0)}-\Q_{(0)} A_{t(0)}^2}
dr
\right]
d\varphi
+{\cal O} (\epsilon^2).
\end{eqnarray}
The existence of the nonzero
${\tilde g}_{r\varphi}\neq 0$ in Eqs. \eqref{ori} and \eqref{trans}
indicates the violation of the circularity conditions,
which explicitly appears at ${\cal O}(\epsilon^2)$.
Up to ${\cal O} (\epsilon)$,
the position of the black hole event horizon in the disformed frame
is determined by a positive root of $\P_{(0)} A_{(0)}-\Q_{(0)} A_{t(0)}^2=0$.
The apparent divergence of  $g_{r\varphi}$ in Eq. \eqref{trans}
at the point where $\P_{(0)}A_{(0)}-\Q_{(0)} A_{t(0)}^2=0$
is just a coordinate artifact.

In the case that
the reference static and spherically symmetric solution
is given by a black hole spacetime 
and
has an event horizon at which $A_{(0)}= D_{(0)}^{-1}=0$,
after the disformal transformation,
the position of the event horizon
is modified 
to the position 
which is fixed by the algebraic equation
 $\P_{(0)}A_{(0)}-\Q_{(0)} A_{t(0)}^2=0$.
Since the number of the positive roots 
of the equation $\P_{(0)}A_{(0)}-\Q_{(0)} A_{t(0)}^2=0$
may be different from 
that of the original equation $A_{(0)}= D_{(0)}^{-1}=0$,
the number of the horizons may be modified
before and after the disformal transformation.
The shift of the position of the event horizon 
happens
in the case that the character the vector field 
is timelike 
with nonzero $A_{t(0)}\neq 0$,
which will be discussed in Secs. \ref{sec4} and \ref{sec5}.

The solution of the vector field 
rewritten in terms of the time coordinate
defined in the disformed frame \eqref{tildet}
is given by 
\begin{eqnarray}
\label{vector_new}
A_\mu dx^\mu
=
A_{t(0)}d{\tilde t}
+\frac{\P_{(0)} A_{(0)}}
         {\P_{(0)}A_{(0)}-\Q_{(0)}  A_{t(0)}^2}
A_{r(0)}
dr
+{\cal O} (\epsilon),
\end{eqnarray}
which will be employed 
to investigate the regularity of the vector field 
at the event and cosmological horizons
of the disformed solutions.

\subsection{The vector disformal transformation with the $U(1)$ gauge symmetry}
\label{sec32}

When we consider the vector-tensor theories with the $U(1)$ gauge symmetry,
instead of the vector disformal transformation \eqref{disformal},
we consider the disformal transformation with the $U(1)$ gauge symmetry given by
\begin{eqnarray}
\label{disformal3}
{\tilde g}_{\mu\nu}
= \R g_{\mu\nu}
+ \W g^{\alpha \beta}  F_{\mu\alpha} F_{\nu\beta},
\end{eqnarray}
where $\R=\R(\F,\G)$ and $\W=\W (\F,\G)$ with 
\begin{eqnarray}
\label{calf_calg}
\F:=
g^{\mu\rho}
g^{\nu\sigma} 
F_{\mu\nu} 
F_{\rho\sigma},
\qquad
\G:= 
g^{\mu\nu}
g^{\alpha\beta}
g^{\rho\sigma}
g^{\gamma\delta}
F_{\mu\alpha}
F_{\rho\beta} 
F_{\sigma\gamma} 
F_{\nu\delta}.
\end{eqnarray}
We choose the functions $\R$ and $\W$,
so that the disformal transformation \eqref{disformal3} is invertible~\cite{Gumrukcuoglu:2019ebp}.
Ref.~\cite{Gumrukcuoglu:2019ebp}
explored the vector-tensor theories 
which are related to the Einstein-Maxwell theory 
under the disformal transformation \eqref{disformal3}.
Ref.~\cite{DeFelice:2019hxb} showed that
Eq. \eqref{disformal3} corresponds to
the most general disformal transformation with the $U(1)$ gauge symmetry
which is built up with $F_{\mu\nu}$
and its dual tensor $F^\ast_{\mu\nu}:=\left(1/2\right)
\epsilon_{\mu\alpha\nu\beta}g^{\alpha\rho} g^{\beta\sigma}F_{\rho\sigma}$,
where $\epsilon_{\alpha\beta\gamma\delta}$ denotes the volume element.
As long as the disformal transformation \eqref{disformal3} is invertible,
a solution $\left(g_{\mu\nu},F_{\mu\nu}\right)$ 
in a class of the vector-tensor theories 
with the $U(1)$ gauge symmetry ${S}\left[{g}_{\mu\nu},F_{\mu\nu}\right]$
is uniquely mapped to 
a solution $\left({\tilde g}_{\mu\nu},F_{\mu\nu}\right)$ 
in a new class obtained by 
${\tilde S}\left[{\tilde g}_{\mu\nu},F_{\mu\nu}\right]
=S\left[g_{\mu\nu}\left({\tilde g}_{\rho\sigma}, F_{\rho\sigma}\right),F_{\mu\nu}\right]$.

Under the disformal transformation \eqref{disformal3},
the metric of the stationary and axisymmetric spacetime \eqref{stationary}
with the vector field \eqref{vector_Ansatz}
is mapped as
\begin{eqnarray}
\label{stationary_gauge}
d{\tilde s}^2
={\tilde g}_{\mu\nu}dx^\mu dx^\nu
&=&
-
\left(
\R A 
-\frac{W A_{t,r}^2}{D}
-\frac{W A_{t,\theta}^2}{E}
\right)
dt^2
\nonumber\\
&+&
\left(
\R B
+\frac{\W A_{\varphi,r}^2}{D}
+\frac{\W A_{\varphi,\theta}^2}{E}
\right)
d\varphi^2
+
2
\left(
\R C
+\frac{\W}{D} A_{t,r}A_{\varphi,r}
+\frac{\W}{E} A_{t,\theta}A_{\varphi,\theta}
\right) dt d\varphi
\nonumber\\
&+&
\left[
\R D
+\frac{\W}{E} 
\left(A_{r,\theta}-A_{\theta,r}\right)^2
+
\frac{\W}
        {AB+C^2}
\left(
-B A_{t,r}^2
+A_{\varphi,r}
\left(
2C A_{t,r}
+A A_{\varphi,r}
\right)
\right)
\right]
dr^2
\nonumber\\
&+&
\left[
\R E
+\frac{\W}{D} 
\left(A_{r,\theta}-A_{\theta,r}\right)^2
+
\frac{\W}
        {AB+C^2}
\left(
-B A_{t,\theta}^2
+A_{\varphi,\theta}
\left(
2C A_{t,\theta}
+A A_{\varphi,\theta}
\right)
\right)
\right]
d\theta^2
\nonumber\\
&+&
\frac{2\W}{AB+C^2}
\left[
-B A_{t,\theta} A_{t,r}
+A A_{\varphi,\theta} A_{\varphi,r}
+C
\left(
 A_{\varphi,\theta}A_{t,r}
+
 A_{\varphi,r}A_{t,\theta}
\right)
\right]
dr d\theta
\nonumber\\
&-&
2\W
\left(
A_{\theta,r}-A_{r,\theta}
\right)
\left(
\frac{A_{t,\theta}} {E}
dt dr
-
\frac{A_{t,r}} {D}
dt d\theta
+
\frac{A_{\varphi,\theta}} {E}
dr d\varphi
-
\frac{A_{\varphi,r}} {D}
d\theta d\varphi
\right),
\end{eqnarray}
where each component in the last line
violates the circularity conditions \eqref{circularity0}.
Imposing the circularity conditions
requires
the vanishing azimuthal component of the magnetic field
\begin{eqnarray}
\label{circularity_case2}
F_{r\theta}=A_{\theta,r}- A_{r,\theta}=0.
\end{eqnarray}
The metric of a slowly rotating solution \eqref{stationary2}
with the vector field \eqref{vector2}
is disformally mapped to 
\begin{eqnarray}
\label{ori_em}
d{\tilde s}^2
=
{\tilde g}_{\mu\nu} dx^\mu dx^\nu
&=&
\left(
\R_{(0)}  -\frac{\W_{(0)} A_{t(0),r}{}^2}{A_{(0)}D_{(0)}}
\right)
\left(
-A_{(0)}dt^2
+D_{(0)} dr^2
\right)
+
\R_{(0)}
r^2
\left(
d\theta^2
+
\sin^2\theta 
d\varphi^2
\right)
\nonumber\\
&-&
2r^2
\sin^2\theta
\left( 
\R_{(0)}
\omega_{(1)}
-\frac{\W_{(0)} (ra_{3(1),r}+2a_{3(1)}) A_{t(0),r}}{rD_{(0)}}
\right)
dt
d\varphi
+
{\cal O} (\epsilon^2),
\end{eqnarray}
where
since $\F=\F_{(0)} (r)+{\cal O} (\epsilon^2)$ and $\G= \G_{(0)}+ {\cal O} (\epsilon^2)$
with 
$\F_{(0)}(r):=-2\left(A_{t(0),r}\right)^2/\left(A_{(0)} D_{(0)}\right)$
and 
$\G_{(0)}(r):=2\left(A_{t(0),r}\right)^4/\left(A_{(0)} D_{(0)}\right)^2$,
\begin{eqnarray}
\R(\F)
&=&
\R \left(\F_{(0)}(r),\G_{(0)} (r)\right) 
+{\cal O} (\epsilon^2)
=:\R_{(0)} (r) 
+{\cal O} (\epsilon^2),
\nonumber\\
\W(\F)
&=&
\W\left(\F_{(0)}(r),\G_{(0)}(r) \right) 
+{\cal O} (\epsilon^2)
=:\W_{(0)} (r)
+{\cal O} (\epsilon^2).
\end{eqnarray}
Since $F_{r\theta}={\cal O} (\epsilon^2)$,
at ${\cal O} (\epsilon)$
all the components  in the last line of Eq.~\eqref{stationary_gauge} vanish.
Thus,
under the $U(1)$-invariant disformal transformation \eqref{disformal3},
up to ${\cal O} (\epsilon)$,
the position of the black hole event horizon
is not modified,
and hence given by a solution of  $A_{(0)}(r)=D_{(0)}(r)^{-1}=0$.
This is in the contrast with 
the case of the disformal transformation
without the $U(1)$ gauge symmetry \eqref{disformal3},
where
the disformal transformation 
can modify the position of the black hole event horizon
even at the leading order of the slow rotation approximation.
We note that beyond the slow rotation approximation
the position of the event horizon (and the other horizons) may be modified.
In Sec. \ref{sec7}, we will check this point explicitly
with the disformed Kerr-Newman solution.

\subsection{The scalar disformal transformation}
\label{sec33}

Finally,
we consider the disformal transformation in the scalar-tensor theories.
With the scalar field $\phi$ itself and its first-order derivative $\partial_\mu \phi$,
the most general scalar disformal transformation is given by  
\begin{eqnarray}
\label{disformal2}
{\tilde g}_{\mu\nu}
= \P_s g_{\mu\nu}
 +\Q_s \partial_\mu \phi \partial_\nu \phi.
\end{eqnarray}
In a class of the shift-symmetric scalar-tensor theories
$\P_s=\P_s({\cal X})$
and 
$\Q_s=\Q_s({\cal X})$,
while
in a class of the scalar-tensor theories without the shift symmetry
$\P_s=\P_s({\cal X},\phi)$
and 
$\Q_s=\Q_s({\cal X},\phi)$.
We choose the functions $\P_s$ and $\Q_s$ in Eq. \eqref{disformal2},
so that the disformal transformation \eqref{disformal2} is invertible,
$\P_s+ {\cal X} \Q_s\neq 0$ and $\P_s\neq 0$
\cite{Zumalacarregui:2013pma}.
As long as the disformal transformation \eqref{disformal2}
is invertible,
a known solution $\left({g}_{\mu\nu},\phi\right)$ 
in a class of the scalar-tensor theories ${\tilde S}\left[{\tilde g}_{\mu\nu},\phi\right]$
is uniquely mapped to a solution $\left({\tilde g}_{\mu\nu},\phi\right)$ 
in a new class given by
${\tilde S}
\left[
{\tilde g}_{\mu\nu},\phi
\right]
=
S
\left[
g_{\mu\nu} \left({\tilde g}_{\rho\sigma},\phi\right), \phi
\right]$.

With the general Ansatz of Eqs. \eqref{stationary} and \eqref{scalar_Ansatz},
after the disformal transformation \eqref{disformal2},
the metric of a stationary and axisymmetric solution \eqref{stationary} is mapped to
\begin{eqnarray}
\label{scalar_disformed}
d{\tilde s}^2
=
{\tilde g}_{\mu\nu}
dx^\mu dx^\nu
&=&
- \left(A\P_s -\Q_s q^2\right)dt^2
+\left(\P_s B + \Q_s m ^2\right)  d\varphi^2
+2
\left(
\P_s C+ \Q_s qm
\right)
dt d\varphi
\nonumber\\
&+&
\left(
\P_s D
+ \Q_s (\partial_r\psi)^2
\right)
dr^2
+ 
\left(
\P_s E
+ \Q_s (\partial_\theta \psi)^2 
\right)
d\theta^2
+
2 \Q_s \partial_r\psi \partial_\theta \psi dr d\theta
\nonumber\\
&+&
2\Q_s
\left(
 q  \partial_r\psi dt dr
+ q \partial_\theta \psi dt d\theta
+ m \partial_r\psi dr d\varphi
+ m \partial_\theta \psi d\theta d\varphi
\right),
\end{eqnarray}
where each component in the last line violates the circularity conditions \eqref{circularity0}.
Thus, 
for nonzero $m$ and $q$
the disformed metric violates the circularity conditions \eqref{circularity0}.
Setting $m=0$ and $\psi=\psi_{(0)} (r)$
in Eq. \eqref{scalar_Ansatz},
\begin{eqnarray}
d{\tilde s}^2
&=&
- \left(A\P_s -\Q_s q^2 \right)dt^2
+\P_s B   d\varphi^2
+2 \P_s C dt d\varphi
+
\left(
\P_s D
+ \Q_s (\partial_r\psi)^2
\right)
dr^2
+
\P_s E
d\theta^2
+ 2\Q_s q \partial_r\psi dt dr,
\end{eqnarray}
which includes the case of the disformed Kerr solution discussed in Refs. \cite{dhostbh6,dhostbh7}.

In the case of a scalar-tensor theory without the shift symmetry,
for the Ansatz \eqref{scalar_Ansatz2},
the disformed metric is given by 
\begin{eqnarray}
d{\tilde s}^2
&=&
- A \P_s dt^2
+ \P_s B  d\varphi^2
+2\P_s C
dt d\varphi
+
\left(
\P_s D
+ \Q_s (\partial_r\psi)^2
\right)
dr^2
+ 
\left(
\P_s E
+ \Q_s (\partial_\theta \psi)^2 
\right)
d\theta^2
+ 
2 \Q_s \partial_r\psi \partial_\theta \psi dr d\theta,
\end{eqnarray}
which satisfies the circularity conditions \eqref{circularity0}.

In the slow rotation limit of a
stationary and axisymmetric solution
in a class of the shift-symmetric scalar-tensor theories,
after the disformal transformation \eqref{disformal2}
the metric \eqref{stationary2} with the scalar field \eqref{qt} is mapped to
\begin{eqnarray}
\label{ori_s}
d{\tilde s}^2
&=&
\P_{s(0)} 
\left[
-
\left(
A_{(0)}
-\frac{\Q_{s(0)} }{\P_{s(0)}}
q^2
\right)
dt^2
+ \left(
D_{(0)}
+\frac{\Q_{s(0)}}{\P_{s(0)}} \left(\partial_r\psi_{(0)}\right)^2\right)dr^2
+2\frac{\Q_{s(0)} }{\P_{s(0)}}
q 
\left(\partial_r\psi_{(0)}\right)
 dt dr
\right.
\nonumber\\
&+&
\left.
r^2
\left(
d\theta^2
+
\sin^2\theta 
d\varphi^2
\right)
-
2r^2
\sin^2\theta
\omega_{(1)}
dt
d\varphi
\right]
+
{\cal O} (\epsilon^2),
\end{eqnarray}
where since 
${\cal X}={\cal X}_{(0)}(r) +{\cal O} (\epsilon^2)$
with ${\cal X}_{(0)}(r):=\left(\partial_r \psi_{(0)} (r)\right)^2/D_{(0)}(r)-q^2 /A_{(0)}(r) $,
\begin{eqnarray}
\P_s ({\cal X})
=\P_{s} \left({\cal X}_0(r)\right) +{\cal O} (\epsilon^2)
=\P_{s (0)} (r) +{\cal O} (\epsilon^2),
\qquad 
\Q_s({\cal X})
=\Q_{s} \left({\cal X}_0(r)\right) +{\cal O} (\epsilon^2)
=\Q_{s (0)} (r) +{\cal O} (\epsilon^2).
\end{eqnarray}
As in the case of the transformation from Eqs. \eqref{ori} to \eqref{trans},
the metric \eqref{ori_s} can be rewritten as 
\begin{eqnarray}
\label{trans_s}
d{\tilde s}^2
&=&
\P_{s(0)}
\left[
-\left(
 A_{(0)}
-\frac{\Q_{s(0)}}{\P_{s(0)}}
q^2\right)d{\tilde t}^2
+
\left(
D_{(0)}+\frac{\Q_{s(0)} A_{(0)} \left(\partial_r\psi_{(0)}\right)^2}
            {\P_{s(0)} A_{(0)}-\Q_{s(0)} q^2}
\right)
dr^2
+r^2
\left(
d\theta^2
+
\sin^2\theta 
d\varphi^2
\right)
\right.
\nonumber\\
&-&
\left.
2\omega_{(1)} 
 r^2 
\sin^2\theta
\left(
 d{\tilde t}
+ \frac{q\Q_{s(0)}\left(\partial_r\psi_{(0)}\right)}
        {\P_{s(0)}A_{(0)}-\Q_{s(0)} q^2}
dr
\right)
d\varphi
\right]
+
{\cal O} (\epsilon^2).
\end{eqnarray}
The nonzero ${\tilde g}_{r\varphi}\neq 0$ in Eq.~\eqref{trans_s}
indicates the violation of the circularity conditions,
which explicitly appears at ${\cal O}(\epsilon^2)$.
In the case that
the reference static and spherically symmetric solution
is a black hole spacetime 
and
has a black hole event horizon 
at which $A_{(0)}= D_{(0)}^{-1}=0$,
the position of the event horizon
is shifted
to the position determined by the solution of $\P_{s(0)}A_{(0)}-\Q_{s(0)} q^2=0$.
The shift of the position of the event horizon 
will happen, in the solution where the scalar field is time dependent
with $q\neq 0$.

On the other hand,
in the slow rotation limit of the stationary and axisymmetric solutions
in the scalar-tensor theories without the shift symmetry,
the metric \eqref{stationary2} with the scalar field \eqref{qt}
is disformally mapped to 
\begin{eqnarray}
\label{ori_s2}
d{\tilde s}^2
&=&
\P_{s(0)} 
\left[
-
A_{(0)}
dt^2
+ \left( 
D_{(0)}
+\frac{\Q_{s(0)}}{\P_{s(0)}} \left(\partial_r\psi_{(0)}\right)^2\right)dr^2
+
r^2 
\left(
d\theta^2
+
\sin^2\theta 
d\varphi^2
\right)
-
2r^2
\sin^2\theta
\omega_{(1)}
dt
d\varphi
\right]
+
{\cal O} (\epsilon^2),
\end{eqnarray}
where since 
$\phi=\psi_{(0)}(r)+{\cal O} (\epsilon^2)$
and 
${\cal X}={\cal X}_{(0)}(r) +{\cal O} (\epsilon^2)$
with ${\cal X}_{(0)}(r):=\left(\partial_r \psi_{(0)} (r)\right)^2/D_{(0)}(r)$,
\begin{eqnarray}
\P_s \left(\phi, {\cal X}\right)
&=&
\P_{s} \left(\psi_{(0)}(r),{\cal X}_0(r)\right) +{\cal O} (\epsilon^2)
=:
\P_{s (0)}  (r) +{\cal O} (\epsilon^2),
\nonumber\\
\Q_s\left(\phi,{\cal X}\right)
&=&
\Q_{s} \left(\psi_{(0)}(r), {\cal X}_0(r)\right) +{\cal O} (\epsilon^2)
=:
\Q_{s (0)} (r) +{\cal O} (\epsilon^2).
\end{eqnarray}
In the case that
the reference static and spherically symmetric solution
is a black hole spacetime 
and
has an event horizon 
where $A_{(0)}= D_{(0)}^{-1}=0$,
the position of the event horizon
remains the same 
before and after the disformal transformation.
This is because
the character of the scalar field is spacelike 
and
does not affect the causal properties of the spacetime.

\subsection{The examples}

In order to discuss more concrete examples, 
in Secs. \ref{sec4}-\ref{sec6}, 
we will consider the following class of the 
generalized Proca theories
with the second-order equations of motion 
\cite{Heisenberg:2014rta,Tasinato:2014eka,DeFelice:2016cri}
\begin{eqnarray}
\label{gp}
S
=\int d^4 x
  \sqrt{-g}
\left[
\frac{m_p^2}{2}
\left(
R-2\Lambda
\right)
-
\frac{1}{4}
{\cal F}
-(m^2g^{\mu\nu}-\beta G^{\mu\nu})
A_\mu A_\nu
\right],
\end{eqnarray}
where $m_p$ and $\Lambda$
represent the reduced Planck mass and the cosmological constant,
$m$ and $\beta$ are the mass and the coupling constant of the vector field $A_\mu$,
$\F$ is defined in Eq. \eqref{calf_calg},
and
$R$ and $G_{\mu\nu}$ are the Ricci scalar and 
Einstein tensor associated with the metric $g_{\mu\nu}$,
respectively.
The static and spherically symmetric black hole solutions
have been obtained 
 in the subclass \eqref{gp} \cite{Chagoya:2016aar,Minamitsuji:2016ydr,Babichev:2017rti},
and 
in other subclasses
of the generalized Proca theories 
\cite{Heisenberg:2017xda,Heisenberg:2017hwb}.

A solution in the vector-tensor theory \eqref{gp}
with the vanishing field strength vanishes $F_{\mu\nu}=0$,
after the replacement of $A_\mu\to \partial_\mu \phi$ and the integration of it,
can be mapped to that in the shift-symmetric scalar-tensor theory
\begin{eqnarray}
\label{st}
S
=\int d^4 x
  \sqrt{-g}
\left[
\frac{m_p^2}{2}
\left(
R-2\Lambda
\right)
-(m^2g^{\mu\nu}-\beta G^{\mu\nu})
\partial_\mu \phi
\partial_\nu \phi
\right],
\end{eqnarray}
which is a class of the shift-symmetric Horndeski theories \cite{Horndeski:1974wa,Deffayet:2009mn,Kobayashi:2011nu}.
The static and spherically symmetric black hole solutions 
in the shift-symmetric scalar-tensor theory \eqref{st}
have also been explored
\cite{Babichev:2013cya,Appleby:2015ysa,Babichev:2016rlq,Rinaldi:2012vy,Anabalon:2013oea,Minamitsuji:2013ura}.

After the discussion on the theories \eqref{gp} and \eqref{st},
we will also discuss the cases of the Einstein-Maxwell theory \eqref{em}
and the Einstein-conformally coupled scalar field theory \eqref{ecs}
in Secs. \ref{sec7} and \ref{sec8},
respectively,
as the examples of 
the vector-tensor theory with the $U(1)$ gauge symmetry 
and 
the scalar-tensor theory without the shift symmetry.

\section{The disformed asymptotically flat black hole solutions}
\label{sec4}

In this section,
we focus on the asymptotically flat black hole solutions in the models \eqref{gp} and \eqref{st}.

\subsection{The stealth Schwarzschild solution}
\label{sec41}

For an arbitrary value of $\beta$ and $m=\Lambda=0$
in the generalized Proca theory \eqref{gp},
there exists
the stealth Schwarzschild solution
with the leading order slow rotation corrections~\cite{Chagoya:2016aar,Minamitsuji:2016ydr}
given by Eqs. \eqref{stationary2} and \eqref{vector2} with 
\begin{eqnarray}
\label{stealth}
A_{(0)}(r) =1-\frac{2M}{r},
\qquad 
D_{(0)}(r)= \frac{1}{A_{(0)}(r)},
\qquad
A_{t(0)}(r)
=q,
\qquad
A_{r(0)}(r)
&=&
\pm
\frac{q}{1-2M/r}
\sqrt{\frac{2M}{r}},
\end{eqnarray}
and 
\begin{eqnarray}
\omega_{(1)}(r)=\frac{2J}{r^3},
\qquad 
a_{3(1)}(r)=0,
\end{eqnarray}
where we set the constant term in $\omega_{(1)}$ to be zero 
because of the asymptotically flat spacetime
and $q$ is the constant.
Since ${\cal Y}=-q^2+{\cal O} (\epsilon^2)$,
up to ${\cal O} (\epsilon)$,
$\P({\cal Y})=\P_{(0)}(-q^2)$ and $\Q({\cal Y})=\Q_{(0)}(-q^2)$ remain constant.

The disformed solution written in terms of the original time coordinate \eqref{ori} is given by 
\begin{eqnarray}
\label{stealth1}
d{\tilde s}^2
&=&
\P_{(0)}
\left[
-
\left(
1-\frac{2M}{r}
-
\frac{\Q_{(0)} q^2}{\P_{(0)}}
\right)
d{t}^2
+
 \frac{ 1-2M/r+2\Q_{(0)} M q^2/(\P_{(0)} r)}
          {\left(1-2M/r\right)^2}
dr^2
\pm 
\frac{2\Q_{(0)} q^2}
       {\P_{(0)} \left(1-2M/r\right)}
\sqrt{\frac{2M}{r}}
dt dr
\right.
\nonumber\\
&+&
\left.
r^2
\left(
d\theta^2
+
\sin^2\theta 
d\varphi^2
\right)
-
\frac{4 J}{r}
\sin^2\theta
d{t}d\varphi
\right]
+{\cal O} 
\left(
\epsilon^2
\right)
\end{eqnarray}
and that rewritten in terms of the new time coordinate \eqref{trans} 
is given by
\begin{eqnarray}
\label{newv0}
d{\tilde s}^2
&=&
\P_{(0)}
\left[
-
\left(
1-\frac{2M}{r}
-
\frac{\Q_{(0)} q^2}{\P_{(0)}}
\right)
d{\tilde t}^2
+\frac{1-\Q_{(0)} q^2/\P_{(0)}}
         {1-2M/r-\Q_{(0)} q^2/\P_{(0)}}
dr^2
+r^2
\left(
d\theta^2
+
\sin^2\theta 
d\varphi^2
\right)
\right.
\nonumber\\
&-&
\left.
\frac{4 J}{r}
\sin^2\theta
d{\tilde t}d\varphi
\mp
\frac{4 \Q_{(0)} J q^2}
        { (r-2M)\left[\P_{(0)} (r-2M)-\Q_{(0)} q^2 r\right]}
\sqrt{2Mr}
\sin^2\theta
drd\varphi
\right]
+{\cal O} 
\left(
\epsilon^2
\right),
\end{eqnarray}
where we assume that $\P_{(0)}>0$ and $\P_{(0)}-\Q_{(0)}q^2>0$
for the invertible disformal transformation up to ${\cal O} (\epsilon)$.

By introducing the rescaled mass parameter 
$\tilde{M}:=  M/(1-\Q_{(0)} q^2/\P_{(0)} )$
and 
the rescaled time coordinate 
$dT:=\sqrt{1-\Q_{(0)} q^2/\P_{(0)} } d{\tilde t}$,
up to ${\cal O} (\epsilon)$,
the disformed metric can be written as  
\begin{eqnarray}
\label{newv01}
d{\tilde s}^2
&=&
\P_{(0)}
\left[
-
\left(
1-\frac{2\tilde M}{r}
\right)
dT^2
+\frac{1}
         {1-2{\tilde M}/r}
dr^2
+r^2
\left(
d\theta^2
+
\sin^2\theta 
d\varphi^2
\right)
\right.
\nonumber\\
&-&
\left.
\frac{4 J}
       {r\sqrt{1-\Q_{(0)} q^2/\P_{(0)}}}
\sin^2\theta
dTd\varphi
\mp
\frac{4 \Q_{(0)} J q^2}
        { (r-2{\tilde M})\left[\P_{(0)} (r-2{\tilde M})+2q^2\Q_{(0)} {\tilde M}\right]}
\sqrt{
\frac{2\P_{(0)} {\tilde M}r}
        {\P_{(0)} -q^2 \Q_{(0)}}
}
\sin^2\theta
drd\varphi
\right]
\nonumber\\
&+&{\cal O} 
\left(
\epsilon^2
\right).
\end{eqnarray}
At ${\cal O} (\epsilon^0)$,
the disformed solution is also the Schwarzschild solution,
but 
the position of the disformed black hole event horizon 
is given by $r=2 {\tilde M}$.
Thus, 
unlike the conformal transformation, 
the disformal transformation 
modifies the position of the event horizon 
and 
the causal properties of the spacetime, 
assuming that
matter is minimally coupled to the metric in each frame.
As $r\to\infty$,
and the off-diagonal components behave as
${\tilde g}_{{\tilde t} \varphi}\sim 1/r$
and 
${\tilde g}_{r \varphi}\sim 1/r^{3/2}$.

At ${\cal O} (\epsilon^0)$,
the vector field expressed in terms of the new time coordinate
defined in the disformed frame \eqref{vector_new}
is given by 
\begin{eqnarray}
\label{newv02}
A_\mu dx^\mu
&=&
q 
\left[
d{\tilde t}
\pm 
\frac{1}
       {1-\frac{2{\tilde M}}
                    {r}
        }
\sqrt{
\frac{2{\tilde M}}
        {r\left (1-\Q_{(0)}q^2/\P_{(0)} \right)}}
dr
\right]
+{\cal O} (\epsilon)
=
\frac{q}
        {\sqrt{1-\Q_{(0)}q^2/\P_{(0)} }}
\left[
dT
\pm 
\frac{\sqrt{
\frac{2{\tilde M}}
        {r}
}}
       {1-\frac{2{\tilde M}} {r}}
dr
\right]
+{\cal O} (\epsilon)
\nonumber\\
&\approx &
\frac{q}
        {\sqrt{1-\Q_{(0)}q^2/\P_{(0)} }}
\left[
dT
\pm 
\frac{dr}
       {1-\frac{2{\tilde M}} {r}}
\right]
+{\cal O} (\epsilon)
=:
\frac{q}
        {\sqrt{1-\Q_{(0)}q^2/\P_{(0)} }}
du_{\pm} 
+{\cal O} (\epsilon),
\end{eqnarray}
where
``$\approx $'' means
that the near-horizon limit $r\to 2{\tilde M}$ is taken,
and  
$u_\pm := T\pm \int  \frac{dr}{1-\frac{2{\tilde M}} {r}}$
represents the null coordinates
which are 
regular 
at the future and past event horizons,
respectively.
Thus,
the positive and negative branch solutions of the vector field
are regular 
at the future and past event horizons,
respectively,
and 
the disformal transformation does not modify the essential properties
of the vector field
from the original one~\cite{Chagoya:2016aar,Minamitsuji:2016ydr}.
In Appendix \ref{app3},
at ${\cal O} (\epsilon^0)$,
we
confirm that 
Eqs. \eqref{newv01} and \eqref{newv02}
are the solutions of the Euler-Lagrange
equations 
obtained 
by varying the action 
of the extended vector-tensor theory 
derived via the disformal transformation
of the generalized Proca theory \eqref{gp}
with $m=\Lambda=0$.

Up to ${\cal O} (\epsilon)$,
the above disformed metric \eqref{stealth1} 
with the replacement of
$\P_{(0)} (-q^2)\to \P_{s(0)} (-q^2)$ and $\Q_{(0)} (-q^2)\to \Q_{s(0)}(-q^2)$
and the scalar field \eqref{qt} with
\begin{eqnarray}
\psi_{(0)} (r)
=
\pm
q 
\int 
\frac{dr}{1-2M/r}
\sqrt{\frac{2M}{r}},
\end{eqnarray}
are the solution in the scalar-tensor theory obtained via the disformal transformation of 
the theory \eqref{st}.

\subsection{The charged stealth Schwarzschild solution}
\label{sec42}

For $\beta=1/4$ and $m=\Lambda=0$
in the generalized Proca theory \eqref{gp},
there exists
the charged stealth Schwarzschild solution 
with the leading order slow rotation corrections 
\cite{Chagoya:2016aar,Minamitsuji:2016ydr},
given by Eqs. \eqref{stationary2} and \eqref{vector2} with 
\begin{eqnarray}
A_{(0)}(r)
&=&1-\frac{2M}{r},
\qquad 
D_{(0)}(r)
=\frac{1}{A_{(0)} (r)},
\nonumber\\
A_{t(0)}(r)
&=&
q+\frac{Q}{r},
\qquad 
A_{r(0)}(r)
=
\pm 
\frac{1}{1-2M/r}
\frac{\sqrt{Q^2+2q (Q+Mq)r}}
       {r},
\end{eqnarray}
and 
\begin{eqnarray}
\omega_{(1)}(r)
=\frac{2J}{r^3},
\qquad
a_{3(1)}(r)=-\frac{JQ}{M r^3},
\end{eqnarray}
where we set the constant term in $\omega_{(1)}$ to be zero,
so that the disformed spacetime is asymptotically flat,
and $M$, $J$, and $Q$ 
represent the mass, angular momentum, and electric charge of the black hole,
respectively.
\footnote{Below,
we will employ the same notation $M$, $J$, and $Q$ 
to describe the mass, angular momentum, and electric charge of the black hole
in the original frame,
respectively,
which are measured at the spatial infinity.}
Since the norm of the vector field ${\cal Y}=-q^2+{\cal O} (\epsilon^2)$,\
up to ${\cal O} (\epsilon)$,
$\P({\cal Y})=\P_{(0)}(-q^2)$ and $\Q({\cal Y})=\Q_{(0)}(-q^2)$ also remain constant.

The disformed solution written in terms of the original time coordinate \eqref{ori}
is given by 
\begin{eqnarray}
\label{cstealth1}
d{\tilde s}^2
&=&
\P_{(0)}
\left[
-
\left(
1-\frac{2M}{r}
-\frac{\Q_{(0)}}{\P_{(0)}}
\left( 
q
+\frac{Q}{r}
\right)^2
\right)
d{t}^2
+\frac{1-\frac{2M}{r}
         +\frac{\Q_{(0)}}{\P_{(0)} r^2}\left(Q^2+2q (Mq+Q)r\right)}
        {\left(1-\frac{2M}{r}\right)^2}
dr^2
\right.
\nonumber\\
&
\pm
&
\frac{2\Q_{(0)} (Q+q r)}{\P_{(0)} r (r-2M)}
\sqrt{
Q^2
+2q \left(Mq+Q\right)r
}
dt dr
+
r^2
\left(
d\theta^2
+
\sin^2\theta 
d\varphi^2
\right)
\nonumber\\
&-&
\left.
\frac{2J\left(2M r+ \frac{\Q_{(0)}}{\P_{(0)}} Q (Q+qr)\right)}
{ Mr^2}
\sin^2\theta
d{t}d\varphi
\mp
\frac{2\Q_{(0)} J Q}
        {\P_{(0)} rM (r-2M)}
\sqrt{
Q^2 +2r q \left(Mq+Q\right)}
\sin^2\theta
drd\varphi
\right]
+
{\cal O} (\epsilon^2),
\end{eqnarray}
and that rewritten in terms of the new time coordinate \eqref{trans} 
is given by
\begin{eqnarray}
\label{cstealth2}
d{\tilde s}^2
&=&
\P_{(0)}
\left[
-
\left(
1-\frac{2M}{r}
-
\frac{\Q_{(0)}}{\P_{(0)}}
\left( 
q
+\frac{Q}{r}
\right)^2
\right)
d{\tilde t}^2
+\frac{1 -\frac{\Q_{(0)} q^2}{\P_{(0)}}}
  {
1-\frac{2M}{r}
-
\frac{\Q_{(0)}}{\P_{(0)}}
\left( 
q
+\frac{Q}{r}
\right)^2}
dr^2
+r^2
\left(
d\theta^2
+
\sin^2\theta 
d\varphi^2
\right)
\right.
\nonumber\\
&-&
\left.
\frac{2J\left(2M r+ \frac{\Q_{(0)}}{\P_{(0)}} Q (Q+qr)\right)}
{ Mr^2}
\sin^2\theta
d{t}d\varphi
\right.
\nonumber \\
&\mp &
\left.
\frac{
2
\left(\frac{\Q_{(0)}}{\P_{(0)}}\right)
 J (Q+qr )\left(2M  r+\frac{\Q_{(0)}}{\P_{(0)}} Q (Q+qr)\right)}
        {M r(r-2M)
\left[
r (r-2M)
-
\left(\frac{\Q_{(0)}}{\P_{(0)}}\right)
 (qr+Q)^2\right]}
\sqrt{2q(Mq+Q)r+Q^2}
\sin^2\theta
drd\varphi
\right]
+
{\cal O} (\epsilon^2),
\end{eqnarray}
where we assume that $\P_{(0)}>0$ and $\P_{(0)}-\Q_{(0)}q^2>0$
for the invertible disformal transformation up to ${\cal O} (\epsilon)$.

By introducing the rescaled mass and charge parameters 
\begin{eqnarray}
\frac{M\P_{(0)}+q Q \Q_{(0)}}
        {\P_{(0)}-q^2 \Q_{(0)}}
={\tilde M},
\qquad
-\frac{\Q_{(0)} Q^2}
        {\P_{(0)}-q^2 \Q_{(0)}}
={\tilde Q}^2,
\end{eqnarray}
and 
the rescaled time coordinate 
$dT:=
\sqrt{1-\Q_{(0)} q^2/\P_{(0)} }
 d{\tilde t}$,
at ${\cal O} (\epsilon^0)$,
the metric of the disformed solution is given by
\begin{eqnarray}
\label{newv03}
d{\tilde s}^2
&=&
\P_{(0)}
\left[
-
\left(
1-\frac{2\tilde M}{r}-\frac{{\tilde Q}^2}{r^2}
\right)
dT^2
+\frac{1}
          {1-\frac{2\tilde M}{r}-\frac{{\tilde Q}^2}{r^2}}
dr^2
+r^2
\left(
d\theta^2
+
\sin^2\theta 
d\varphi^2
\right)
\right]
+
{\cal O} 
\left(
\epsilon
\right).
\end{eqnarray}
At ${\cal O} (\epsilon^0)$,
the disformed solution is given by the RN solution with
the mass $\tilde{M}$ and charge ${\tilde Q}$,
and 
the position of the black hole event horizon
is modified as 
$r={\tilde M}+\sqrt{{\tilde M}^2-{\tilde Q}^2}$.
There is also the inner Cauchy horizon at 
$r={\tilde M}-\sqrt{{\tilde M}^2-{\tilde Q}^2}$.
In this particular case,
the disformal transformation maps 
the Schwarzschild solution to the RN solution.
Thus,
unlike the conformal transformation,
the disformal transformation modifies the causal properties of the 
spacetime
in terms of the position of the event horizon and the number of the horizons,
assuming that
matter is minimally coupled to the metric in each frame.
As $r\to\infty$,
${\tilde g}_{{\tilde t} \varphi}\sim 1/r$
and 
${\tilde g}_{r \varphi}\sim 1/r^{3/2}$.

At ${\cal O} (\epsilon^0)$,
the vector field expressed in terms of the new time coordinate
defined in the disformed frame \eqref{vector_new}
is given by 
\begin{eqnarray}
\label{newv04}
A_\mu dx^\mu
&=&
\left(q+\frac{Q}{r}\right)
d{\tilde t}
\pm 
\frac{1}
       {1-\frac{2{\tilde M} }  {r}+\frac{{\tilde Q}^2}{r^2}}
\frac{
\sqrt{
Q^2 
+2q r
 \left(1-\frac{\Q_{(0)}}{\P_{(0)}}q^2\right) 
 \left(\tilde{M}q+Q\right)
}}
       {r\left(1-\frac{\Q_{(0)}}{\P_{(0)}} q^2\right)}
dr
+{\cal O} (\epsilon)
\nonumber\\
&=&
\frac{1}
       {\sqrt{1-\frac{\Q_{(0)}}{\P_{(0)}} q^2}}
\left[
\left(q+\frac{Q}{r}\right)
dT
\pm 
\frac{
\sqrt{
Q^2 
+2q r
 \left(1-\frac{\Q_{(0)}}{\P_{(0)}}q^2\right) 
 \left(\tilde{M}q+Q\right)}}
   {r\left(1-\frac{2{\tilde M} } {r}+\frac{{\tilde Q}^2}{r^2}\right)
       \sqrt{1-\frac{\Q_{(0)}}{\P_{(0)}} q^2}}
dr
\right]
+{\cal O} (\epsilon).
\end{eqnarray}
In the limit of the black hole event horizon 
$r\to {\tilde M}+\sqrt{{\tilde M}^2-{\tilde Q}^2}$,
the vector field can be approximated by
\begin{eqnarray}
A_\mu dx^\mu
\approx
\frac{q+\frac{Q}{r}}
       {\sqrt{1-\frac{\Q_{(0)}}{\P_{(0)}} q^2}}
\left[
dT
\pm 
\frac{dr}
   {1-\frac{2{\tilde M} } {r}+\frac{{\tilde Q}^2}{r^2}}
\right]
+{\cal O} (\epsilon)
=:
\frac{q+\frac{Q}{r}}
       {\sqrt{1-\frac{\Q_{(0)}}{\P_{(0)}} q^2}}
du_{\pm}
+{\cal O} (\epsilon),
\end{eqnarray}
where $u_+$ and $u_-$
represent the null coordinates
which are regular at the future and past event horizons,
respectively.
Thus,
the positive and negative branch solutions of the vector field
are regular 
at the future and past event horizons,
respectively,
and
the disformal transformation does not modify the essential properties
of the vector field
from the original one~\cite{Chagoya:2016aar,Minamitsuji:2016ydr}.
In Appendix \ref{app3},
at ${\cal O} (\epsilon^0)$,
we
confirm that 
Eqs. \eqref{newv03} and \eqref{newv04}
are the solutions of the Euler-Lagrange
equations of the extended vector-tensor theory 
derived via the disformal transformation
of the generalized Proca theory \eqref{gp} with $m=\Lambda=0$.

Since $F_{tr}\neq  0$,
there is no corresponding solution 
in the scalar-tensor theory obtained via the disformal transformation of the theory \eqref{st}.

\section{The disformed Schwarzschild-(anti-) de SItter solutions with the slow rotation corrections}
\label{sec5}

In this section,
we focus on asymptotically-(anti-) de Sitter black holes in the models \eqref{gp} and \eqref{st}.

\subsection{The Schwarzschild- (anti-) de Sitter solutions}

For an arbitrary value of $\beta$ in the generalized Proca theory \eqref{gp},
there exists
the Schwazschild-(anti-) de Sitter solution
with the leading order slow rotation corrections 
given by Eqs. \eqref{stationary2} and \eqref{vector2} with 
\cite{Chagoya:2016aar,Minamitsuji:2016ydr}
\begin{eqnarray}
A_{(0)}(r)
&=&
1-\frac{2M}{r}+\frac{m^2}{3\beta}r^2,
\qquad
D_{(0)}(r)
=
\frac{1}{A_{(0)} (r)},
\nonumber\\
A_{t(0)}(r)
&=&
\frac{m_p}{m}
\sqrt{
\frac{m^2+\beta\Lambda}{2\beta}},
\qquad
A_{r(0)}(r)
=
\pm 
\frac{\sqrt{1-A_{(0)}(r)}}
         {A_{(0)}(r)}
A_{t(0)}(r),
\end{eqnarray}
where we assume that $m^2+\beta\Lambda\leq 0$ and $\beta<0$ (or $m^2+\beta\Lambda\geq 0$ and $\beta>0$),
and 
\begin{eqnarray}
\omega_{(1)}(r)
=
\omega_0
+\frac{2J}{r^3},
\qquad
a_{3(1)}(r)=0.
\end{eqnarray}
Here, we keep the nonzero $\omega_0\neq 0$,
because the spacetime is not asymptotically flat.
Since the norm of the vector field ${\cal Y}=-\frac{m_p^2 (m^2+\beta\Lambda)}{2m^2\beta} +{\cal O} (\epsilon^2)$,
up to ${\cal O} (\epsilon)$,
$\P({\cal Y})=\P_{(0)}\left(-\frac{m_p^2 (m^2+\beta\Lambda)}{2m^2\beta}\right)$ 
and 
$\Q({\cal Y})=\Q_{(0)} \left(-\frac{m_p^2 (m^2+\beta\Lambda)}{2m^2\beta}\right)$
remain constant.

The disformed solution written in terms of the original time coordinate \eqref{ori} is given by 
\begin{eqnarray}
\label{sds1}
d{\tilde s}^2
&=&
\P_{(0)}
\left[
-
\left(
1-\frac{2M}{r}+\frac{m^2r^2}{3\beta}
-
\frac{\Q_{(0)}}{\P_{(0)}}
\frac{m_p^2 (m^2+\beta \Lambda)}
       {2m^2\beta}
\right)
d{t}^2
\right.
\nonumber\\
&+&
 \frac{ 6m^2 r\beta \left(m^2 r^3+3(r-2M)\beta\right)
         -3\frac{\Q_{(0)}}{\P_{(0)}} m_p^2 r (m^2r^3-6M\beta) (m^2+\beta\Lambda)}
          {2\left(m^3 r^3+3m\beta (r-2M)\right)^2}
dr^2
\nonumber\\
&
\pm &
\left.
\frac{\Q_{(0)} m_p^2 (m^2+\beta \Lambda)}
       {\P_{(0)} \left(
m^4 r^3
+3m^2 (r-2M)\beta
\right)}
\sqrt{
-
\frac{3 r
\left(m^2 r^3-6M\beta\right)}
       {\beta}
}
dt dr
+
r^2
\left(
d\theta^2
+
\sin^2\theta 
d\varphi^2
\right)
-2r^2
\left(
\frac{2J}{r^3}
+\omega_0
\right)
\sin^2\theta
d{\tilde t}d\varphi
\right]
\nonumber\\
&+&
{\cal O} 
\left(
\epsilon^2
\right),
\end{eqnarray}
and that rewritten in terms of the new time coordinate \eqref{trans}
is given by
\begin{eqnarray}
\label{newv1}
d{\tilde s}^2
&=&
\P_{(0)}
\left[
-
\left(
1-\frac{2M}{r}+\frac{m^2r^2}{3\beta}
-
\frac{\Q_{(0)}}{\P_{(0)}}
\frac{m_p^2 (m^2+\beta \Lambda)}
       {2m^2\beta}
\right)
d{\tilde t}^2
+\frac{ 1
-\frac{\Q_{(0)}}{\P_{(0)}}
\frac{m_p^2 (m^2+\beta\Lambda)}
        {2m^2\beta}
}
         {1-\frac{2M}{r}+\frac{m^2r^2}{3\beta}
-\frac{\Q_{(0)}}{\P_{(0)}}
\frac{m_p^2 (m^2+\beta \Lambda)}
         {2m^2\beta}
         }
dr^2
\right.
\nonumber\\
&+&
r^2
\left(
d\theta^2
+
\sin^2\theta 
d\varphi^2
\right)
-
2r^2
\left(
\frac{2J}{r^3}
+\omega_0
\right)
\sin^2\theta
d{\tilde t}d\varphi
\nonumber\\
&\mp&
\left.
\left(
1-\frac{2M}{r}+\frac{m^2r^2}{3\beta}
-
\frac{\Q_{(0)}}{\P_{(0)}}
\frac{m_p^2 (m^2+\beta \Lambda)}
       {2m^2\beta}
\right)^{-1}
\frac{\Q_{(0)} m_p^2 \beta (m^2+\beta \Lambda) (2J+r^3\omega_0)}
        {\P_{(0)}\beta \left(m^4r^3+3m^2 (r-2M)\beta\right) }
\sqrt{\frac{3(-m^2r^3+6M\beta)}{r\beta}}
\sin^2\theta
drd\varphi
\right]
\nonumber\\
&
+&{\cal O} 
\left(
\epsilon^2
\right),
\end{eqnarray}
where we assume that $\P_{(0)}>0$ and $\P_{(0)}-m_p^2\Q_{(0)}\left(m^2+\Lambda \beta \right)/ (2m^2\beta)>0$
for the invertible disformal transformation up to ${\cal O} (\epsilon)$.
In the case of $m=\pm \sqrt{-\beta \Lambda}$,
the disformed metric remains the same as 
the Schwarzschild-(anti-) de Sitter solutions 
besides the overall constant factor $\P_{(0)}>0$.
By introducing the rescaled mass parameter
and the effective cosmological constant 
\begin{eqnarray}
\tilde{M}:= 
\frac{M}
{ 1
-\frac{\Q_{(0)}}{\P_{(0)}}
\frac{m_p^2 (m^2+\beta\Lambda)}
        {2m^2\beta}},
\qquad 
{\tilde \Lambda}
=
-
\frac{2m^4 \P_{(0)}}
       {2m^ 2\beta \P_{(0)}-m_p^2\Q_{(0)}\left(m^2+\beta\Lambda \right)},
\end{eqnarray}
and the time coordinate
$dT:=\sqrt{ 1
-\frac{\Q_{(0)}}{\P_{(0)}}
\frac{m_p^2 (m^2+\beta\Lambda)}
        {2m^2\beta}} d{\tilde t}$,
\begin{eqnarray}
d{\tilde s}^2
&=&
\P_{(0)}
\left[
-
\left(
1-\frac{2\tilde M}{r}
-\frac{{\tilde \Lambda}r^2}{3}
\right)
dT^2
+\frac{1}
         {1-2{\tilde M}/r-\frac{{\tilde \Lambda}r^2}{3}}
dr^2
+r^2
\left(
d\theta^2
+
\sin^2\theta 
d\varphi^2
\right)
\right]
+{\cal O} 
\left(
\epsilon
\right).
\end{eqnarray}
Thus, 
at ${\cal O} (\epsilon^0)$,
the disformed solution is also the Schwarzschild-(anti)-de Sitter solution,
and 
for ${\tilde \Lambda}>0$
has  
the cosmological and black hole event horizons.
Unlike the conformal transformation, 
the disformal transformation modifies the position of the event horizon,
assuming that
matter is minimally coupled to the metric in each frame.

At ${\cal O} (\epsilon^0)$,
the vector field expressed in terms of the new time coordinate
defined in the disformed frame \eqref{vector_new}
is given by 
\begin{eqnarray}
A_\mu dx^\mu
&=&
A_t
\left[
d{\tilde t}
\pm 
\frac{\P_{(0)} \sqrt{1-A_{(0)}}}
       {\P_{(0)}A_{(0)}-\Q_{(0)} A_{t(0)}^2}
dr
\right]
+{\cal O} (\epsilon)
\nonumber\\
&\approx& 
\frac{A_{t(0)}}
       {\sqrt{1-\frac{\Q_{(0)}}{\P_{(0)}} A_{t(0)}^2}}
\left[
dT
\pm 
\frac{dr}
        {1-\frac{2{\tilde M}}{r}-\frac{{\tilde \Lambda}r^2}{3}}
\right]
+{\cal O} (\epsilon)
=:
\frac{A_{t(0)}}
       {\sqrt{1-\frac{\Q_{(0)}}{\P_{(0)}} A_{t(0)}^2}}
du_{\pm}
+{\cal O} (\epsilon),
\end{eqnarray}
where 
``$\approx $'' means
that the near-horizon limit is taken,
and  
$u_+$ ($u_-$)
represents the null coordinate
which is regular 
at the future event and past cosmological horizons
(the past event and future cosmological horizons).
Thus, 
the positive (negative) branch solution of the vector field
is regular 
at the future event and past cosmological horizons
(the past event and future cosmological horizons).
The disformal transformation does not modify the essential properties
of the vector field
from the original one~\cite{Minamitsuji:2016ydr}.

Up to ${\cal O} (\epsilon)$,
the above disformed metric \eqref{sds1} 
with the replacement of
\begin{eqnarray}
\P_{(0)}
\left(-\frac{m_p^2 (m^2+\beta\Lambda)}{2m^2\beta}\right)
&\to&
\P_{s(0)} 
\left(-\frac{m_p^2 (m^2+\beta\Lambda)}{2m^2\beta}\right),
\nonumber\\
\Q_{(0)} 
\left(-\frac{m_p^2 (m^2+\beta\Lambda)}{2m^2\beta}\right)
&\to&
 \Q_{s(0)}
\left(-\frac{m_p^2 (m^2+\beta\Lambda)}{2m^2\beta}\right),
\end{eqnarray}
and the scalar field \eqref{qt} with
\begin{eqnarray}
\psi_{(0)} (r)
=
\pm
\frac{m_p}{m} 
\int 
\frac{dr}{
1-\frac{2M}{r}+\frac{m^2}{3\beta}r^2}
\sqrt{
-\frac{(m^2r^3-6\beta M) (m^2+\beta\Lambda) }
         {6\beta^2r }
},
\end{eqnarray}
are also the solution in the scalar-tensor theory obtained via the disformal transformation of 
the theory \eqref{st}.

\subsection{The charged Schwarzschild- (anti-) de Sitter solution}

For $\beta=1/4$ and $4m^2+\Lambda\geq 0$
in the generalized Proca theory \eqref{gp},
there exists
the charged Schwazschild-(anti-) de Sitter solution 
with the leading order slow rotation corrections 
given by Eqs. \eqref{stationary2} and \eqref{vector2} with 
\cite{Chagoya:2016aar,Minamitsuji:2016ydr}
\begin{eqnarray}
A_{(0)}(r)
&=&
1
-\frac{2M}{r}
+\frac{4m^2 r^2}{3},
\qquad 
D_{(0)}(r)
=
\frac{1}{A_{(0)} (r)},
\nonumber\\
A_{t(0)}(r)
&=&
\frac{m_p\sqrt{4m^2+\Lambda}}{\sqrt{2} m}
+\frac{Q}{r},
\qquad
A_{r(0)}(r)
=
\pm 
\frac{1}{A_{(0)} (r)}
\sqrt{
A_{t(0)}(r)^2
-
\frac{m_p^2\left(4m^2+\Lambda\right)}{2 m^2}
A_{(0)}(r)
},
\end{eqnarray}
and 
\begin{eqnarray}
\omega_{(1)}(r)
=
\omega_0+ \frac{2J}{r^3},
\qquad
a_{3(1)}(r)=-\frac{JQ}{M r^3}.
\end{eqnarray}
Since the norm of the vector field ${\cal Y}=-\frac{m_p^2 (4m^2+\Lambda)}
         {2m^2}
+{\cal O} (\epsilon^2)$,
up to ${\cal O} (\epsilon)$,
$\P({\cal Y})=\P_{(0)}\left(-\frac{m_p^2 (4m^2+\Lambda)}{2m^2}\right)$ 
and 
$\Q({\cal Y})=\Q_{(0)}\left(-\frac{m_p^2 (4m^2+\Lambda)}{2m^2}\right)$ 
also remain constant.
Thus, the disformed solution written in terms of the original time coordinate \eqref{ori}
is given by 
\begin{eqnarray}
\label{sds2}
d{\tilde s}^2
&=&
\P_{(0)}
\left[
-
\left(
1
-\frac{2M}{r}
+\frac{4}{3}m^2 r^2
-\frac{\Q_{(0)}}{\P_{(0)}}
\left( 
\frac{m_p \sqrt{4m^2+\Lambda}}{\sqrt{2}m}
+\frac{Q}{r}
\right)^2
\right)
dt^2
\right.
\nonumber\\
&+&
\frac{1}{3\left(r-2M+\frac{4m^2r^3}{3}\right)^2}
\nonumber\\
&\times&
\left(
r
\left(
3r
-6M 
+4m^2 r^3
\right)
+\frac{\Q_{(0)}}{\P_{(0)}}
\left(
  3Q^2
+\frac{3m_p Q r \sqrt{2(4m^2+\Lambda)}}
          {m}
+\frac{m_p^2 r (3M-2m^2r^3)\left(4m^2+\Lambda\right) } 
          {m^2}
\right)
\right)
dr^2
\nonumber\\
&
\pm 
&
\frac{2\Q_{(0)}\left(\frac{Q}{r}+\frac{m_p\sqrt{4m^2+\Lambda}}{\sqrt{2}m}\right)}
        {\P_{(0)}\left[\frac{4m^3 r^3}{3} + m (r-2M)\right]}
\sqrt{
m^2Q^2
+m m_p Q r \sqrt{2(4m^2+\Lambda)}
+\frac{m_p^2 r} {3}
  \left(3M-2m^2r^3\right)
  (4m^2+\Lambda)
}
dt dr
\nonumber\\
&+&
r^2
\left(
d\theta^2
+
\sin^2\theta 
d\varphi^2
\right)
\nonumber\\
&-&
\frac{1}{m Mr^2}
\left[
2m M r\left(2J+r^3\omega_0\right)
+
\frac{\Q_{(0)} J Q}{\P_{(0)}}
\left(
2mQ
+
m_p r
\sqrt{2(4m^2+\Lambda)}
\right)
\right]
\sin^2\theta
d{t}d\varphi
\nonumber\\
&\mp&
\left.
\frac{6\Q_{(0)} J Q
        \sqrt{
      m^2 Q^2
+ mm_p Q r\sqrt{2\left(4m^2+\Lambda\right)}
+\frac{m_p^2r}{3}  (3M-2m^2r^3) (4m^2+\Lambda)}
          }
        {\P_{(0)}\left(
         -6m M^2 r+3m M r^2 +4m^3 M r^4
        \right)}
\sin^2\theta
drd\varphi
\right]
+
{\cal O} (\epsilon^2),
\end{eqnarray}
and that rewritten in terms of the new time coordinate \eqref{trans}
is given by
\begin{eqnarray}
\label{newv2}
d{\tilde s}^2
&=&
\P_{(0)}
\left[
-
\left(
1
-\frac{2M}{r}
+\frac{4m^2r^2}{3}
-
\frac{\Q_{(0)}}{\P_{(0)}}
\left( 
\frac{Q}{r}
+\frac{m_p\sqrt{4m^2+\Lambda}}
          {\sqrt{2}m}
\right)^2
\right)
d{\tilde t}^2
\right.
\nonumber\\
&+&
\left(
1
-\frac{2M}{r}
+\frac{4m^2r^2}{3}
-
\frac{\Q_{(0)}}{\P_{(0)}}
\left( 
\frac{Q}{r}
+\frac{m_p\sqrt{4m^2+\Lambda}}
          {\sqrt{2}m}
\right)^2
\right)^{-1}
\left(
1
-\frac{\Q_{(0)} m_p^2 (4m^2+\Lambda)}
          {2m^2 \P_{(0)}}
\right)
dr^2
\nonumber\\
&+&
r^2
\left(
d\theta^2
+
\sin^2\theta 
d\varphi^2
\right)
\nonumber\\
&-&
\frac{1}{m Mr^2}
\left[
2m M r\left(2J+r^3\omega_0\right)
+
\frac{\Q_{(0)} J Q}{\P_{(0)}}
\left(
2mQ
+
m_p r
\sqrt{2(4m^2+\Lambda)}
\right)
\right]
d{\tilde t}d\varphi
\nonumber\\
&
\mp
&
\left(
1
-\frac{2M}{r}
+\frac{4m^2r^2}{3}
-
\frac{\Q_{(0)}}{\P_{(0)}}
\left( 
\frac{Q}{r}
+\frac{m_p\sqrt{4m^2+\Lambda}}
          {\sqrt{2}m}
\right)^2
\right)^{-1}
\left( 
\frac{Q}{r}
+\frac{m_p\sqrt{4m^2+\Lambda}}
          {\sqrt{2}m}
\right)
\nonumber\\
&\times&
\left(
\omega_0
+\frac{2J}{r^3}
+
\frac{JQ\Q_{(0)}}
         {M \P_{(0)} r^3}
\left(
  \frac{Q}{r}
+
\frac{m_p\sqrt{4m^2+\Lambda}}{\sqrt{2}m}
\right)
\right)
\nonumber\\
&\times& 
\left.
\frac{2\Q_{(0)} r
\sqrt{m^2Q^2+ m m_p Q r\sqrt{2(4m^2+\Lambda)}
       +\frac{m_p^2 r}{3} (3M-2m^2r^3) (4m^2+\Lambda) }
}
{m\P_{(0)} 
\left(1 -\frac{2M}{r} +\frac{4m^2r^2}{3}\right) } 
\sin^2\theta
drd\varphi
\right]
+
{\cal O} (\epsilon^2),
\end{eqnarray}
where we assume that $\P_{(0)}>0$ and $\P_{(0)}-m_p^2\Q_{(0)}\left(4m^2+\Lambda \right)/ (2m^2)>0$
for the invertible disformal transformation up to ${\cal O} (\epsilon)$.

By introducing the rescaled mass and charge parameters,
the effective cosmological constant 
\begin{eqnarray}
\tilde{M}
&:=& 
\frac
{m
\left(
2mM \P_{(0)}
+m_p Q \Q_{(0)}
 \sqrt{2 \left(4m^2+\Lambda\right)}
\right)}
{2m^ 2 \P_{(0)}-m_p^2\Q_{(0)}\left(4m^2+\Lambda \right)},
\qquad 
{\tilde \Lambda}
:=
-
\frac{8m^4 \P_{(0)}}
       {2m^ 2 \P_{(0)}-m_p^2\Q_{(0)}\left(4m^2+\Lambda \right)}
<0,
\nonumber\\
{\tilde \Q}
&:=&
\sqrt{
-
\frac{2m^2 \Q_{(0)}}
       {2m^ 2\P_{(0)}-m_p^2\Q_{(0)}\left(4m^2+\Lambda \right)}}Q,
\end{eqnarray}
and the time coordinate
$dT:=\sqrt{ 1
-\frac{\Q_{(0)}}{\P_{(0)}}
\frac{m_p^2 (4m^2+\Lambda)}
        {2m^2}} d{\tilde t}$,
\begin{eqnarray}
d{\tilde s}^2
&=&
\P_{(0)}
\left[
-
\left(
1-\frac{2\tilde M}{r}
-\frac{{\tilde \Lambda}r^2}{3}
+\frac{{\tilde Q}^2}{r^2}
\right)
dT^2
+\frac{1}
         {1-2\frac{\tilde M}{r}-\frac{{\tilde \Lambda}r^2 }{3}+\frac{{\tilde Q}^2}{r^2}}
dr^2
+r^2
\left(
d\theta^2
+
\sin^2\theta 
d\varphi^2
\right)
\right]
+{\cal O} 
\left(
\epsilon
\right).
\end{eqnarray}
Thus, 
at ${\cal O} (\epsilon^0)$,
the disformed solution is 
given by the RN-anti-de Sitter solution
which contains maximally two horizons, i.e., the black hole event and inner Cauchy horizons.
Unlike the conformal transformation, 
the disformal transformation modifies the position of the event horizon,
assuming that
matter is minimally coupled to the metric in each frame.

At ${\cal O} (\epsilon^0)$,
the vector field expressed in terms of the new time coordinate
defined in the disformed frame \eqref{vector_new}
is given by 
\begin{eqnarray}
A_\mu dx^\mu
&=&
A_{t(0)}
\left[
d{\tilde t}
\pm 
\frac{\P_{(0)}}
       {\P_{(0)}A_{(0)}-\Q_{(0)} A_{t(0)}^2}
\sqrt{
A_{t(0)}^2-q^2 A_{(0)}
}
dr
\right]
+{\cal O} (\epsilon)
\nonumber\\
&\approx& 
\frac{A_{t(0)}}
       {\sqrt{1-\frac{\Q_{(0)}}{\P_{(0)}} A_{t(0)}^2}}
\left[
dT
\pm 
\frac{dr}
        {1-\frac{2{\tilde M}}{r}-\frac{{\tilde \Lambda}r^2}{3}+\frac{{\tilde Q}^2}{r^2}}
\right]
+{\cal O} (\epsilon)
=:
\frac{A_{t(0)}}
       {\sqrt{1-\frac{\Q_{(0)}}{\P_{(0)}} A_{t(0)}^2}}
du_{\pm}
+{\cal O} (\epsilon),
\end{eqnarray}
where 
``$\approx $'' means that the near-horizon limit is taken,
and  
$u_+$ and $u_-$
represent the null coordinates
which are 
regular 
at the future and past event horizons,
respectively.
Thus,  
the positive and negative branch solutions of the vector field
are regular at the future and past event horizons, respectively,
and
the disformal transformation does not modify the essential properties
of the vector field
from the original one~\cite{Minamitsuji:2016ydr}.

Since $F_{tr}\neq  0$,
there is no corresponding solution 
in the scalar-tensor theory obtained via the disformal transformation of the theory \eqref{st}.

\section{The disformed asymptotically locally anti-de Sitter solutions with the slow rotation corrections}
\label{sec6}

Finally, 
we focus on the asymptotically locally anti-de Sitter
black hole solutions in the models \eqref{gp} and \eqref{st}.

\subsection{The asymptotically locally anti-de Sitter solutions}

For an arbitrary value of $\beta>0$ in the generalized Proca theory \eqref{gp},
there exist the black hole solutions with the leading order slow rotation corrections 
given by Eqs. \eqref{stationary2} and \eqref{vector2} with 
\cite{Chagoya:2016aar,Minamitsuji:2016ydr}
\begin{eqnarray}
\label{loc_ads}
A_{(0)}(r)
&=&
\frac{1}
       {3mr \beta (m^2 -\beta \Lambda)^2}
\left[
m^7 r^3 
-3mr \beta^3 \Lambda^2
+m^3 r \beta^2 \Lambda 
   \left(-6+r^2\Lambda\right)
+m^5 \beta 
\left(9r-2r^3\Lambda-24M\right)
\right.
\nonumber\\
&+&
\left.
3\beta^{\frac{3}{2}}
(m^2+\beta\Lambda)^2
{\rm arctan} 
\left(
\frac{mr}{\sqrt{\beta}}
\right)
\right],
\nonumber\\
D_{(0)} (r)
&=&
\frac{m^4\left(m^2r^2+\beta (2-r^2\Lambda)\right)^2 }
       { \left(m^2-\beta\Lambda\right)^2 (m^2r^2+\beta)^2 A_{(0)}(r)},
\nonumber\\
A_{t(0)}(r)
&=&
0,
\qquad
A_{r(0)}(r)
=
\pm
m^2 m_p r 
\sqrt{
-
\frac{(m_p^2+\beta\Lambda) \left(2\beta+r^2 (m^2-\beta \Lambda)\right)^2}
        {2\beta (m^2 r^2+\beta)^3 (m^2-\beta\Lambda)^2A_{(0)}(r)}
},
\end{eqnarray}
and 
\begin{eqnarray}
\omega_{(1)}(r)=\omega_0+ \frac{2J}{r^3},
\qquad 
a_{3(1)}(r)=0,
\end{eqnarray}
where we assume $\beta>0$ and $m^2+\beta\Lambda<0$,
and $A_{(0)}(r)\sim D_{(0)}(r)^{-1}\sim  m^2r^2/(3\beta)$ as $r\to \infty$,
which are asymptotically locally anti-de Sitter. 
For $\Lambda<0$,
the equation $A_{(0)}=0$ has a single positive root,
and hence there is only the black hole event horizon \cite{Minamitsuji:2016ydr}.

Since the norm of the vector field ${\cal Y}=-\frac{m_p^2r^2\left(m^2+\beta\Lambda\right)}{2\beta (m^2r^2+\beta)}+{\cal O} (\epsilon^2)$,
up to ${\cal O}(\epsilon)$,
$\P=\P_{(0)} \left(-\frac{m_p^2r^2\left(m^2+\beta\Lambda\right)}{2\beta (m^2r^2+\beta)}\right)$
and 
$\Q=\Q_{(0)}\left(-\frac{m_p^2r^2\left(m^2+\beta\Lambda\right)}{2\beta (m^2r^2+\beta)}\right)$ 
are the pure functions of $r$.
Thus, 
the disformed solution can be obtained as 
\begin{eqnarray}
\label{locads2}
d{\tilde s}^2
&=&
\P_{(0)}
\left[
-
\frac{m^4 
\left(m^2 r^2 +\beta (2-r^2\Lambda)\right)^2 }
       {\left(m^2r^2+\beta\right)^2 \left(m^2-\beta\Lambda\right)^2 D_{(0)}(r)}
dt^2
+
\left(
1
-\frac{m_p^2 \Q_{(0)}  r^2 (m^2+\beta\Lambda)}
         {2 \P_{(0)}\beta (m^2r^2+\beta)}
\right)
D_{(0)}(r)
dr^2
+
r^2
\left(
d\theta^2
+
\sin^2\theta 
d\varphi^2
\right)
\right.
\nonumber\\
&-& 
\left.
2
  r^2
\left(
\omega_0+\frac{2J}{r^3}
\right)
\sin^2\theta
d{t}d\varphi
\right]
+{\cal O} 
\left(
\epsilon^2
\right),
\end{eqnarray}
where $D_{(0)}(r)$ is defined in Eq. \eqref{loc_ads}.
The disformed solution
also contains only the black hole event horizon
located at the point where $A_{(0)}=0$.
Thus,
the disformal transformation does not modify 
the position of the event horizon,
since the character of the vector field is spacelike 
and 
does not affect the causal properties of the solution.
The solution of the vector field remains the same 
before and after the disformal transformation,
and 
$A_{r(0)}$
diverges as $\left(r-r_h\right)^{-\frac{1}{2}}$
in the vicinity of the event horizon.

The disformed metric \eqref{locads2}
with the replacement 
of 
\begin{eqnarray}
&&
\P_{(0)}\left(-\frac{m_p^2r^2\left(m^2+\beta\Lambda\right)}{2\beta (m^2r^2+\beta)}\right)
\to
\P_{s(0)}\left(-\frac{m_p^2r^2\left(m^2+\beta\Lambda\right)}{2\beta (m^2r^2+\beta)}\right),
\\
&&
\Q_{(0)}\left(-\frac{m_p^2r^2\left(m^2+\beta\Lambda\right)}{2\beta (m^2r^2+\beta)}\right)
\to
\Q_{s(0)}\left(-\frac{m_p^2r^2\left(m^2+\beta\Lambda\right)}{2\beta (m^2r^2+\beta)}\right),
\end{eqnarray}
and the scalar field \eqref{qt} 
with 
\begin{eqnarray}
\phi
=\pm
m_p r
\int  dr
\sqrt{
-\frac{\left(m^2+\beta \Lambda\right) D_{(0)}(r) }
         {2\beta (m^2r^2+\beta)}},
\end{eqnarray}
are the solutions in the scalar-tensor theory obtained via the disformal transformation of the theory \eqref{st}
\cite{Rinaldi:2012vy,Anabalon:2013oea,Minamitsuji:2013ura,Maselli:2015yva}.

\subsection{The charged asymptotically locally anti-de Sitter solutions}

For $\beta=1/4$ in the generalized Proca theory \eqref{gp},
there exist the charged  locally anti-de Sitter black hole solutions 
with the leading order slow rotation corrections 
given by Eqs. \eqref{stationary2} and \eqref{vector2} 
with 
\cite{Chagoya:2016aar,Minamitsuji:2016ydr}
\begin{eqnarray}
\label{loc_ads3}
A_{(0)}(r)
&=&
\frac{1}
       {6mr\left (\Lambda-4m^2\right)^2}
\left[
-6\Lambda^2 mr
+128 m^7 r^3
-32 m^5 (24M+2\Lambda r^3-9r)
+8\Lambda m^3 r\left(\Lambda r^2-6\right)
\right.
\nonumber\\
&+&
\left.
3(\Lambda+4m^2)^2
{\rm arctan} 
\left(
2mr
\right)
\right],
\nonumber\\
D_{(0)}
(r)
&=&
\frac{16 m^4\left(4m^2 r^2 -\Lambda r^2 +2)\right)^2 }
       { \left(\Lambda-4m^2\right)^2 \left(4m^2r^2+1\right)^2 A_{(0)} (r)},
\nonumber\\
A_{t(0)}(r)
&=&
\frac{Q}{r},
\qquad
A_{r(0)}(r)
=
\pm
\frac{\sqrt{D_{(0)}(r)}}
       {r\sqrt{A_{(0)}(r)}
\sqrt{1+4m^2 r^2}}
\sqrt{
 Q^2 (1+4m^2 r^2)
-2m_p^2 (\Lambda+4m^2) r^4 A_{(0)} (r)
},
\end{eqnarray}
and 
\begin{eqnarray}
\omega_{(1)}(r)=\omega_0+ \frac{2J}{r^3},
\qquad 
a_{3(1)}(r)=-\frac{JQ}{Mr^3}.
\end{eqnarray}
Here, we keep the nonzero $\omega_0\neq 0$,
because the spacetime is not asymptotically flat.

For $m=\pm \frac{\sqrt{-\Lambda}}{2}$ ($\Lambda<0$),
the above solution \eqref{loc_ads3}
reduces to the Schwarzschild-anti-de Sitter solution
\begin{eqnarray}
A_{(0)}(r)
&=&
\frac{1}{D_{(0)}(r)}
=
1-\frac{2M}{r}-\frac{\Lambda r^2}{3},
\qquad
A_{t(0)}(r)
=\frac{Q}{r},
\qquad
A_{r(0)}(r)
=
\pm \frac{D_{(0)}(r) Q}{r}.
\end{eqnarray}
For simplicity, we focus on this particular case of $m=\pm \frac{\sqrt{-\Lambda}}{2}$.
Since the norm of the vector field ${\cal Y}={\cal O} (\epsilon^2)$,
up to ${\cal O}(\epsilon)$,
we find that
$\P=\P_{(0)} \left(0\right)={\rm const}$
and 
$\Q=\Q_{(0)}\left(0\right)={\rm const}$.
Hence, the disformed solution can be obtained as 
\begin{eqnarray}
\label{dis_loc1}
d{\tilde s}^2
&=&
\P_{(0)} 
\left[
-
\left(
1-\frac{2M}{r}-\frac{\Lambda r^2}{3}
-\frac{\Q_{(0)} Q^2}{\P_{(0)} r^2}
\right)
dt^2
+
\frac{9\left(\Q_{(0)}/\P_{(0)}\right) Q^2-3
    r(6M-3r+r^3\Lambda)}
        { (6M-3r+r^3\Lambda)^2}
dr^2
\right.
\nonumber\\
&-&
\left.
\frac{6Q^2 \Q_{(0)}}
         {\P_{(0)} \left(6Mr-3r^2+r^4\Lambda\right)}
dr dt
+
r^2
\left(
d\theta^2
+
\sin^2\theta 
d\varphi^2
\right)
\right.
\nonumber\\
&-&
\left. 
2\sin^2\theta 
\frac{2JMr
          +M \omega_0 r^4 
          +JQ^2 \left(\Q_{(0)}/\P_{(0)}\right) }
       {Mr^2}
d{t}d\varphi
+
\frac{6\Q_{(0)} J Q^2 \sin^2\theta}
       {\P_{(0)}\left(6M^2 r-3Mr^2 +M r^4\Lambda\right)}
drd\varphi
\right]
\nonumber\\
&+&
{\cal O} 
\left(
\epsilon^2
\right),
\end{eqnarray}
where $D_{(0)}(r)$ is defined in Eq. \eqref{loc_ads3},
which after the coordinate transformation \eqref{tildet}
reduces to 
\begin{eqnarray}
\label{dis_loc2}
d{\tilde s}^2
&=&
\P_{(0)}
\left[
-
\left(
1-\frac{2M}{r}-\frac{\Lambda r^2}{3}
-\frac{\Q_{(0)} Q^2}{\P_{(0)} r^2}
\right)
d{\tilde t}^2
+
\left(
1-\frac{2M}{r}-\frac{\Lambda r^2}{3}
-\frac{\Q_{(0)} Q^2}{\P_{(0)} r^2}
\right)^{-1}
dr^2
\right.
\nonumber\\
&+&
r^2
\left(
d\theta^2
+
\sin^2\theta 
d\varphi^2
\right)
- 
2\sin^2\theta 
\frac{2JMr 
      +M r^4\omega_0 r^4
      +JQ^2 \left(\Q_{(0)}/\P_{(0)}\right)}
       {Mr^2}
d{\tilde t}d\varphi
\nonumber\\
&+&
\left.
\frac{2
 \Q_{(0)} Q^2 \left(J \left(1-\frac{r^2\Lambda}{3}\right)+Mr^2\omega_0\right)}
        {M \left(1-\frac{2M}{r}-\frac{r^2}{3}\Lambda\right)
 \left(
\Q_{(0)} Q^2
+\P_{(0)} r 
\left(2M-r+\frac{r^3}{3}\Lambda\right)
 \right)
}
\sin^2\theta
drd\varphi
\right]
+
{\cal O} 
\left(
\epsilon^2
\right).
\end{eqnarray}
Thus, at ${\cal O} (\epsilon^0)$,
the reference Schwarzschild-anti-de Sitter solution
is disformally 
mapped to the RN-anti-de Sitter solution.
The existence of the nonzero electric charge $Q\neq 0$
modifies the number of the horizons
and the position of the black hole event horizon.

With Eq. \eqref{tildet},
at ${\cal O} (\epsilon)$,
the solution of the vector field expressed in terms of the new time coordinate ${\tilde t}$ 
is given by 
\begin{eqnarray} 
  A_\mu dx^\mu
=\frac{Q}{r}
\left[
d{\tilde t}
\pm 
\frac{dr}
       {1-\frac{2M}{r}-\frac{\Lambda r^2}{3}
-\frac{\Q_{(0)} Q^2}{\P_{(0)} r^2}}
\right]
=\frac{Q}{r}
du_{\pm},
\end{eqnarray}
where $u_+$ and $u_-$ represent the null coordinates 
which are regular at the future and past event horizons,
respectively.
Thus, 
the positive and negative branch solutions of the vector field 
are regular at the future and past event horizons, respectively.
We note that 
in this particular choice of $\Lambda+4m^2=0$,
the character of the vector field is null 
in both the frames before and after the disformal transformation,
${\cal Y}=0$ and $\bar{\cal Y}=0$. 
Even for $\Lambda+4m^2\neq 0$,
the positive and negative branch solutions of the vector field
are regular at the future and past event horizons,
respectively.

Since $F_{tr}\neq  0$,
there is no corresponding solution 
in the scalar-tensor theory obtained via the disformal transformation of the scalar-tensor theory \eqref{st}.

\section{The disformed Kerr-Newman solution}
\label{sec7}

In this section, we consider the disformal transformation
of the black hole solutions in the Einstein-Maxwell theory
\begin{eqnarray}
\label{em}
S
=\int d^4 x
  \sqrt{-g}
\left(
\frac{m_p^2}{2}
R
-\frac{1}{4}
{\cal F}
\right),
\end{eqnarray}
which is the typical and simplest example
of the vector-tensor theories with the $U(1)$ gauge symmetry.
As the unique static and spherically symmetric solution in the theory \eqref{em},
there is the RN solution 
given by Eqs. \eqref{stationary2} and \eqref{vector2}
with 
\begin{eqnarray}
\label{rn}
A_{(0)}(r)
=1-\frac{2M}{r}+\frac{Q^2}{2m_p^2 r^2},
\qquad 
D_{(0)}(r)
=\frac{1}{A_{(0)} (r)},
\qquad
A_{t(0)}(r)
=\frac{Q}{r},
\end{eqnarray}
and $A_{r(0)} (r)$ remains an arbitrary function of $r$ because of the gauge symmetry.
Up to ${\cal O}(\epsilon)$, 
the slow rotation corrections to the RN solution are given by 
\begin{eqnarray}
\label{kn}
\omega_{(1)}(r)=
\frac{2J}{r^3}
\left(
1
-\frac{Q^2}{4m_p^2 M r}
\right),
\qquad 
a_{3(1)}(r)=-\frac{JQ}{M r^3}.
\end{eqnarray}
The slow rotation corrections to the Schwarzschild solution
are obtained 
by smoothly taking the limit of $Q\to 0$.
We recall that
$M$, $J$, and $Q$
represent the mass, angular momentum,
and the electric charge of the black hole,
respectively.

Beyond the slow rotation approximation, 
as the unique stationary and axisymmetric black hole solution
in the Einstein-Maxwell theory \eqref{em},
there is 
the Kerr-Newman solution 
which in the Boyer–Lindquist form  
is given by Eqs.~\eqref{stationary} and \eqref{vector_Ansatz} with 
\begin{eqnarray}
\label{kerr_newman1}
&&
A=1
-\frac{2Mr}{\rho^2}
\left(
1-\frac{Q^2}{4m_p^2 M r}
\right),
\quad 
B= 
\frac{\sin^2\theta}{\rho^2}
\left[
\left(r^2+\frac{J^2}{M^2}\right)^2
-\frac{J^2}{M^2}
\Delta_Q \sin^2\theta
\right],
\nonumber\\
&&
C=
-\frac{2J r }{\rho^2}
\left(
1
-\frac{Q^2}{4m_p^2 M r}
\right)
\sin^2\theta,
\quad
D= \frac{\rho^2}{\Delta_Q},
\quad 
E=\rho^2,
\end{eqnarray}
and 
\begin{eqnarray}
\label{kerr_newman2}
A_t= \frac{Q r}{\rho^2},
\qquad
A_\varphi
= -\frac{J Q r}{M \rho^2}
   \sin^2\theta,
\end{eqnarray}
where we have defined 
\begin{eqnarray}
\label{kerr_newman3}
\Delta_Q(r):= r^2+\frac{J^2}{M^2}
-2Mr\left(1-\frac{Q^2}{4m_p^2 M r }\right),
\qquad 
\rho(r,\theta):= \sqrt{r^2+\frac{J^2}{M^2}\cos^2\theta}.
\end{eqnarray}
The positions of the black hole event and inner Cauchy horizons
are given by 
\begin{eqnarray}
\label{kn_horizon}
r=
M+ \sqrt{M^2-\left(\frac{J^2}{M^2}+\frac{Q^2}{2m_p^2}\right)},
\qquad 
M- \sqrt{M^2-\left(\frac{J^2}{M^2}+\frac{Q^2}{2m_p^2}\right)},
\end{eqnarray}
respectively.
In the limit of the small dimensionless spin $J/M^2\ll 1$,
\begin{eqnarray}
C\to 
-\frac{2J }{r}
\left(
1
-\frac{Q^2}{4m_p^2 M r}
\right)
\sin^2\theta,
\qquad 
A_\varphi
\to  -\frac{J Q }{M r}
   \sin^2\theta,
\end{eqnarray}
which reproduce the slow rotation corrections \eqref{kn}.

\subsection{The slow rotation limit}

We consider the disformal transformation \eqref{disformal3} of the RN solution \eqref{rn} 
with the first-order slow rotation corrections \eqref{kn}.
Assuming that $\R$ and $\W$ in Eq.~\eqref{disformal3} are the regular functions of $\F$ and $\G$,
up to ${\cal O} (\epsilon)$ of the slow rotation approximation,
 $\R$ and $\W$ can be expanded as 
\begin{eqnarray}
\R
(\F)
&=&
\R_{(0)}(r)
+
{\cal O} 
\left(\epsilon^2\right),
\qquad 
\R_{(0)}(r)
= 
1
+
\sum_{m+n\geq 1}^\infty 
\R_{m,n}
\left(\F_{(0)} (r)\right)^m
\left(\G_{(0)} (r)\right)^n
:= 
1
+
\sum_{j\geq 1}^\infty 
{\tilde \R}_{j}
\left(\frac{Q^2}{r^4}\right)^j,
\nonumber\\
\W(\F)
&=&
\W_{(0)}(r)
+
{\cal O} 
\left(\epsilon^2\right),
\qquad 
\W_{(0)}(r)
=1
+
\sum_{m+n\geq 1}^\infty 
\W_{m,n}
\left(\F_{(0)} (r)\right)^m
\left(\G_{(0)} (r)\right)^n
= 
1
+
\sum_{j\geq 1}^\infty 
{\tilde \W}_{j}
\left(\frac{Q^2}{r^4}\right)^j,
\end{eqnarray}
where
${\R}_{m,n}$ and  ${\W}_{m,n}$ ($m,n=0,1,2,\cdots$),
and 
$\tilde{\R}_j$ and $\tilde{\W}_j$ ($j=0,1,2,\cdots$) are constants.
Here, 
we have used
$\F=\F_{(0)} (r) +{\cal O} (\epsilon^2)$
and 
$\G=\G_{(0)} (r) +{\cal O} (\epsilon^2)$
with 
$\F_{(0)}(r)=-\frac{2Q^2}{r^4}$
and 
$\G_{(0)}(r)=2\left(\frac{Q^2}{r^4}\right)^2$.
Thus, 
the metric of the disformed RN solution 
with the leading order slow rotation corrections
\eqref{disformal3} is given by 
\begin{eqnarray}
d{\tilde s}^2
&=&
\left(
\R_{(0)} -\frac{\W_{(0)} Q^2}{r^4}
\right)
\left[
-\left(1-\frac{2M}{r} +\frac{Q^2}{2m_p^2 r^2}\right)
dt^2
+
\frac{ dr^2}
{1-\frac{2M}{r}  +\frac{Q^2}{2m_p^2 r^2}}
\right]
\nonumber\\
&+&
\R_{(0)}
r^2
\left(
d\theta^2
+
\sin^2\theta 
d\varphi^2
\right)
\nonumber\\
&-&
2
\sin^2\theta
\left( 
\frac{2J\R_{(0)}}{r}
-
\frac{JQ^2\R_{(0)}}
       {2 M m_p^2 r^2}
+\frac{JQ^2 \W_{(0)} \left(Q^2+2m_p^2 r (r-2M)\right)}
         {2M m_p^2r^6}
\right)
dt
d\varphi
+
{\cal O} (\epsilon^2).
\end{eqnarray}
In the limit of $r\to \infty$,
as $\R_{(0)}\to 1$ and $\W_{(0)}\to 1$,
the disformed metric reduces to
\begin{eqnarray}
d{\tilde s}^2
&\to&
-dt^2
+dr^2
+r^2
\left(
d\theta^2
+
\sin^2\theta 
d\varphi^2
\right)
-
\frac{4J}{r}
\sin^2\theta
dt
d\varphi
+
{\cal O} (\epsilon^2).
\end{eqnarray}
Thus, 
unlike the examples of the disformal transformation 
without the $U(1)$ gauge symmetry,
up to ${\cal O} (\epsilon)$,
the position of the black hole event horizon in the slow rotation limit 
$r=M+\sqrt{M^2-\frac{Q^2}{2m_p^2}}$
is not modified by the disformal transformation \eqref{disformal3}.
This is just the confirmation of the general argument 
in Sec. \ref{sec32}.
In the next subsection, 
we will confirm whether 
the position of the black hole event horizon 
is modified or not,
going beyond the slow-rotation approximation.

\subsection{The disformed Kerr-Newman solution}
\label{disf_kn}

Since $A_r=A_\theta=0$, 
the metric of the disformed Kerr-Newman solution
reduces to the form of  Eq.~\eqref{stationary_gauge} 
with the vanishing terms in the last line,
which definitively satisfies the circularity conditions \eqref{circularity0}. 
Since the explicit form of the disformed solution 
obtained after the substitution of Eqs. \eqref{kerr_newman1}-\eqref{kerr_newman3}
is involved, 
we do not show it explicitly.
In this subsection, 
we investigate the basic properties of the disformed Kerr-Newman
solution in the vicinity of the event horizon numerically
and discuss the behaviors in the large distance limit.
For simplicity, 
we assume that 
$\R=1$ and $\W=\W_{0} ={\rm const}$.

Substituting Eq. \eqref{stationary} into Eq. \eqref{disformal3}
with Eqs. \eqref{kerr_newman1}-\eqref{kerr_newman3},
we obtain the disformed Kerr-Newman solution
satisfying the circularity conditions \eqref{circularity0},
whose structure is given by 
\begin{eqnarray}
\label{circular_disform}
d{\tilde s}^2
&=&
-{\tilde A}dt^2
+{\tilde B} d\varphi^2
+2 {\tilde C} dt d\varphi
+
{\tilde  D} dr^2
+ 
{\tilde E} d\theta^2
=
-\left(
{\tilde A}
+
\frac{{\tilde C}^2}
        {{\tilde B}}
\right)
dt^2
+{\tilde B} 
\left(
d\varphi
+\frac{{\tilde C}}{{\tilde B}} dt
\right)^2
+
{\tilde  D} 
dr^2
+ 
{\tilde E} d\theta^2.
\end{eqnarray}
If the black hole event horizon exists,
its position is determined by the equations
$
{\tilde A}
+
{\tilde C}^2/{\tilde B}
=
{\tilde  D}^{-1}
=0$.
In the pure Kerr case of $Q=0$,
the disformed solutions coincide 
with the original solutions
for any value of the dimensionless spin parameter $J/M^2\lesssim 1$
and the disformed solution is given by the Kerr solution
with the same $M$ and $J$,
and hence
the position of the black hole event horizon remains
the same as the Kerr case $r=M+ \sqrt{M^2-J^2/M^2}$.
In the case $Q\neq 0$,
although 
the disformed solution 
is no longer given by the Kerr-Newman metric,
the position of the black hole event horizon
remains the same as 
that in the original frame shown in Eq. \eqref{kn_horizon}
for any value of the polar angle $\theta$.
However,
for a sufficiently large positive or negative value of $\W_0$,
singularities can be formed 
at the finite values of $r$ inside or outside the event horizon,
whose position crucially depends
on the values of $\W_0$ and $Q$.

\begin{figure}[h]
\unitlength=1.1mm
\begin{center}\
  \includegraphics[height=4.5cm,angle=0]{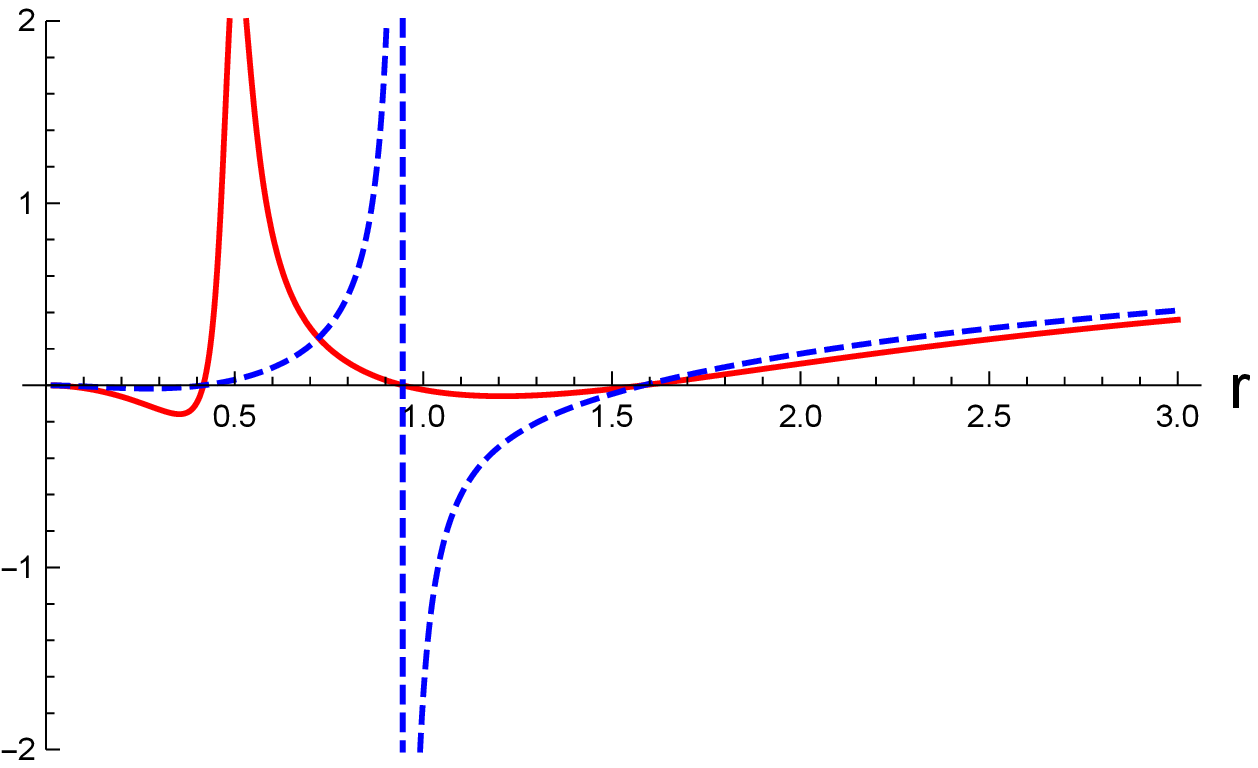}  
  \includegraphics[height=4.5cm,angle=0]{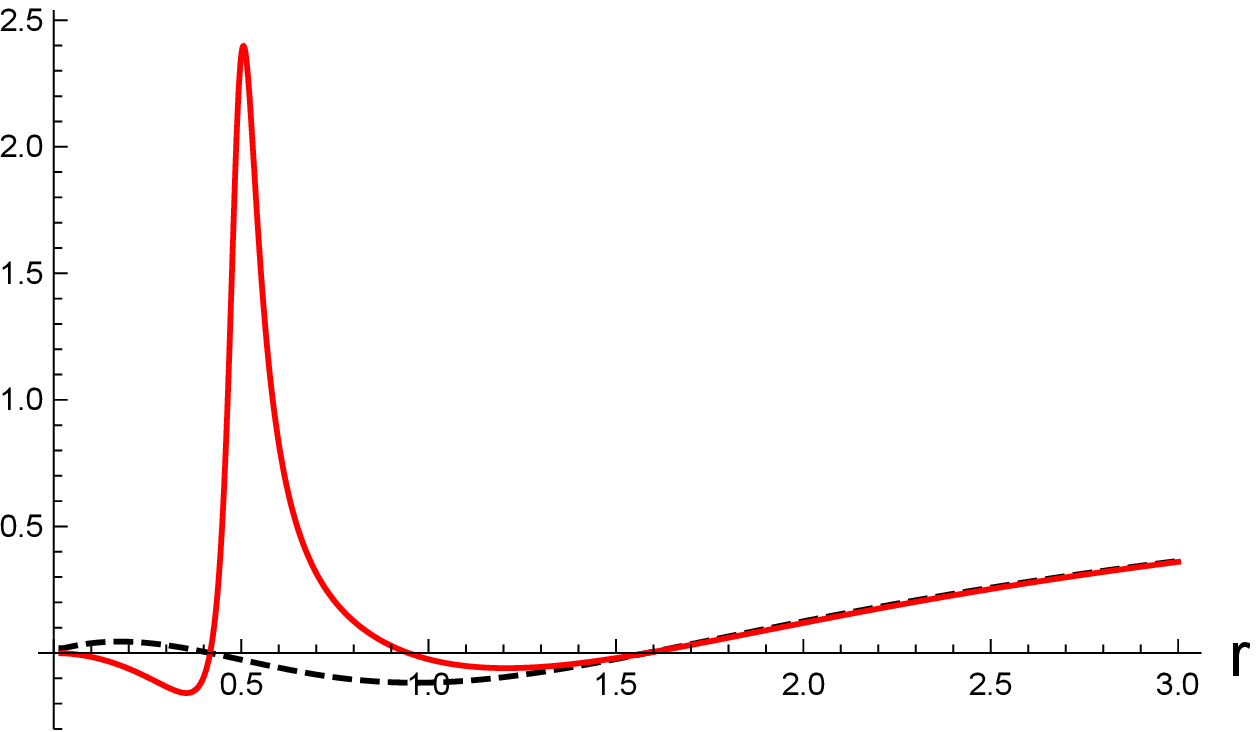}
  \includegraphics[height=4.5cm,angle=0]{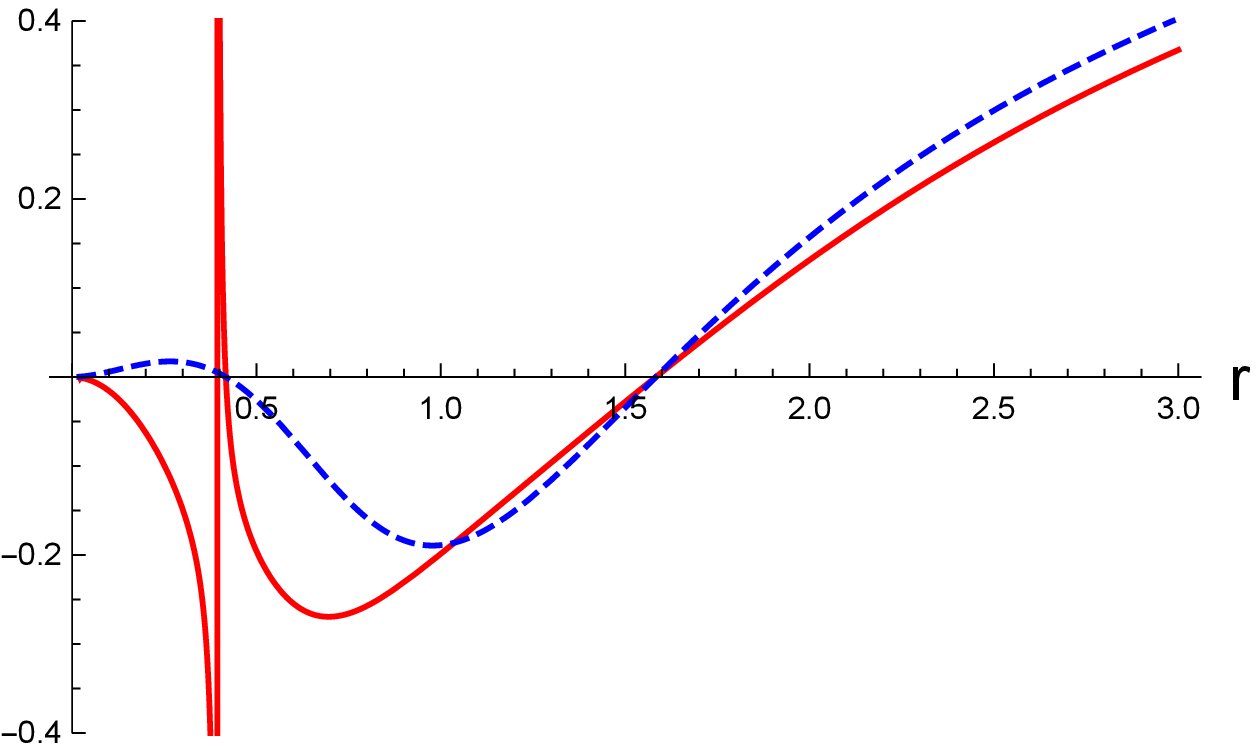}
  \includegraphics[height=4.5cm,angle=0]{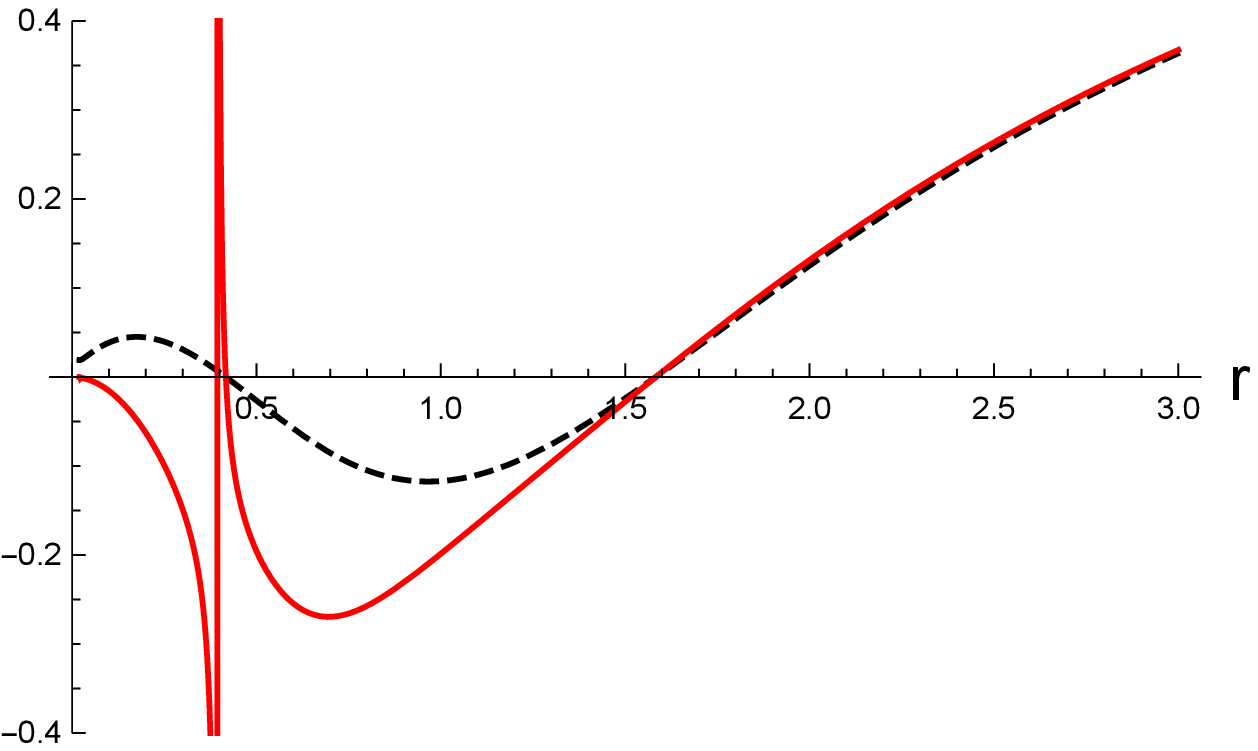}
\caption{
In the left panels,
${\tilde A}+{\tilde C}^2/{\tilde B}$ (the disformed Kerr-Newman solution, the red-solid curves)
and
${\tilde D}^{-1}$ (the disformed Kerr-Newman solution, the blue-dashed curves)
are shown as the functions of $r$.
In the right panels,
${\tilde A}+{\tilde C}^2/{\tilde B}$ (the disformed Kerr-Newman solution, the red-solid curves)
and
${A}+{C}^2/{B}$ (the Kerr-Newman solution, the black-dashed curves)
are shown as the functions of $r$.
We choose $\W_0=20.0$ and $\W_0=-20.0$ in the top and bottom panels, respectively,
and set the other parameters to $J=0.8$, $Q=0.2$, and $\theta=\pi/2$ in the units of $m_p=M=1$.
The black hole event horizon is located at $r=1.5831$
in the units of $m_p=M=1$.
}
  \label{disfig}
\end{center}
\end{figure}

In Fig. \ref{disfig},
we show a few examples of the disformed Kerr-Newman solution
in the vicinity of the black hole event horizon.
The metric functions in the disformed Kerr-Newmans solution 
are shown as the functions of $r$.
In the left panels,
${\tilde A}+{\tilde C}^2/{\tilde B}$ (the disformed Kerr-Newman solution, the red-solid curves)
and
${\tilde D}^{-1}$ (the disformed Kerr-Newman solution, the blue-dashed curves)
are shown as the functions of $r$.
In the right panels,
${\tilde A}+{\tilde C}^2/{\tilde B}$ (the disformed Kerr-Newman solution, the red-solid curves)
and
${A}+{C}^2/B$ (the Kerr-Newman solution, the black-dashed curves)
are shown as the functions of $r$.
We choose $\W_0=20.0$ and $\W_0=-20.0$ in the top and bottom panels, respectively,
and set the other parameters to $J=0.8$, $Q=0.2$, and $\theta=\pi/2$ in the units of $m_p=M=1$.
\begin{figure}[h]
\unitlength=1.1mm
\begin{center}
  \includegraphics[height=4.5cm,angle=0]{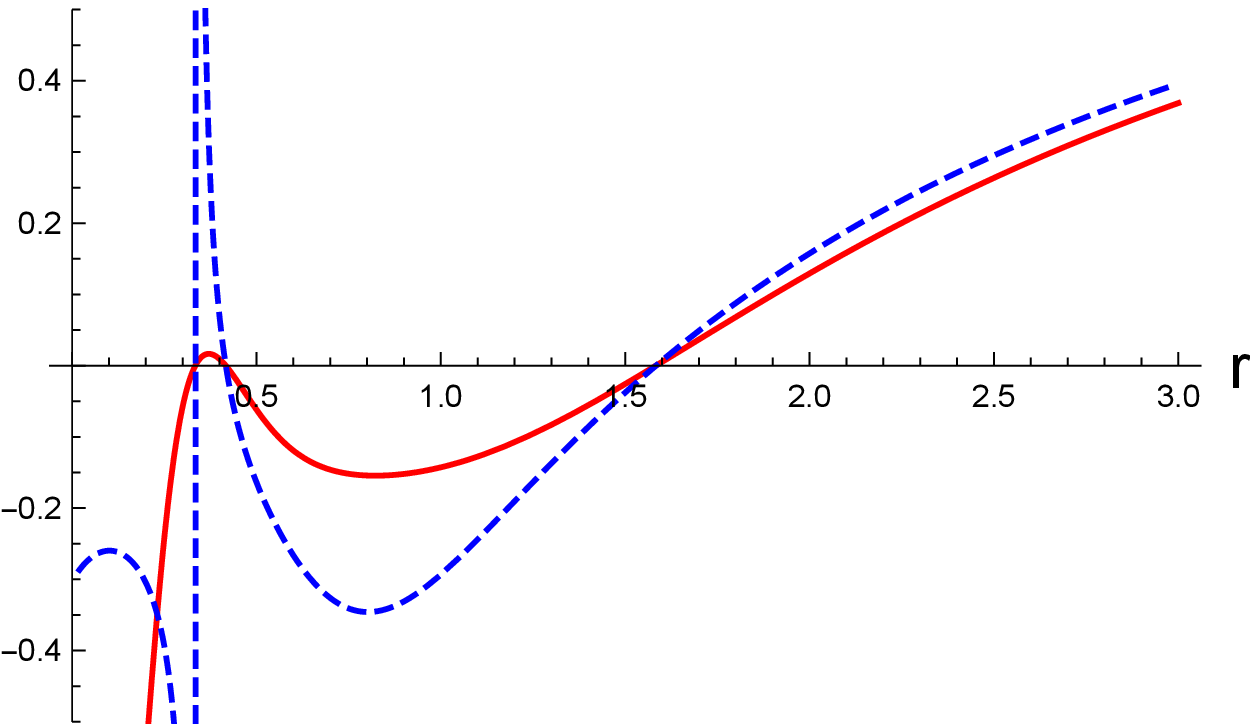}  
  \includegraphics[height=4.5cm,angle=0]{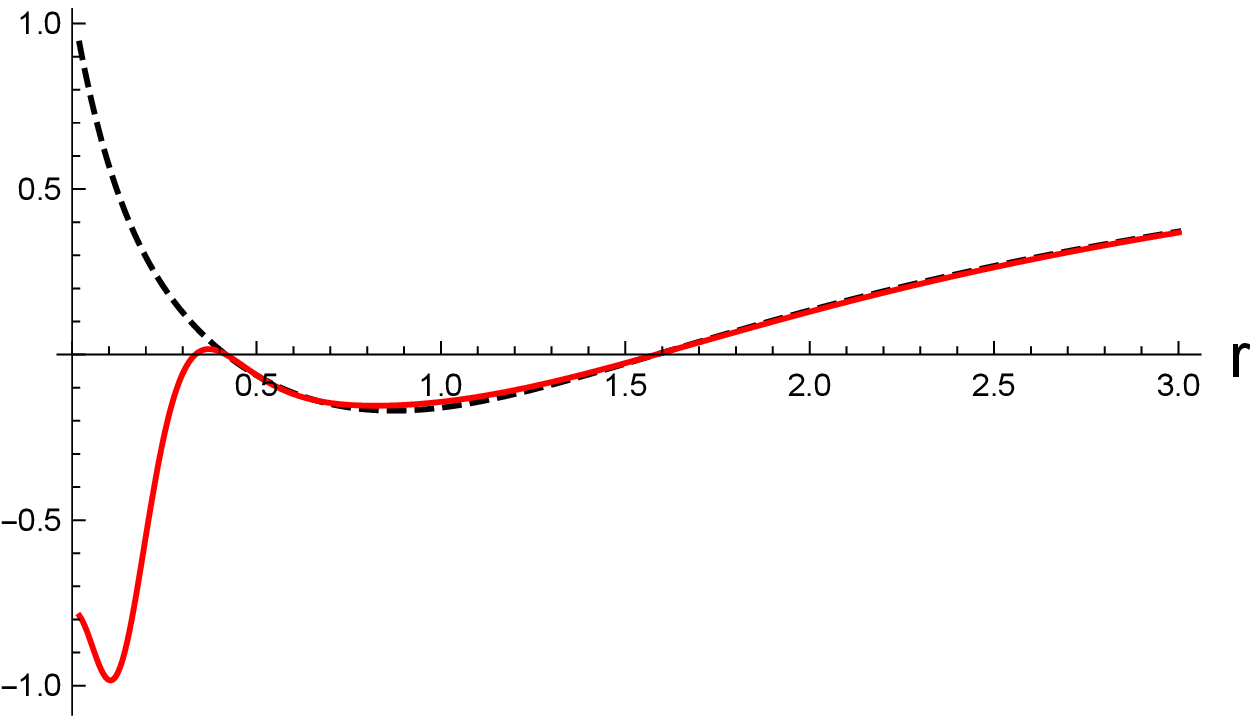}
  \includegraphics[height=4.5cm,angle=0]{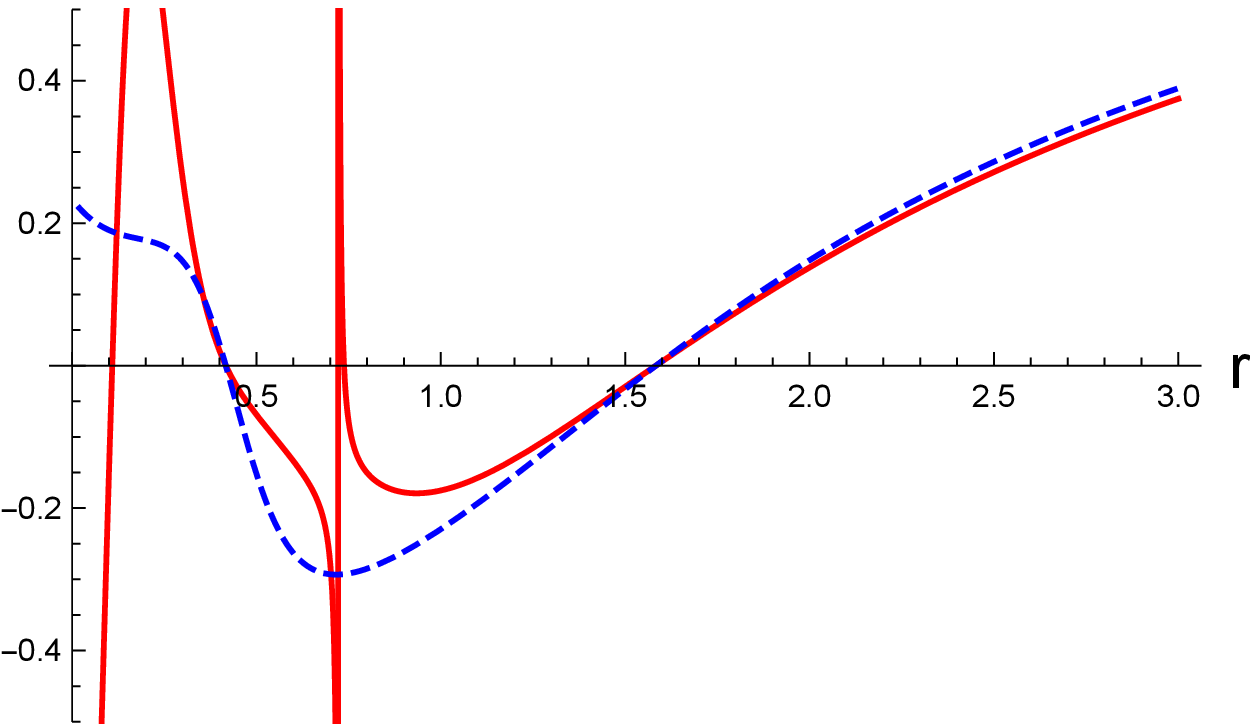}
  \includegraphics[height=4.5cm,angle=0]{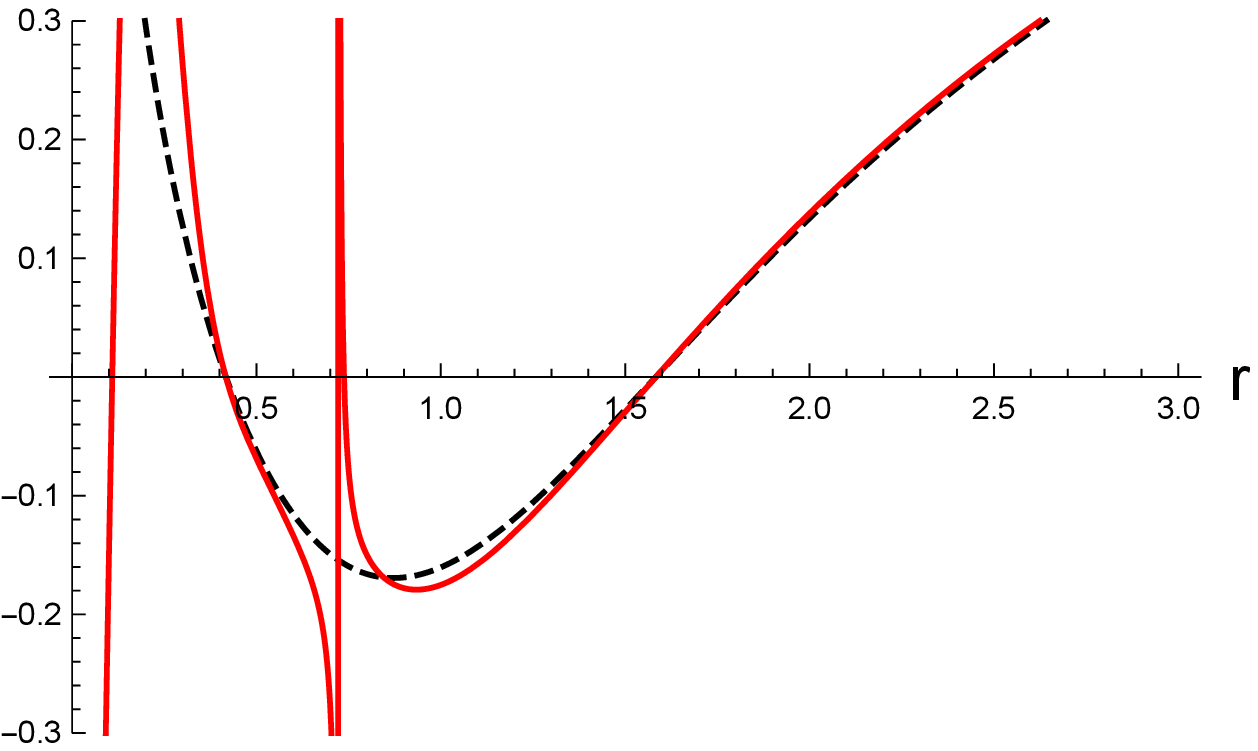}
\caption{
The same plots as Fig. \ref{disfig} for 
the different angle $\theta=\pi/4$.
}
  \label{disfig2}
\end{center}
\end{figure} 
The same plots as Fig. \ref{disfig} for the different angle $\theta=\pi/4$
are shown in Fig \ref{disfig2}.
In these examples,
the black hole event horizon is located  at $r=1.5831$
in the units of $m_p=M=1$
irrespective of the polar angle $\theta$,
and there are several singularities
at the finite values of $r$ inside the black hole event horizon.
The value of $\W_0$ does not affect the position of the event horizon,
which remains the same as that in the original solution before the disformal transformation.
Outside the event horizon, 
the difference between the disformed Kerr-Newman and original Kerr-Newman solutions
quickly approaches zero.

\begin{figure}[h]
\unitlength=1.1mm
\begin{center}
  \includegraphics[height=4.5cm,angle=0]{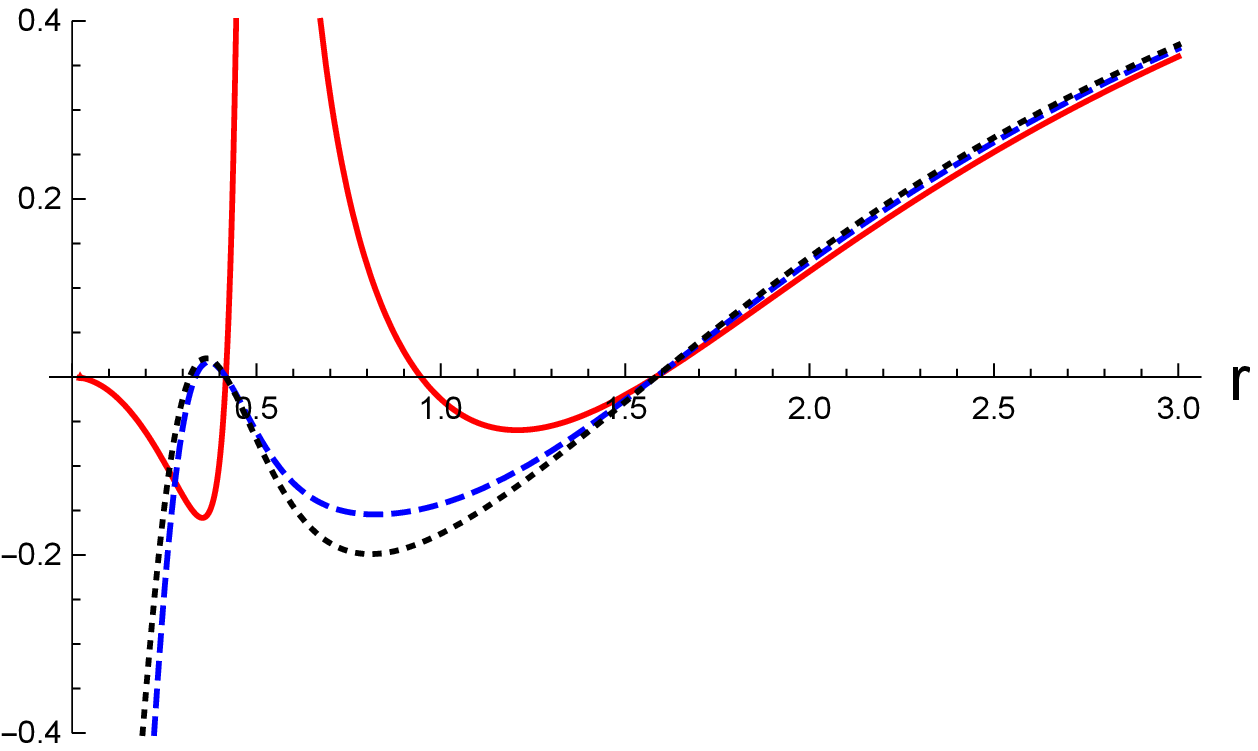}  
  \includegraphics[height=4.5cm,angle=0]{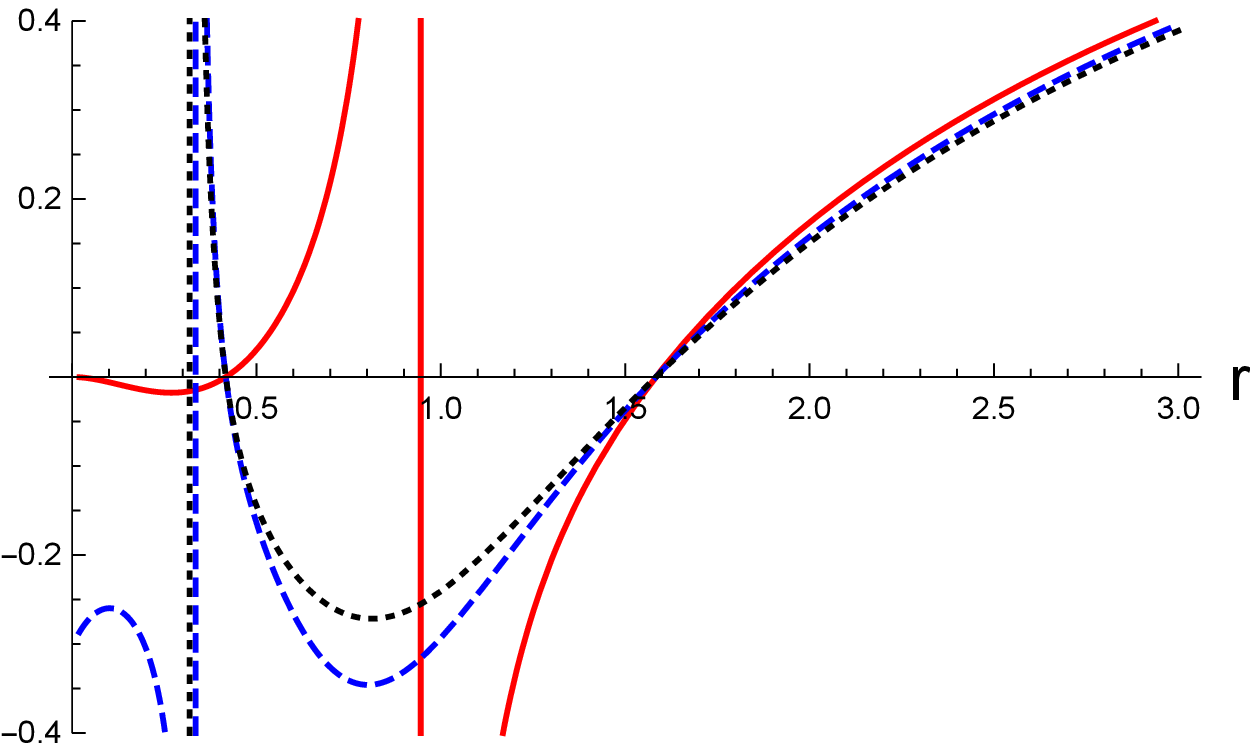}
  \includegraphics[height=4.5cm,angle=0]{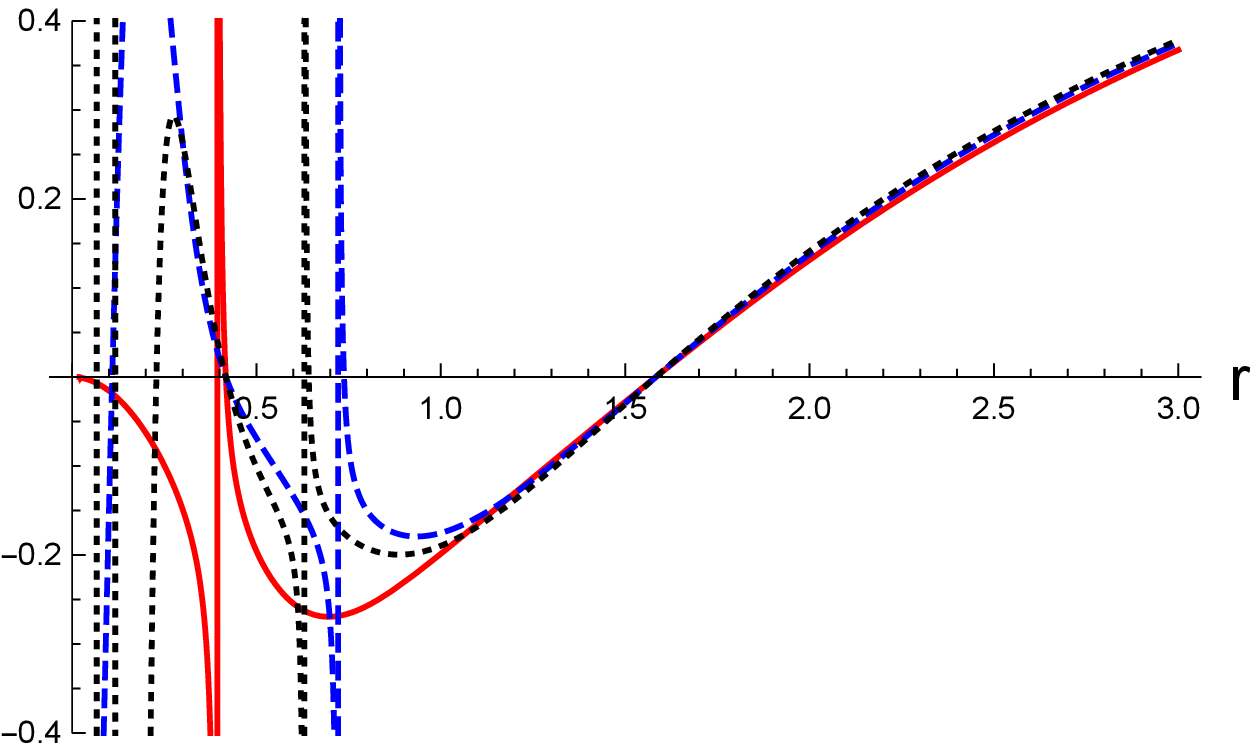}
  \includegraphics[height=4.5cm,angle=0]{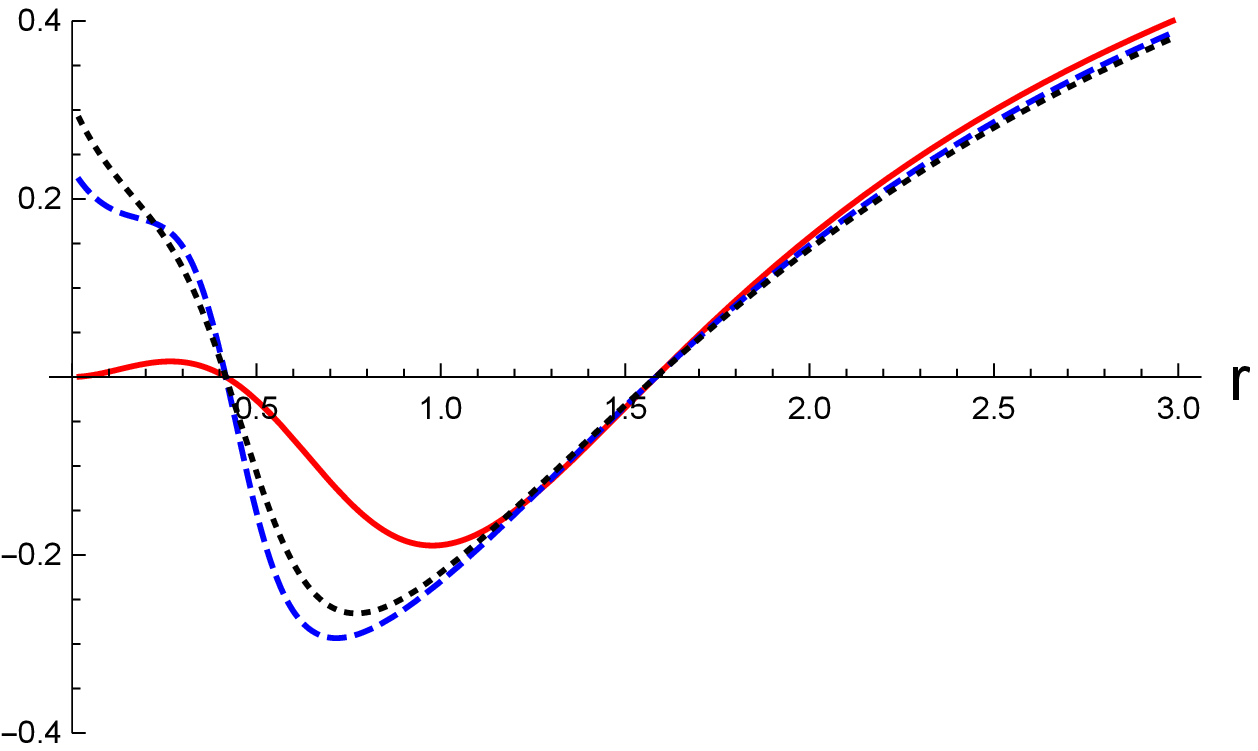}
\caption{
In the left panels,
${\tilde A}+{\tilde C}^2/{\tilde B}$
(the disformed Kerr-Newman solution)
is shown as the function of $r$
for $\theta=\pi/2, \pi/4, \pi/6$
by the red-solid, blue-dashed, and black-dotted curves,
respectively.
In the right panels,
${\tilde D}^{-1}$ (the disformed Kerr-Newman solution)
is shown as the function of $r$
for $\theta=\pi/2, \pi/4, \pi/6$
by the red-solid, blue-dashed, and black-dotted curves,
respectively.
We choose $\W_0=20.0$ and $\W_0=-20.0$ in the top and bottom panels, respectively,
and set the other parameters to $J=0.8$ and $Q=0.2$ in the units of $m_p=M=1$.
In these examples,
the black hole event horizon is located at $r=1.5831$ in the units of $m_p=M=1$.
}
  \label{disfig3}
\end{center}
\end{figure}

\begin{figure}[h]
\unitlength=1.1mm
\begin{center}
  \includegraphics[height=4.5cm,angle=0]{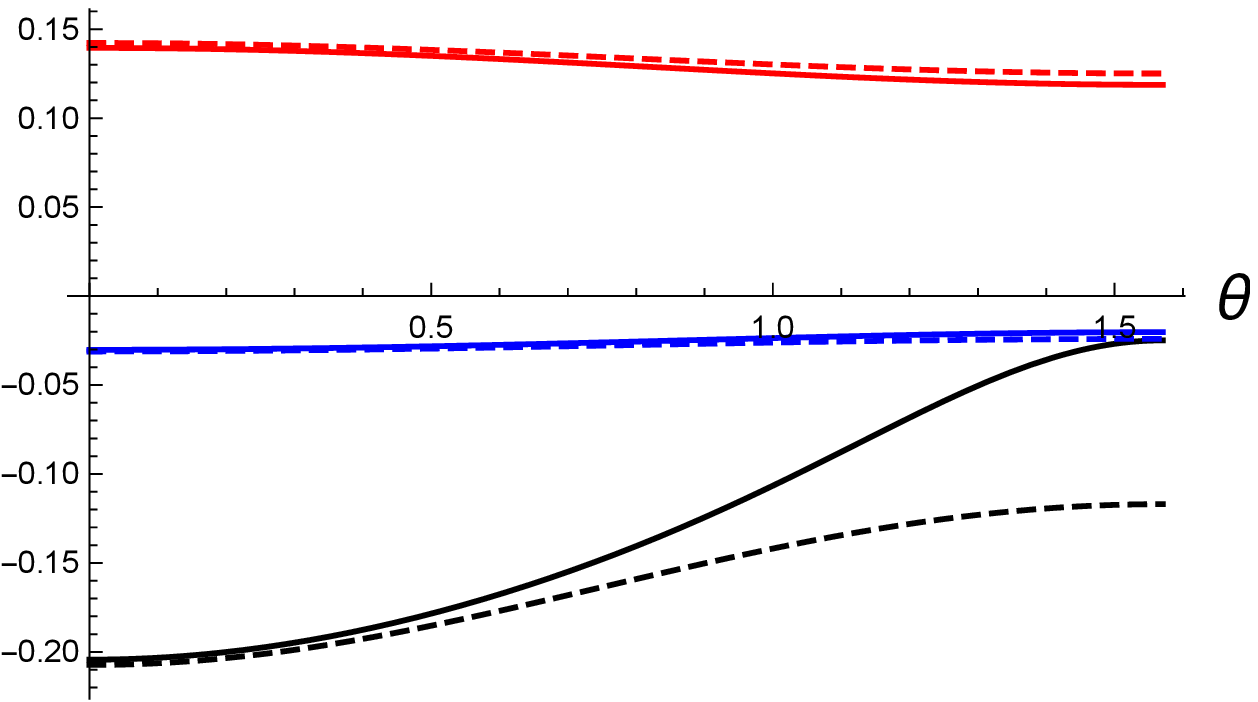}  
  \includegraphics[height=4.5cm,angle=0]{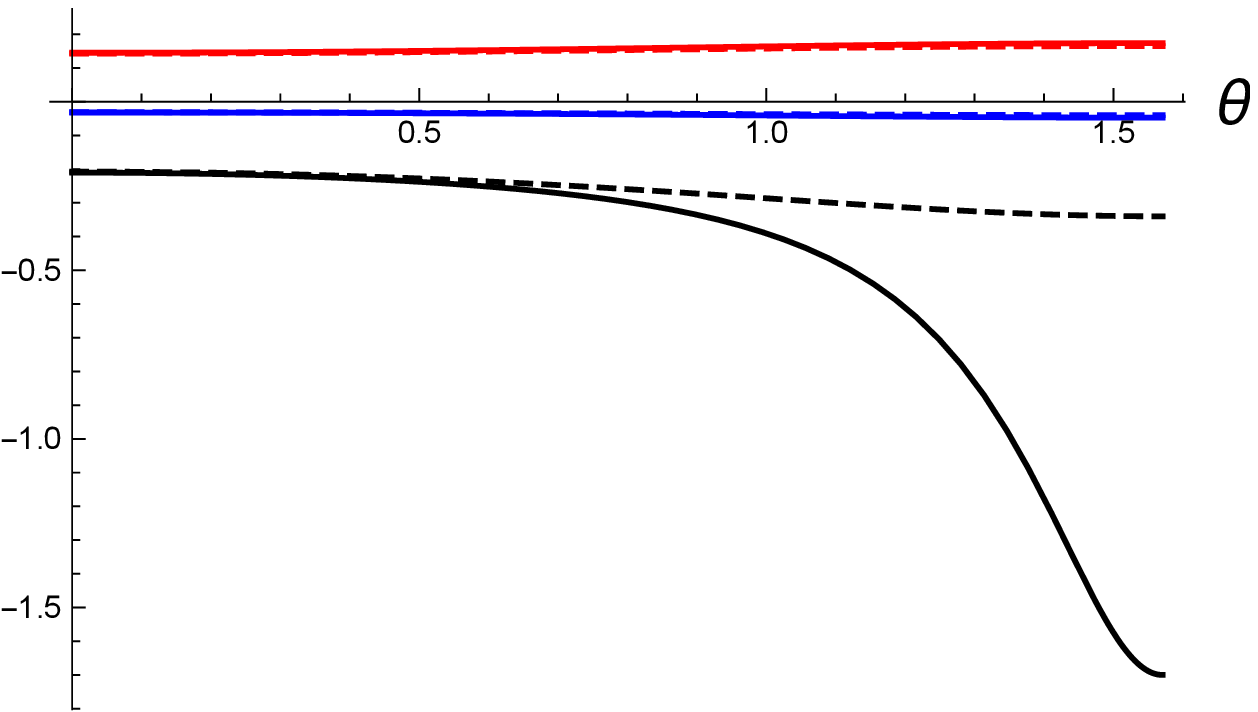}
  \includegraphics[height=4.5cm,angle=0]{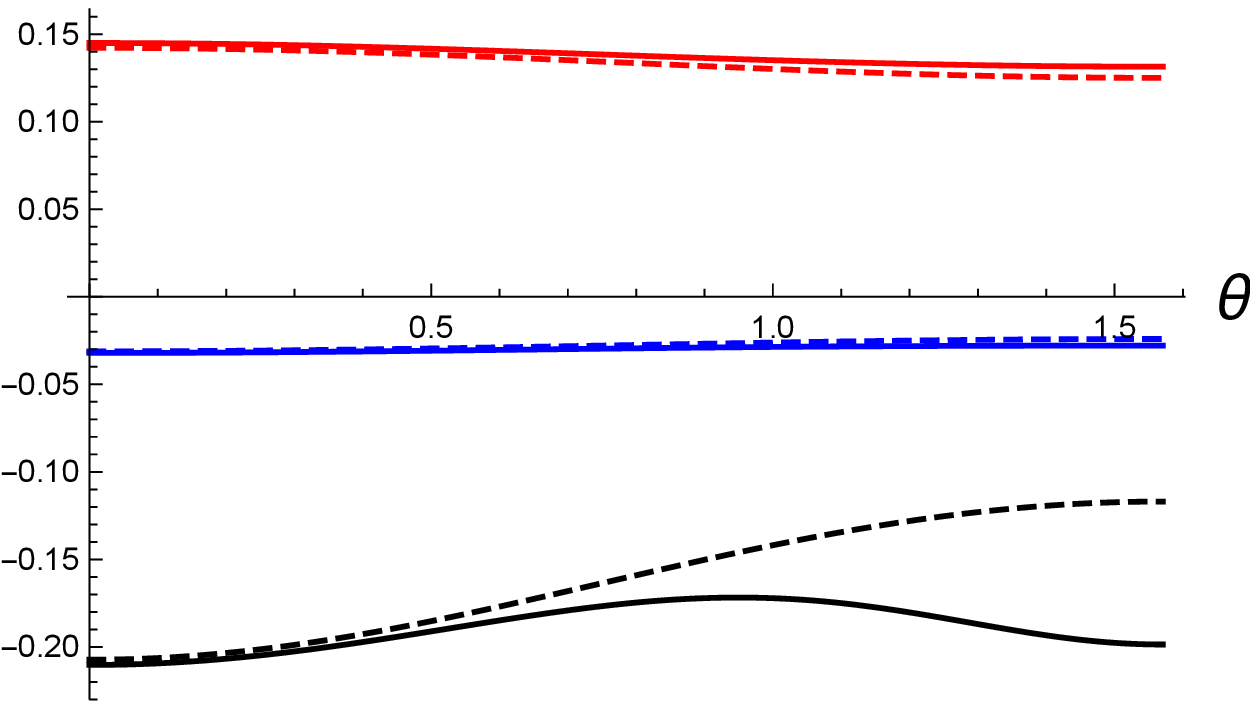}
  \includegraphics[height=4.5cm,angle=0]{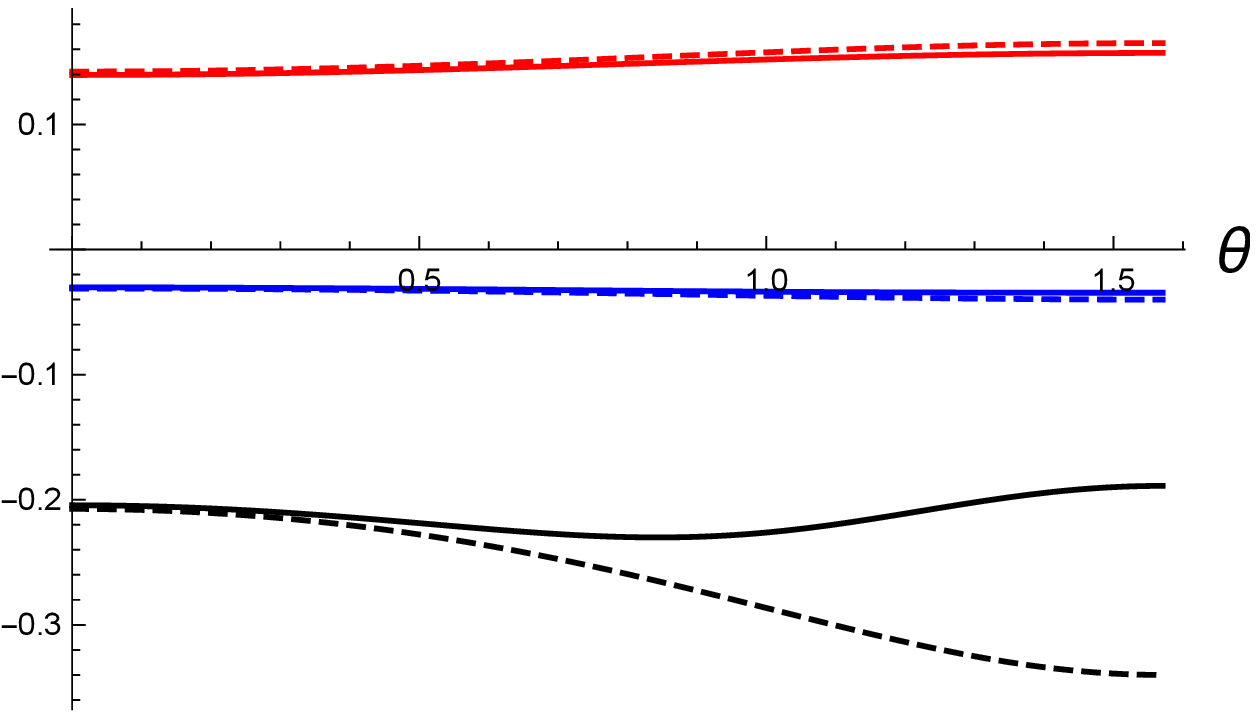}
\caption{
In the left panels,
${\tilde A}+{\tilde C}^2/{\tilde B}$ (the disformed Kerr-Newman solution, the solid curves)
and 
${A}+{C}^2/B$ (the Kerr-Newman solution, the dashed curves)
are shown as the functions of $\theta$
respectively.
In the right panels,
${\tilde D}^{-1}$ (the disformed Kerr-Newman solution, the solid curves)
and 
${D}^{-1}$ (the Kerr-Newman solution, the dashed curves)
are shown as the functions of $\theta$.
The red, blue, and black curves correspond to 
the positions of  $r=2.0, 1.5, 1.0$ in the units of $m_p=M=1$, respectively.
For the disformed Kerr-Newman solution, 
we choose $\W_0=20.0$ and $\W_0=-20.0$ in the top and bottom panels, respectively,
and set the other parameters to $J=0.8$ and $Q=0.2$ in the units of $m_p=M=1$.
In these examples,
the black hole event horizon is located at $r=1.5831$ in the units of $m_p=M=1$.
We note that the horizontal axis denotes $\theta$ between $0$ and $\pi/2$,
as the metric components are symmetric across the equatorial plane $\theta=\pi/2$. 
}
  \label{disfig4}
\end{center}
\end{figure}

In Fig, \ref{disfig3},
${\tilde A}+{\tilde C}^2/{\tilde B}$ and ${\tilde D}^{-1}$
(the disformed Kerr-Newman solution, 
see Eq. \eqref{circular_disform})
for several different polar angles
with the fixed other parameters
are shown.
In the left panels,
${\tilde A}+{\tilde C}^2/{\tilde B}$ (the disformed Kerr-Newman solution)
is shown as the function of $r$
for $\theta=\pi/2, \pi/4, \pi/6$
by the red-solid, blue-dashed, and black-dotted curves,
respectively.
In the right panels,
${\tilde D}^{-1}$ (the disformed Kerr-Newman solution) 
is shown as the function of $r$
for $\theta=\pi/2, \pi/4, \pi/6$
by the red-solid, blue-dashed, and black-dotted curves,
respectively.
We choose $\W_0=20.0$ and $\W_0=-20.0$ in the top and bottom panels, respectively,
and set the other parameters to $J=0.8$ and $Q=0.2$ in the units of $m_p=M=1$.
In these examples,
the black hole event horizon is located at $r=1.5831$ in the units of $m_p=M=1$,
irrespective of $\theta$.
In Fig. \ref{disfig4},
the angular dependence of the metric functions 
in the Kerr-Newman and disformed Kerr-Newman solutions
is shown as the function of $\theta$.
We note that the horizontal axis denotes $\theta$ between $0$ and $\pi/2$,
as the metric components are symmetric across the equatorial plane $\theta=\pi/2$. 
In the left panels,
${\tilde A}+{\tilde C}^2/{\tilde B}$ (the disformed Kerr-Newman solution, the solid curves)
and 
${A}+{C}^2/{B}$ (the Kerr-Newman solution, the dashed curves)
are shown as the functions of $\theta$
respectively.
In the right panels,
${\tilde D}^{-1}$ (the disformed Kerr-Newman solution, the solid curves)
and 
${D}^{-1}$ (the Kerr-Newman solution, the dashed curves)
are shown as the functions of $\theta$.
The red, blue, and black curves correspond to 
the positions of  $r=2.0, 1.5, 1.0$ in the units of $m_p=M=1$, respectively.
For the disformed Kerr-Newman solution, 
we choose $\W_0=20.0$ and $\W_0=-20.0$ in the top and bottom panels, respectively,
and set the other parameters to $J=0.8$ and $Q=0.2$ in the units of $m_p=M=1$.
In both Figs. \ref{disfig3} and \ref{disfig4},
the black hole event horizon is located at $r=1.5831$ in the units of $m_p=M=1$.
In these examples,
the singularities are hidden by the event horizon. 
We find that the position of the event horizon does not depend on
the polar angle $\theta$,
and 
the spacetime structure inside the horizon and 
the position of the singularities 
depend on $\theta$.
The deviation from the Kerr-Newman solution is 
more significant in the vicinity of the equatorial plane $\theta=\pi/2$
and for smaller values of $r$.

The leading order contributions to the disformed Kerr-Newman metric
in the large distance limit $r\to \infty$
are given by 
\begin{eqnarray}
{\tilde g}_{tt}
&=&
-1 
+\frac{2M}{r}
-\frac{Q^2}{2m_p^2 r^2}
-\frac{2J^2 \cos^2\theta}{M r^3}
+ {\cal O} \left(\frac{1}{r^4}\right),
\nonumber\\
{\tilde g}_{rr}
&=&
1 +\frac{2M}{r}
+\frac{1}{r^2}
\left(
4M^2
-\frac{Q^2}{2m_p^2}
-\frac{J^2}{M^2}
\sin^2\theta
\right)
+\frac{M}{r^3}
\left(
-\frac{3J^2}{M^2}
+8M^2
+\frac{J^2}{M^2}
\cos(2\theta)
-\frac{2Q^2}{m_p^2}
\right)
+ {\cal O} \left(\frac{1}{r^4}\right),
\nonumber\\
{\tilde g}_{\theta\theta}
&=&r^2
\left(
1+\frac{J^2}{M^2 r^2}\cos^2\theta
+{\cal O} \left(\frac{1}{r^4}\right)
\right),
\nonumber\\
{\tilde g}_{\varphi\varphi}
&=&
r^2 \sin^2\theta
\left(
1
+
\frac{J^2}{ M^2 r^2} 
+
\frac{2J^2 \sin^2\theta}{M r^3}
+
{\cal O} \left(\frac{1}{r^4}\right)
\right),
\nonumber\\
{\tilde g}_{t\varphi}
&=&
-
\frac{2J \sin^2\theta}{r}
\left(
1
-\frac{Q^2}{4M m_p^2 r}
-\frac{J^2 \cos^2\theta}{M^2 r^2}
+\frac{Q^2}{4Mm_p^2 r^3}
 \left(
 \frac{J^2}{M^2}
\cos^2\theta
+2m_p^2 \W_0
\right)
+{\cal O} 
\left(
\frac{1}{r^4}
\right)
\right).
\end{eqnarray}
Thus,
for a generic dimensionless spin parameter  $J/M^2\lesssim 1$, 
the leading order disformal corrections,
which depend on $\W_0$,
appear in the frame-dragging component ${\tilde g}_{t\varphi}$
at ${\cal O} \left(1/r^4\right)$,
which is suppressed by the factor of $1/r^3$
in comparison with the leading order Kerr term $-(2J \sin^2\theta)/r$.
We note in the case that $\W$ is given by some positive powers of $\F$ and $\G$,
the disformal correction to the frame-dragging component ${\tilde g}_{t\varphi}$
is suppressed by inverse powers of $r$ higher than $r^{-3}$.
Thus, 
the disformed black hole solution
may be able to be distinguished 
from the original Kerr-Newman solution
only via the direct measurements such as black hole shadows.
The detailed studies of the observational properties of the disformed Kerr-Newman solution
will be left for future work.

\section{The disformed solutions in the Einstein-conformally coupled scalar field theory}
\label{sec8}

Finally,
as the example of scalar-tensor theories without the shift symmetry,
we consider the Einstein-conformally coupled scalar field theory
\begin{eqnarray}
\label{ecs}
S
=
\int d^4x
\sqrt{-g}
\left(
\frac{m_p^2}{2}R
-\frac{1}{2}
\left(
{\cal X}
+\frac{1}{6}R\phi^2
\right)
\right).
\end{eqnarray}
There exists the BBMB solution
\cite{bbmb1,bbmb2}
with the leading order slow rotation corrections 
given by Eqs. \eqref{stationary2} and \eqref{qt} 
with 
\begin{eqnarray}
\label{bbmb}
A_{(0)}(r)
&=&
\left(
1-\frac{M}{r}
\right)^2,
\qquad 
D_{(0)}(r)
=\frac{1}
        {\left(1-\frac{M}{r}\right)^2}
\qquad 
\psi_{(0)} (r)
=
\pm 
\frac{\sqrt{6} m_p M}
        {r-M},
\end{eqnarray}
and
\begin{eqnarray}
\label{bbmb1}
\omega_{(1)}(r)
=-\frac{3J}{16M^3r^4}
\left[
2M
\left(r-M\right)
\left(
r^2+2Mr-2M^2
\right)
+ r^4
{\rm ln}
\left(
1-\frac{2M}{r}
\right)
\right],
\end{eqnarray}
where we set the constant term to be zero,
so that the disformed spacetime is asymptotically flat.
In the large distance limit $r\to\infty$, 
\begin{eqnarray}
\omega_{(1)}(r)
= \frac{2J}{r^3}+\frac{6JM^2}{5r^5}+{\cal O} \left(\frac{1}{r^6}\right).
\end{eqnarray}
Thus, the leading term of ${\cal O}\left(1/r^3\right)$
agrees with Eq. \eqref{kn} with $\omega_0=0$,
and the difference from the slow rotation limit of the Kerr solution
($Q=0$ in the case of Eq. \eqref{kn})
appears at ${\cal O}\left(1/r^5\right)$.

We consider the disformal transformation \eqref{disformal2}
where $\P$ and $\Q$ are the functions only of $\phi$ or ${\cal X}$, separately.
First, 
in the case of $\P=\P_1$ and $\Q=\Q_1/\phi^2$
where $\P_1>0$ and $\Q_1>0$ are constants,
the disformed BBMB solution
with the leading order slow rotation corrections
is given by  
\begin{eqnarray}
\label{disf_bbmb3}
d{\tilde s}^2
&=&
\P_1
\left[
-
\left(1-\frac{M}{r}\right)^2
d{\tilde t}^2
+
\frac{1+\Q_1/(\P_1 r^2)}
        {\left(1-M/r\right)^2}
dr^2
+
r^2
\left(
d\theta^2
+\sin^2\theta
d\varphi^2
\right)
\right.
\nonumber\\
&+&
\left.
\frac{3J}{8M^3r^2}
\sin^2\theta
\left(
2M
\left(r-M\right)
\left(
r^2+2Mr-2M^2
\right)
+ r^4
{\rm ln}
\left(
1-\frac{2M}{r}
\right)
\right)
dt d\varphi
\right]
+{\cal O} (\epsilon^2).
\end{eqnarray}
As $r\to \infty$,
the metric solution asymptotically approaches
the flat Minkowski spacetime.
Second, 
in the case of $\P=\P_2$ and $\Q=\Q_2/{\cal X}$
where $\P_2>0$ and $\Q_2>0$ are constant,
the disformed BBMB solution with the leading order slow rotation corrections 
is given by 
\begin{eqnarray}
\label{disf_bbmb2}
d{\tilde s}^2
&=&
\P_2
\left[
-
\left(1-\frac{M}{r}\right)^2
d{\tilde t}^2
+
\frac{1+\Q_2/\P_2}
        {\left(1-M/r\right)^2}
dr^2
+
r^2
\left(
d\theta^2
+\sin^2\theta
d\varphi^2
\right)
\right.
\nonumber\\
&+&
\left.
\frac{3J}{8M^3r^2}
\sin^2\theta
\left(
2M
\left(r-M\right)
\left(
r^2+2Mr-2M^2
\right)
+ r^4
{\rm ln}
\left(
1-\frac{2M}{r}
\right)
\right)
dt d\varphi
\right]
+{\cal O} (\epsilon^2).
\end{eqnarray}
As $r\to \infty$, besides the overall conformal factor $\P_2$,
the metric approaches
$d{\tilde s}^2
\to
-d{\tilde t}^2
+
\left(1+\Q_2/\P_2\right)
dr^2
+
r^2
\left(
d\theta^2
+\sin^2\theta
d\varphi^2
\right)
=-d{\tilde t}^2
+
d{\tilde r}^2
+
{\tilde r}^2/\left(1+\Q_2/\P_2\right)
\left(
d\theta^2
+\sin^2\theta
d\varphi^2
\right)$,
where ${\tilde r}:= \sqrt{1+\Q_2/\P_2}dr$.
Thus 
the deficit solid angle  is given by $4\pi\left(\Q_2/\P_2\right)/\left(1+\Q_2/\P_2\right)$.

In both the cases of Eqs. \eqref{disf_bbmb3} and \eqref{disf_bbmb2},
up to ${\cal O} (\epsilon)$,
the degenerate black hole event horizon
is located at $r=M$.
Thus, 
at ${\cal O} (\epsilon)$,
the disformal transformation 
does not modify 
the position of the black hole event horizon,
since the character of the scalar field $\phi$ is spacelike.
This confirms 
the general properties
of the disformed black hole solutions
within the slow rotation approximation in the scalar-tensor theories 
without the shift symmetry discussed in Sec.~\ref{sec33}.
The physical properties of the disformed solution
crucially depend
on the choice of the functions $\P$ and $\Q$.
As discussed in Sec. \ref{sec33},
the circularity conditions~\eqref{circularity0}
are always satisfied even beyond the slow rotation approximation.
The scalar field in the disformed BBMB solution 
blows up at the event horizon $r=M$
as in the case of the original solution,
irrespective of $\P$ and $\Q$.

\section{Conclusions}
\label{sec9}

We have investigated the disformal transformation of stationary and axisymmetric solutions 
in a broad class of the vector-tensor and scalar-tensor theories.
We have started from the most general form of
the stationary and axisymmetric solutions 
obeying the integrability conditions \eqref{circularity2},
for which the circularity conditions \eqref{circularity0} have to be satisfied.
We have shown
that 
after the disformal transformations \eqref{disformal} and \eqref{disformal3}
in the vector-tensor theories without and with the $U(1)$ gauge symmetry,
respectively,
and the resultant stationary and axisymmetric solutions
\eqref{vector_disformed} and \eqref{stationary_gauge}
 in general do not satisfy
the circularity conditions \eqref{circularity0},
except for the cases that
Eqs. \eqref{circularity_case1} and \eqref{circularity_case2} are satisfied.

Using the fact that 
a solution in a class of the vector-tensor theories without the $U(1)$ gauge symmetry
and 
with the vanishing field strength $F_{\mu\nu}=\partial_\mu A_\nu-\partial_\nu A_\mu=0$,
after the replacement of $A_\mu\to \partial_\mu\phi$ and the integration of it, 
is uniquely mapped to
that in the corresponding class of the shift-symmetric scalar-tensor theories with the scalar field $\phi$,
we have shown that in the shift-symmetric scalar-tensor theories
the general form of the scalar field 
in the stationary and axisymmetric spacetime is given by Eq.~\eqref{scalar_Ansatz},
and 
the disformed stationary and axisymmetric metric \eqref{scalar_disformed}
does not satisfy the circularity conditions for  $q\neq 0$ or $m\neq 0$,
namely,
in the case that the scalar field depends on the time $t$ or the azimuthal angle $\varphi$.
On the other hand,
in the scalar-tensor theories without the shift symmetry,
the scalar field cannot depend on these coordinates
and then 
the disformed stationary and axisymmetric solution
always satisfies the circularity conditions.
Thus, 
any violation of the circularity conditions 
may indicate
the existence of the vector field  
or 
the scalar field with the unbroken shift symmetry,
irrespective of the detailed structure of the action
of the vector-tensor or scalar-tensor theories.

Second,
as the concrete examples,
we have considered 
the generalized Proca theory with the nonminimal coupling  to the Einstein tensor $G^{\mu\nu}A_\mu A_\nu$ \eqref{gp}.
As the particular case, 
a solution with the vanishing field strength $F_{\mu\nu}=0$,
after the replacement of $A_\mu \to \partial_\mu \phi$ and the integration of it,
could be mapped to that in the shift-symmetric scalar-tensor theory
with the nonminimal derivative coupling to the Einstein tensor $G^{\mu\nu}\partial_\mu \phi \partial_\nu\phi$ \eqref{st}.
We have started from 
the reference static and spherically symmetric solutions in these theories
and then derived the leading order slow rotation corrections 
to the metric functions and the vector or scalar field.
Up to the leading order slow rotation corrections to the reference solutions,
in the generalized Proca theory \eqref{gp}
we have obtained the nonzero values of the $(t,\varphi)$ component of the metric
and the azimuthal component of the vector field,
while 
in the shift-symmetric scalar-tensor theory \eqref{st}
the scalar field remains the same as that in the static and spherically symmetric solution \eqref{qt}.
The disformal transformations of Eqs. \eqref{disformal} and \eqref{disformal3}
induced the $(t,r)$ and $(r,\varphi)$ components of the metric
and 
the $\varphi$ component of the vector field,
up to the first order of the slow rotation approximation,
respectively,
where the former could be eliminated by the redefinition of the timelike coordinate \eqref{tildet}.

After the disformal transformation \eqref{disformal},
the generalized Proca theory 
is mapped to a class of the extended vector-tensor theories
without the Ostrogradsky ghosts \cite{Kimura:2016rzw}.
At  the zeroth order of the slow rotation approximation,
via the disformal transformation 
the stealth Schwarzschild solution
obtained for any value of the coupling parameter
and
the charged stealth Schwarzschild solution
obtained for the particular value of the coupling constant
are mapped 
to 
the stealth Schwarzschild solution
and 
the RN solution with the maximally two horizons,
respectively,
where 
the position of the black hole event horizon
was modified.
In the near-horizon limit, 
the vector field expressed
in terms of the new time coordinate defined in the disformed frame \eqref{tildet}
is proportional to the null coordinates
which are 
regular 
at the future and past event horizons,
respectively.
Thus,
the positive and negative branch solutions 
of the vector field
are regular at the future and past event horizons,
respectively,
and 
the disformal transformation does not modify the essential properties
of the vector field from the original one~\cite{Minamitsuji:2016ydr}.
The similar properties could also be verified
for 
the Schwarzschild-(anti-) de Sitter solutions and 
the charged Schwarzschuld- (anti-) de Sitter solutions,
where 
the vector field written in terms of the time coordinate 
defined in the disformed frame 
is regular
at either the future or past event horizon
(and
at either the past or future cosmological horizon,
if the cosmological horizons exist).

In the case of the disformal transformation
with the $U(1)$ gauge invariance \eqref{disformal3},
we showed
at the leading order of the slow rotation approximation, 
the disformal transformation does not modify 
the position of the black hole event horizon,
in contrast to the case of the disformal transformation
without the $U(1)$ gauge symmetry \eqref{disformal}.
We have explicitly confirmed this 
for the reference RN solution.
We also obtained the disformal transformation
of the Kerr-Newman solution
for an arbitrary dimensionless spin parameter.
We then showed that
the position of the black hole event horizon is not modified  
by the disformal transformation,
although the too large disformal contribution
leads to singularities at the finite values of the radial coordinate.
Moreover, 
we saw
that 
in the large distance limit
the leading disformal contribution 
appears at the frame-dragging component
of the disformed Kerr-Newman metric,
but is suppressed by the inverse powers of the radial coordinate.

Finally,
as the example of the scalar-tensor theories without the shift symmetry,
we have considered the disformal transformation
of the BBMB solution \eqref{bbmb}
with the leading order slow rotation corrections \eqref{bbmb1}.
We confirmed that 
the disformed BBMB solution with the slow rotation corrections 
satisfies the circularity conditions,
which is the general property of 
the disformed stationary and axisymmetric solutions  
in the scalar-tensor theories without the shift symmetry.

Before closing this paper, 
we would like to emphasize again
that 
as long as the disformal transformation is invertible,
there should be always one-to-one correspondence
between the original theory
and the theory obtained after the disformal transformation,
and they should share the equivalent solution space,
in the case that the contribution of matter is absent
\cite{Zumalacarregui:2013pma,Achour:2016rkg,Kimura:2016rzw,Gumrukcuoglu:2019ebp}.
Even in the case that the theory obtained via the disformal transformation
is apparently higher derivative
while the original theory gives rise to the second-order Euler-Lagrange equations,
the higher-derivative Euler-Lagrange equations in the disformed theory
should be degenerate
and reduce to the second-order system 
after the suitable redefinition of the variables.
In such a case, 
the two theories
should be just the two different mathematical descriptions of the same theory
and have no physical difference at all.
Hence, 
the order of the disformal transformation
and the variation of the action 
should be able to be exchanged,
and 
the disformed solutions
should always satisfy the Euler-Lagrange equations 
obtained 
by varying the action of the theory 
derived via the disformal transformation.
We note that
since we assumed that
matter is minimally coupled to 
the metric in each frame,
we have regarded the two theories as the physically independent ones;
e.g., 
the trajectories of particles freely falling into a black hole are different,
although we have neglected the contribution of matter 
in deriving the solutions.

In Appendix \ref{app3},
we have explicitly demonstrated that
the disformal transformation of the asymptotically flat solutions
discussed in Sec. \ref{sec4}
satisfies
the Euler-Lagrange equations
of the extended vector-tensor theories 
derived via the disformal transformation
of the generalized Proca theories \eqref{gp}.
The same conclusions should be 
obtained 
for the vector disformal transformation with the $U(1)$ gauge symmetry 
and 
the scalar disformal transformation without the shift symmetry,
as long as 
these disformal transformations are invertible.

On the other hand,
our main purpose 
was to show 
to which a new metric 
a disformal transformation
maps 
a general stationary and axisymmetric solution
satisfying the circularity conditions,
assuming that the transformation is invertible.
The explicit examples shown  in Secs. \ref{sec4}-\ref{sec8}
were just to confirm the general properties of the disformed solution
mentioned in Sec. \ref{sec3}.
Our main result
is that
if the future observations could measure 
any violation of the circularity conditions, 
it would indicate the existence 
of the vector field without the $U(1)$ symmetry
or
the scalar field with the shift symmetry
in the vicinity of the black hole,
irrespective of the detailed structure of the action
of the scalar-tensor or vector-tensor theories
obtained via the disformal transformation.
Therefore,
in this paper,
instead of the derivation of the explicit action 
of the scalar-tensor and vector-tensor theories
obtained via the disformal transformations,
we have focused on the disformal transformations
at the level of the solutions.

There are several issues left for future studies.
First,
in the case that the disformed stationary and axisymmetric solutions do not satisfy the circularity conditions,
it will be important to see the consequences on astrophysical phenomena 
and the implications on observations such as black hole shadows and gravitational waves
(see Refs. \cite{Psaltis:2020lvx,Psaltis:2018xkc}
for the constraints on the deviations from the Kerr solution 
with the latest data of black hole shadows by the Event Horizon Telescope).
Second,
it will also be important
to explore more explicit stationary and axisymmetric solutions
with the nontrivial field profiles of the scalar or vector field
beyond the slow rotation
and 
the properties of their disformed solutions.
We hope to come back to these issues in our future publications.

\acknowledgments{
M.M.~was supported by the Portuguese national fund 
through the Funda\c{c}\~{a}o para a Ci\^encia e a Tecnologia
in the scope of the framework of the Decree-Law 57/2016 
of August 29, changed by Law 57/2017 of July 19,
and the CENTRA through the Project~No.~UIDB/00099/2020.
}

\appendix

\section{The disformed Kerr solution}
\label{app1}

In this appendix, we review the disformed Kerr solution
obtained from the stealth Kerr solution in the DHOST theories \cite{dhostbh8,dhostbh9}.

In the coordinate system \eqref{stationary},
the Kerr solution in the Boyer–Lindquist form
is described by Eqs. \eqref{kerr_newman1}--\eqref{kerr_newman3}
with $Q=0$ and $\Delta(r) := \Delta_{Q=0} (r)$.
We assume that the scalar field does not depend on $\theta$ and $\varphi$,
\begin{eqnarray}
\phi= qt + \psi (r),
\end{eqnarray}
and consider the scalar disformal transformation \eqref{disformal2}
with 
$\P=1$ and $\Q=-\frac{\D_0}{q^2}={\rm const}$.
The disformed solution is then given by 
\begin{eqnarray}
\label{disform_int}
d{\tilde s}^2
&=&
- 
\left(A+\D_0 \right)dt^2
+B   d\varphi^2
+2 C dt d\varphi
+
\left(
D
-\frac{\D_0 \left(\partial_r\psi\right)^2}{q^2} 
\right)
dr^2
+ 
E
d\theta^2
-\frac{2\D_0 \left(\partial_r\psi \right)}{q} 
dt dr.
\end{eqnarray}
In a subclass of the Class-Ia DHOST theories \cite{Langlois:2015cwa,Achour:2016rkg,BenAchour:2016fzp,Langlois:2018dxi}, 
the stealth Kerr solution was obtained \cite{dhostbh6,dhostbh7,dhostbh8,dhostbh9},
where $\psi (r)$ is given by 
\begin{eqnarray}
\psi (r)
=
q
\int 
dr
\frac{\sqrt{2M r \left(r^2+\frac{J^2}{M^2}\right)}}
        {\Delta},
\end{eqnarray}
for which the kinetic term of the scalar field \eqref{kin} is given by 
${\cal X}=-q^2$.

The disformed metric \eqref{disform_int}
then reduces to 
\begin{eqnarray}
d{\tilde s}^2
&=&
- 
\left(
1+ \D_0-\frac{2Mr}{\rho^2}
\right)dt^2
+
\frac{\sin^2\theta}{\rho^2}
\left[
\left(r^2+\frac{J^2}{M^2}\right)^2
-\frac{J^2}{M^2}
\Delta \sin^2\theta
\right]
d\varphi^2
 -\frac{4J r  \sin^2\theta}{\rho^2}
 dt d\varphi
\nonumber\\
&+&
\left(
\frac{\rho^2}{\Delta}
-\frac{2\D_0 M r (r^2+\frac{J^2}{M^2})}{\Delta^2}
\right)
dr^2
+ 
\rho^2
d\theta^2
-\frac{2\sqrt{2} \D_0\sqrt{M r \left(r^2+\frac{J^2}{M^2}\right)}}{\Delta}
 dt dr.
\end{eqnarray}
Introducing the rescaled time coordinate and mass, 
${\tilde t}=\sqrt{1+\D_0} t$ and ${\tilde M}=M/(1+\D_0)$,
respectively,
\begin{eqnarray}
d{\tilde s}^2
&=&
- 
\left(
1-\frac{2{\tilde M}r}{\rho^2}
\right)
d{\tilde t}^2
+
\frac{\sin^2\theta}{\rho^2}
\left[
\left(r^2+\frac{J^2}{M^2}\right)^2
-\frac{J^2}{M^2}
\Delta \sin^2\theta
\right]
d\varphi^2
 -\frac{4J r  \sin^2\theta}
             {\sqrt{1+\D_0} \rho^2}
 d{\tilde t} d\varphi
\nonumber\\
&+&
\left(
\frac{\rho^2}{\Delta}
-\frac{2\D_0 (1+\D_0) \tilde{M} r \left(r^2+\frac{J^2}{M^2}\right)}{\Delta^2}
\right)
dr^2
+ 
\rho^2
d\theta^2
-\frac{2\sqrt{2} \D_0\sqrt{{\tilde M} r \left(r^2+\frac{J^2} {M^2}\right)}}{\Delta}
 d{\tilde t} dr,
\end{eqnarray}
which does not reduce to the circular form of Eq. \eqref{stationary}
because of the nonzero $(\tilde{t},r)$ component of ${\tilde g}_{\mu\nu}$.
The physical properties of the disformed Kerr metric,
e.g., the position of the black hole event horizon,
were discussed in Refs. \cite{dhostbh8,dhostbh9}.

\section{The solutions of the equations of motion in the disformed generalized Proca theory}
\label{app3}

As long as the disformal transformation is invertible,
there should be one-to-one correspondence
between the original theory 
and the theory obtained after the disformal transformation,
and they should share the equivalent solution space,
in the case that the contribution of matter is absent.
Even in the case that the theory obtained via the disformal transformation
is apparently higher derivative
while the original theory gives rise to the second-order Euler-Lagrange equations,
the higher-derivative Euler-Lagrange equations in the disformed theory
should be degenerate
and reduce to the second-order system 
after the suitable redefinition of the variables.
The two theories
should be the two different mathematical descriptions of the equivalent theory,
as long as the contribution of matter is absent.
Thus, 
the order of the disformal transformation
and the variation of the action 
should be able to be exchanged,
and 
the disformed solutions
should satisfy the Euler-Lagrange equations 
obtained 
by varying the action of the theory 
derived via the disformal transformation.

In this appendix,
we demonstrate
that 
the disformed asymptotically flat solutions
discussed in Sec. \ref{sec4}
satisfy the Euler-Lagrange equations  
obtained
by varying the action of the extended vector-tensor theory
derived via the disformal transformation \eqref{disformal}
of the generalized Proca theory \eqref{gp} with $m=\Lambda=0$,
as long as the disformal transformation is invertible.
Here, 
we employ the conventions employed in Ref. \cite{Kimura:2016rzw},
where
the correspondence with 
the conventions in the main text 
(see Eqs. \eqref{disformal} and \eqref{disform_el})
is given by 
\begin{eqnarray}
{\tilde g}_{\mu\nu}\to g_{\mu\nu},
\qquad  
{g}_{\mu\nu}\to {\bar g}_{\mu\nu},
\qquad 
\P\to 1,
\qquad 
\Q\to -\Gamma.
\end{eqnarray}

We consider the generalized Proca theory \eqref{gp} 
with $m=\Lambda=0$,
rewritten in terms of the new conventions as
\begin{eqnarray}
\label{gp0}
S
=\int d^4 x
\sqrt{-{\bar g}}
\left[
\frac{m_p^2}{2}
{\bar R}
-\frac{1}{4}\bar{\F}
+\beta {\bar G}^{\mu\nu}
{A}_\mu { A}_\nu
\right],
\end{eqnarray}
where
we attach ``bar''
to the quantities defined 
in the frame before the disformal transformation (see Eq. \eqref{disform_el}).
Following the general formulation of the generalized Proca theory \cite{Heisenberg:2014rta,DeFelice:2016cri},
the theory \eqref{gp0} is given by the choice of
\begin{eqnarray}
{\bar G}_4({\bar {\cal Y}})= \frac{m_p^2}{2}-\frac{\beta}{2} \bar{\cal Y},
\end{eqnarray}
and the others to be zero,
where $\bar{\cal Y}={\bar g}^{\mu\nu}A_\mu A_\nu$.
Moreover, 
following the general form of the extended vector-tensor theories \cite{Kimura:2016rzw},
the theory \eqref{gp0}
can be described by 
\begin{eqnarray}
S=
\int  d^4x\sqrt{-\bar{g}}
\left[
{\bar f}(\bar{\cal Y})\bar{R}
+{\bar C}^{\mu\nu\rho\sigma}
{\bar \nabla}_\mu A_\nu 
{\bar \nabla}_\rho A_\sigma
\right],
\end{eqnarray}
with
\begin{eqnarray}
\label{defc4}
C^{\mu\nu\alpha\beta}
&:=&
\alpha_1 ({\cal Y})
g^{\mu (\rho} g^{\sigma)\nu}
+
\alpha_2 ({\cal Y})
g^{\mu\nu}g^{\rho\sigma}
+
\frac{\alpha_3 ({\cal Y})}{2}
\left(
   A^\mu A^\nu g^{\rho\sigma}
+ A^\rho A^\sigma g^{\mu\nu} 
\right)
+
\frac{\alpha_4 ({\cal Y})}{2}
\left(
   A^\mu A^{(\rho} g^{\sigma)\nu}
+
   A^\nu A^{(\rho} g^{\sigma)\mu}
\right)
\nonumber\\
&+&
\alpha_5({\cal Y})
A^\mu A^\nu A^\rho A^\sigma
+
\alpha_6 ({\cal Y})
g^{\mu[\rho} g^{\sigma]\nu}
+
\frac{\alpha_7 ({\cal Y})}{2}
\left(
     A^\mu A^{[\rho} g^{\sigma]\nu}
  - A^\nu A^{[\rho} g^{\sigma]\mu}
\right)
+
\frac{\alpha_8({\cal Y})}{4}
\left(
  A^\mu A^{\rho} g^{\sigma\nu}
-A^\nu A^{\sigma} g^{\mu\rho}
\right),
\nonumber\\
\end{eqnarray}
and ${\bar C}^{\mu\nu\rho\sigma}$
with the replacement to 
the barred quantities
($g^{\mu\nu}\to {\bar g}^{\mu\nu}$, $A^\mu:=g^{\mu\nu}A_\nu \to {\bar A}^\mu:={\bar g}^{\mu\nu}A_\nu$, and ${\cal Y}\to \bar{\cal Y}$),
where
the coefficients in ${\bar C}^{\mu\nu\rho\sigma}$
are given by 
\begin{eqnarray}
&&
{\bar f}
={\bar G}_{4} (\bar{\cal Y}),
\qquad 
{\bar\alpha}_1
=
-{\bar\alpha}_2
=
2{\bar G}_{4,\bar{\cal Y}},
\qquad
{\bar \alpha}_6
=-2{\bar G}_{4,\bar{\cal Y}}-1,
\qquad
{\bar \alpha}_3
=
{\bar \alpha}_4
=
{\bar \alpha}_5
=
{\bar \alpha}_7
=
{\bar \alpha}_8
=0.
\end{eqnarray}
We consider the disformal transformation
from the frame ${\bar g}_{\mu\nu}$
to the new frame $g_{\mu\nu}$, given by 
\begin{eqnarray}
\label{disform_el}
g_{\mu\nu}
=
{\bar g}_{\mu\nu}
-\Gamma (\bar{\cal Y})A_\mu A_\nu,
\end{eqnarray}
which maps the generalized Proca theory written in terms of ${\bar g}_{\mu\nu}$
to a class of the extended vector-tensor theories
without the Ostrogradsky ghosts
 \cite{Kimura:2016rzw},
given by 
\begin{eqnarray}
\label{evt}
S=
\int  d^4x\sqrt{-g}
\left[
f({\cal Y})R
+C^{\mu\nu\rho\sigma}
\nabla_\mu A_\nu 
\nabla_\rho A_\sigma
\right],
\end{eqnarray}
where $C^{\mu\nu\rho\sigma}$ is defined in Eq. \eqref{defc4}
with the coefficients $\alpha_i$
related to ${\bar \alpha}_i$ ($i=1,2,\cdots, 8$) by 
\begin{eqnarray}
f&=& 
{\bar G}_4 (\bar{\cal Y})
\sqrt{\Gamma {\cal Y}+1},
\\
\alpha_1
&=&-\alpha_2
=
\frac{2{\bar G}_{4,\bar{\cal Y}}+\Gamma {\bar G}_4 (\Gamma {\cal Y}+1)}
       {(\Gamma {\cal Y}+1)^{3/2}},
\\
\alpha_3
&=&
-\alpha_4
=
-\frac{2\Gamma_{\cal Y}
            \left({\bar G}_4 (\Gamma {\cal Y}+1)-2 Y {\bar G}_{4,\bar{\cal Y}}\right)}
          {(\Gamma {\cal Y}+1)^{3/2}},
\\
\alpha_6
&=&
{\bar \alpha}_6 
\sqrt{\Gamma {\cal Y}+1}
-\Gamma {\bar G}_4
\sqrt{\Gamma {\cal Y}+1},
\\
\alpha_7
&=&
\frac{2 \left(
           2(\Gamma+{\cal Y}\Gamma_{\cal Y}){\bar G}_{4,\bar{\cal Y}}
         +{\bar G}_4 (\Gamma {\cal Y}+1) (\Gamma^2-\Gamma_{\cal Y})
          \right)}
        {(\Gamma {\cal Y}+1)^{3/2}}
-
\frac{2\Gamma {\bar\alpha}_6}{\sqrt{\Gamma {\cal Y}+1}},
\\
\alpha_5
&=&
\alpha_8
=0.
\end{eqnarray}
The inverse transformation to Eq. \eqref{disform_el} is given by 
\begin{eqnarray}
g^{\mu\nu}
&=&
{\bar g}^{\mu\nu}
+
\frac{\Gamma}
        {1- \Gamma\bar{\cal Y}}
\bar{A}^\mu 
\bar{A}^\nu,
\end{eqnarray}
where
${\cal Y}=\frac{{\bar{\cal Y}}}{1- \Gamma {\bar {\cal Y}}}$,
respectively.
For simplicity, we assume the constant disformal transformation 
\begin{eqnarray}
\Gamma({\cal Y})=\Gamma_0,
\end{eqnarray}
and the invertibility of the transformation $\Gamma_0 \bar{\cal Y}\neq 1$.

We consider the static and spherically symmetric Ansatz
for the metric and the vector field 
\begin{eqnarray}
\label{Ansatz}
ds^2
&=&
g_{\mu\nu}dx^\mu dx^\nu
=
-f(r)dt^2
+\frac{dr^2}{h(r)}
+r^2
\left(
  d\theta^2
+\sin^2\theta d\varphi^2
\right),
\\
\label{Ansatz2}
A_\mu
&=&
A_0(r)dt
+
A_1(r)dr.
\end{eqnarray}
Plugging the Ansatz \eqref{Ansatz}-\eqref{Ansatz2} into the action \eqref{evt}
and 
varying it with respect to $f$, $h$, $A_0$, and $A_1$,
we obtain a set of the four independent Euler-Lagrangian equations.

\subsection{On the disformal transformation of the stealth Schwarzschild solution}

For our purpose, 
we impose the further simplification of $A_0(r)=q=const$ and $\bar{\cal Y}=-q^2$
(and hence ${\cal Y}=-\frac{q^2}{1+ \Gamma_0 q^2}$),
and hence
\begin{eqnarray}
\label{step_a1}
A_1(r)
=
\pm 
q\frac{\sqrt{1+q^2{\Gamma}_0 -f}}
         {\sqrt{\left(1+q^2\Gamma_0\right)fh}}.
\end{eqnarray}
We also assume the proportionality of  $f$ and $h$, $f=\alpha h$
with $\alpha$ being a constant.
With these assumptions, 
the Euler-Lagrange equations with respect to $A_0$ and $A_1$
reduce to the first-order  degenerate equation of $h$,
\begin{eqnarray}
h'+\frac{h}{r}
+
\frac{\alpha\left(m_p^2\Gamma_0-\beta (2+q^2\Gamma_0)\right)
-q^2\Gamma_0 (1+q^2\Gamma_0)
\left(
-m_p^2\Gamma_0 +\beta (4+3q^2\Gamma_0)
 \right) }
        {r\alpha (1+q^2\Gamma_0)\left(-m_p^2\Gamma_0+\beta (2+3q^2\Gamma_0)\right)}
=0,
\end{eqnarray}
whose solution is given by the form of 
\begin{eqnarray}
\label{hc0}
h (r)=
-\frac{\alpha\left(m_p^2\Gamma_0-\beta (2+q^2\Gamma_0)\right)
-q^2\Gamma_0 (1+q^2\Gamma_0)
\left(
-m_p^2\Gamma_0 +\beta (4+3q^2\Gamma_0)
 \right) }
        {\alpha (1+q^2\Gamma_0)\left(-m_p^2\Gamma_0+\beta (2+3q^2\Gamma_0)\right)}
-\frac{2{\tilde M}}{r},
\end{eqnarray}
where ${\tilde  M}$ denotes the integration constant.
In order to avoid the appearance of the deficit solid angle,
we impose the constant part of $h$ to be unity,
which fixes
\begin{eqnarray}
\label{hc1}
\alpha= 1+q^2\Gamma_0.
\end{eqnarray}
Eq. \eqref{hc0} with Eq. \eqref{hc1}
also satisfies the Euler-Lagrange equations for $f$ and $h$.
Thus,  all the Euler-Lagrange equations are satisfied.
In this way, 
we finally obtain the solution for the metric and the vector field, respectively,
given by 
\begin{eqnarray}
\label{disf_stealth1}
ds^2
=
g_{\mu\nu} dx^\mu dx^\nu
=
-
\left(1+q^2 \Gamma_0 \right) 
\left(
1-\frac{2{\tilde M}}{r}
\right)dt^2
+\frac{dr^2}
         {1-\frac{2{\tilde M}}{r}}
+r^2 (d\theta^2+\sin^2\theta d\varphi^2),
\end{eqnarray}
and 
\begin{eqnarray}
\label{disf_stealth2}
A_0=q,
\qquad 
A_1
=
\pm
\frac{q}
       {1-\frac{2{\tilde M}}{r}}
\sqrt{
\frac{2{\tilde M}}
       {r\left(1+q^2\Gamma_0\right)}}.
\end{eqnarray}
The solution \eqref{disf_stealth1}-\eqref{disf_stealth2}
is equivalent
to that obtained after the disformed stealth Schwarzschild solution given 
by Eqs. \eqref{newv0}-\eqref{newv02}
obtained in Sec. \ref{sec41},
after the replacement of
${\tilde g}_{\mu\nu}\to {g}_{\mu\nu}$,
${\tilde t}\to t$,
${g}_{\mu\nu}\to {\bar g}_{\mu\nu}$,
$\P_{(0)}\to 1$,
and 
$\Q_{(0)}\to -\Gamma_0$.

\subsection{On the disformal transformation of the charged stealth Schwarzschild solution}

Following the same procedure, 
we can also confirm 
that  
the disformal transformation of the charged stealth Schwarzschild black hole solution
discussed in Sec. \ref{sec42}
is also the solution of the extended vector-tensor theory
\eqref{evt}
derived via the disformal transformation.

We impose the simplification of $A_0(r)=q+Q/r$ and $\bar{\cal Y}=-q^2$
with $q$ and $Q$ being constants,
and hence
\begin{eqnarray}
\label{step_a2}
A_1(r)
=
\pm 
\frac{\sqrt{
 \left(Q+qr\right)^2\left(1+q^2{\Gamma}_0 \right)-q^2 r^2f}}
         {r\sqrt{\left(1+q^2\Gamma_0\right)fh}}.
\end{eqnarray}
Again, we assume the proportionality of  $f$ and $h$, $f=\alpha h$
with $\alpha$ being a constant.
With these assumptions, 
the Euler-Lagrange equations with respect to $A_0$ and $A_1$
reduce to the first-order  degenerate equation of $h$,
whose solution is given by
\begin{eqnarray}
\label{hc02}
h (r)
&=&
-\frac{\alpha\left(m_p^2\Gamma_0-\beta (2+q^2\Gamma_0)\right)
-q^2\Gamma_0 (1+q^2\Gamma_0)
\left(
-m_p^2\Gamma_0 +\beta (4+3q^2\Gamma_0)
 \right) }
        {\alpha (1+q^2\Gamma_0)\left(-m_p^2\Gamma_0+\beta (2+3q^2\Gamma_0)\right)}
-\frac{2{\tilde M}}{r}
\nonumber\\
&+&
\frac{
Q^2\Gamma_0 \left(-1-2m_p^2\Gamma_0 +\beta (8+6q^2\Gamma_0)\right)
}
{2\alpha\left( -m_p^2 \Gamma_0 +\beta (2+3q^2\Gamma_0)\right) 
r^2},
\end{eqnarray}
where ${\tilde  M}$ denotes the integration constant.
In order to avoid the appearance of the deficit solid angle,
we impose the constant part of $h$ to be unity,
which fixes $\alpha$ as Eq. \eqref{hc1}.

Substituting Eq. \eqref{hc02} with Eq. \eqref{hc1}
into the Euler-Lagrange equations for $f$ and $h$
results in the degenerate algebraic equation
\begin{eqnarray}
0&=&
\left(\beta-\frac{1}{4}\right)
\nonumber\\
&\times&
\left[
4r\left(r-2{\tilde M}\right)\beta
+
\left(
Q^2(-1+8\beta)-2r (r-2{\tilde M}) (m_p^2-5q^2\beta)
\right)
\Gamma_0
-2\left(Q^2+q^2 r(r-2{\tilde M})\right) (m_p^2-3q^2\beta)\Gamma_0^2
\right].
\nonumber\\
\end{eqnarray}
Here, we choose the branch 
\begin{eqnarray}
\beta=\frac{1}{4}.
\end{eqnarray}
Then, all the Euler-Lagrange equations are satisfied,
and the solution for the metric and vector field
is explicitly given by 
\begin{eqnarray}
\label{disf_stealth12}
ds^2
=
g_{\mu\nu} dx^\mu dx^\nu
&=&
-
\left(1+q^2 \Gamma_0 \right) 
\left(
1-\frac{2{\tilde M}}{r} +\frac{{Q}^2\Gamma_0}{\left(1+\Gamma_0 q^2\right)r^2}
\right)dt^2
+\frac{dr^2}
         {1-\frac{2{\tilde M}}{r} +\frac{{Q}^2\Gamma_0}{\left(1+\Gamma_0 q^2\right)r^2}}
\nonumber\\
&+&
r^2 (d\theta^2+\sin^2\theta d\varphi^2),
\end{eqnarray}
and 
\begin{eqnarray}
A_0(r)=q+\frac{Q}{r},
\qquad 
A_1(r)
=
\pm 
\frac{1}
       {1-\frac{2{\tilde M}}{r} +\frac{{Q}^2\Gamma_0}{\left(1+\Gamma_0 q^2\right)r^2}}
\frac{\sqrt{{Q}^2+2q r\left(\tilde{M} q+ {Q} \right)\left(1+\Gamma_0 q^2\right) }}
        {r\left(1+\Gamma_0q^2 \right)},
\end{eqnarray}
which agree with Eq. \eqref{newv03}-\eqref{newv04}
after the replacement of
${\tilde g}_{\mu\nu}\to {g}_{\mu\nu}$,
${\tilde t}\to t$,
${g}_{\mu\nu}\to {\bar g}_{\mu\nu}$,
$\P_{(0)}\to 1$,
and 
$\Q_{(0)}\to -\Gamma_0$.

These examples
demonstrated that
the disformed solutions
satisfy 
the Euler-Lagrange equations 
obtained
by varying the the action
of the vector-tensor theories
derived 
via the disformal transformation of the original theory,
as long as the disformal transformation is invertible.
The same conclusion
should be obtained 
for the scalar disformal transformation
and
the vector disformal transformation with the $U(1)$ gauge symmetry,
given by 
Eqs. \eqref{disformal2} and \eqref{disformal3},
respectively,
as long as these disformal transformations are invertible.

\section{The stationary and axisymmetric solutions and circularity conditions}
\label{app2}

In this appendix,
we review 
the conditions under which 
the most general stationary and axisymmetric spacetime
Eq. \eqref{stationary0}
with the two commuting Killing vector fields
$\xi^\mu=(\partial/\partial t)^\mu$
and 
$\sigma^\mu=(\partial/\partial \varphi)^\mu$
reduce to the form of Eq. \eqref{stationary_ast}
with the integrable two-surfaces orthogonal to $\xi^\mu$ and $\sigma^\mu$.

There is the theorem stating that
the two-surfaces orthogonal to $\xi^\mu$ and $\sigma^\mu$
are integrable,
if 
$\xi^\mu$ and $\sigma^\mu$ satisfy (see e.g., Refs. \cite{circ1,circ2,waldbook})

\begin{description}

\item[(i)]
Each of 
$\xi_{[\mu} \sigma_{\nu} \nabla_{\alpha} \xi_{\beta] }$
and 
$\xi_{[\mu} \sigma_{\nu} \nabla_{\alpha} \sigma_{\beta]}$
vanishes at least at a point in the spacetime.

\item[(ii)]
The circularity conditions \eqref{circularity0},
which we recapitulate here 
\begin{eqnarray}
\xi^{\mu} R_{\mu} {}^{[\nu} \xi^{\alpha} \sigma^{\beta]}=0,
\qquad 
\sigma^{\mu} R_{\mu} {}^{[\nu} \xi^{\alpha}  \sigma^{\beta]}=0,
\end{eqnarray}
are satisfied.

\end{description}

Frobenius's theorem
states that 
the necessary and sufficient conditions 
for integrability are given by Eq. \eqref{circularity2},
which we recapitulate here
\begin{eqnarray}
\sigma_{[\mu} \xi_{\nu}\nabla_{\alpha} \xi_{\beta]}=0,
\qquad 
\sigma_{[\mu} \xi_{\nu}\nabla_{\alpha} \sigma_{\beta]}=0,
\end{eqnarray}
or equivalently,
\begin{eqnarray}
\label{dual2}
\sigma^\mu \omega_\mu=0,
\qquad 
\xi^\mu \rho_\mu=0,
\end{eqnarray}
where 
$\omega_\alpha:=\epsilon_{\alpha\beta\mu\nu}  \xi^\beta \nabla^\mu \xi^\nu$,
$\rho_\alpha:=\epsilon_{\alpha\beta\mu\nu}  \sigma^\beta \nabla^\mu \sigma^\nu$,
and 
$\epsilon_{\alpha\beta\mu\nu}$ denotes the volume element.

Here, we confirm that the assumptions (i) and (ii) are necessary for the integrability conditions \eqref{circularity2}.
First, the assumption (i) ensures that  
each of 
the products 
$\sigma^\mu \omega_\mu$
and 
$\xi^\mu \rho_\mu$
vanishes at least at a point in the spacetime (e.g., on the symmetry axis).
They vanish in the entire spacetime,
if
\begin{eqnarray}
\label{derivatives}
\nabla_\nu (\sigma^\mu \omega_\mu)
=
\pounds_\sigma  \omega_\nu
+2\sigma^\mu \nabla_{[\nu} \omega_{\mu]}=0,
\qquad
\nabla_\nu (\xi^\mu \rho_\mu)
=
\pounds_\xi  \rho_\nu
+2\xi^\mu \nabla_{[\nu} \rho_{\mu]}=0.
\end{eqnarray}
Since $[\xi,\psi]^\mu=0$,
the diffeomorphism group $\chi_\varphi$ generated by $\sigma^\mu$
leaves $\xi^\mu$ invariant,
and similarly,
the diffeomorphism group $\chi_t$ generated by $\xi^\mu$
does $\sigma^\mu$ invariant.
Thus, any tensor constructed by $\xi^\mu$
and their derivatives 
is invariant under $\chi_\varphi$,
and hence $\omega_\mu$ is invariant under $\chi_\psi$,
i.e., $\pounds_\sigma  \omega_\nu=0$.
Similarly,
any tensor constructed by $\sigma^\mu$
and their derivatives 
is also invariant under $\chi_t$,
and hence $\rho^\mu$ is invariant under $\chi_t$,
i.e., $\pounds_\xi  \rho_\nu=0$.
Thus, Eq. \eqref{derivatives} reduces to 
\begin{eqnarray}
\nabla_\nu (\sigma^\mu \omega_\mu)
=
2\sigma^\mu \nabla_{[\nu} \omega_{\mu]},
\qquad
\nabla_\nu (\xi^\mu \rho_\mu)
=
2\xi^\mu \nabla_{[\nu} \rho_{\mu]}.
\end{eqnarray}
After some algebra, we obtain 
\begin{eqnarray}
\nabla_{[\nu} \omega_{\mu]}
=-\epsilon_{\nu\mu\alpha\beta} \xi^\alpha R^{\beta}_{\kappa} \xi^\kappa,
\qquad  
\nabla_{[\nu} \rho_{\mu]}
=-\epsilon_{\nu\mu\alpha\beta} \sigma^\alpha R^{\beta}_{\kappa} \sigma^\kappa.
\end{eqnarray}
Then, 
under the assumption (ii), i.e., Eq. \eqref{circularity0},
\begin{eqnarray}
\nabla_\nu (\sigma^\mu \omega_\mu)
=
-2\epsilon_{\nu\mu\alpha\beta} \sigma^\mu \xi^\alpha R^{\beta}{}_{\kappa} \xi^\kappa
=0,
\qquad
\nabla_\nu (\xi^\mu \rho_\mu)
=
-2 \epsilon_{\nu\mu\alpha\beta} \xi^\mu\sigma^\alpha R^{\beta}{}_{\kappa} \sigma^\kappa
=0,
\end{eqnarray}
which ensures that Eq. \eqref{dual2} holds in the entire spacetime
and hence the existence of the integrable orthogonal two-surfaces.
This
ensures that 
the circularity conditions \eqref{circularity0} are necessary
for the integrability conditions \eqref{circularity2}. 


For the general stationary and axisymmetric spacetime metric \eqref{stationary0},
the possible terms which violate the circularity conditions \eqref{circularity0}
are given by 
the components of the Ricci tensor 
$R_t{}^r$,
$R_t{}^\theta$,
$R_\varphi {}^r$,
and
$R_\varphi {}^\theta$,
and 
all these components of the Ricci tensor vanish,
if the metric form is given by Eq. \eqref{stationary_ast} or \eqref{stationary}.
In general relativity,
the condition \eqref{circularity0}
holds 
in the vacuum spacetime where $R_{\mu\nu}=\lambda g_{\mu\nu}$ with $\lambda$ being constant,
or 
in the spacetime filled with the perfect fluids
with the four-velocities
spanned by the two Killing vector fields $\xi^\mu$ and $\sigma^\mu$,
$u^\mu= c_1 \xi^\mu+ c_2 \sigma^\mu$
with $c_1$ and $c_2$ being constants.
On the other hand,
in the other gravitational theories
with the nontrivial scalar or vector field degrees of freedom,
the components $R_{\mu\nu}$ does not necessarily vanish
even in the vacuum spacetimes
(or in the spacetime with the perfect fluids)
and the circularity conditions \eqref{circularity0} may not hold.

\bibliography{disformal_refs}
\end{document}